\documentclass[11pt]{article}
\pdfoutput=1 
\usepackage{amssymb}
\usepackage{amsmath}
\usepackage{amstext}
\usepackage{graphicx,epsfig}
\usepackage{epsfig}
\usepackage{verbatim} 
\usepackage{fancybox}
\usepackage{color}
\usepackage{ulem}
\usepackage{enumitem}
\usepackage{subfigure}
\usepackage{bbm}
\usepackage{parskip}
\usepackage{dsfont}
\usepackage[numbers,sort&compress]{natbib}
\usepackage{ytableau}
\usepackage{framed}
\usepackage{float}

\usepackage{tikz}
\usetikzlibrary{decorations}
\pgfdeclaredecoration{complete sines}{initial}
{
    \state{initial}[
        width=+0pt,
        next state=upsine,
        persistent precomputation={\pgfmathsetmacro\matchinglength{
            \pgfdecoratedinputsegmentlength / int(\pgfdecoratedinputsegmentlength/\pgfdecorationsegmentlength)}
            \setlength{\pgfdecorationsegmentlength}{\matchinglength pt}
        }] {}
    \state{upsine}[width=\pgfdecorationsegmentlength,next state=downsine]{
        \pgfpathsine{\pgfpoint{0.25\pgfdecorationsegmentlength}{0.5\pgfdecorationsegmentamplitude}}
        \pgfpathcosine{\pgfpoint{0.25\pgfdecorationsegmentlength}{-0.5\pgfdecorationsegmentamplitude}}
    }
    \state{downsine}[width=\pgfdecorationsegmentlength,next state=upsine]{
        \pgfpathsine{\pgfpoint{0.25\pgfdecorationsegmentlength}{-0.5\pgfdecorationsegmentamplitude}}
        \pgfpathcosine{\pgfpoint{0.25\pgfdecorationsegmentlength}{0.5\pgfdecorationsegmentamplitude}}
}
    \state{final}{}
}

\newcommand{\Comment}[1]{{}}
\definecolor{darkblue}{rgb}{0.15,0.35,0.55}
\definecolor{comment}{rgb}{1,0.4,0.4}
\definecolor{reddish}{rgb}{0.65, 0.2, 0.2}
\definecolor{darkred}{rgb}{0.7,0.3,0.3}
\definecolor{darkgreen}{RGB}{65, 180, 44}
\definecolor{greyish}{rgb}{.90,.90,.90}
\definecolor{greyish2}{rgb}{.96,.96,.96}
\definecolor{darkblue2}{rgb}{0.3,0.3,0.9}
\usepackage{pifont}%
\usepackage{xcolor,colortbl}

\usepackage{tcolorbox}
\usepackage[linktocpage=true]{hyperref}

\hypersetup{
colorlinks=true,
citecolor=darkblue,
linkcolor=reddish,
urlcolor=darkblue,
pdfauthor={},
pdftitle={},
pdfsubject={}
}

\newcommand{\be}{\begin{equation}}
\newcommand{\ee}{\end{equation}}
\newcommand{\bea}{\begin{eqnarray}}
\newcommand{\eea}{\end{eqnarray}}
\newcommand{\beas}{\begin{eqnarray*}}
\newcommand{\eeas}{\end{eqnarray*}}

\def\({\left(}
\def\){\right)}

\newcommand{\rd}{{\rm d}}

\def\gsim{ \lower .75ex \hbox{$\sim$} \llap{\raise .27ex \hbox{$>$}} }
\def\lsim{ \lower .75ex \hbox{$\sim$} \llap{\raise .27ex \hbox{$<$}} }

\def\xyma{\xymatrix@M.7em}
\def\xymas{\xymatrix@M.1em}
\newcommand{\ba}{\begin{eqnarray}}
\newcommand{\ea}{\end{eqnarray}}

\makeatletter
\renewcommand\@dotsep{200}
\makeatother

\title{}
\author{}

\numberwithin{equation}{section}

\begin{document}
%
\renewcommand{\thefootnote}{\fnsymbol{footnote}}
~
%
%
\begin{center}
{\huge \bf{Vortices and waves in light dark matter}}
\end{center} 

\vspace{1.2truecm}
\thispagestyle{empty}
\centerline{{\Large Lam Hui,${}^{\,a,}$\footnote{\href{mailto:lh399@columbia.edu}{\texttt{lh399@columbia.edu}}} 
Austin Joyce,${}^{\,a,b,}$\footnote{\href{mailto:a.p.joyce@uva.nl}{\texttt{a.p.joyce@uva.nl}}} 
Michael J. Landry,${}^{\,a,}$\footnote{\href{mailto:ml2999@columbia.edu}{\texttt{ml2999@columbia.edu}}} 
and Xinyu Li${}^{\,a,c,d,}$\footnote{\href{mailto:xli@cita.utoronto.ca}{\texttt{xli@cita.utoronto.ca}}}
}}
\vspace{.6cm}

\centerline{{\it ${}^a$Center for Theoretical Physics, Department of Physics,}}
\centerline{{\it Columbia University, New York, NY 10027, USA}}
\vspace{.2cm}

\centerline{{\it $^b$Delta-Institute for Theoretical Physics,}}
\centerline{{\it University of Amsterdam, Amsterdam, 1098 XH, The Netherlands}}

\vspace{.2cm}
\centerline{{\it ${}^c$Canadian Institute for Theoretical
    Astrophysics,}}
\centerline{{\it University of Toronto, 60 St George St, Toronto, ON M5R 2M8, Canada}} 
\vspace{.2cm}
\centerline{{\it ${}^d$Perimeter Institute,}}
\centerline{{\it 31 Caroline Street North, Waterloo, Ontario, Canada, N2L 2Y5, Canada}}

 \vspace{.75cm}
\begin{abstract}
\noindent
In a galactic halo like the Milky Way,
bosonic dark matter particles lighter than about $30$ eV have a de
Broglie wavelength larger than the average inter-particle separation and 
are therefore well described as a set of
classical waves.
This applies to, for instance, the QCD axion as well as to
lighter axion-like particles such as fuzzy dark matter. 
We show that the interference of waves inside a halo inevitably leads
to vortices, locations where chance destructive interference takes the
density to zero. 
The phase of the wavefunction has non-trivial winding around these points.
This can be interpreted as a non-zero velocity circulation,
so that vortices are sites where the fluid velocity has a non-vanishing curl. 
Using analytic arguments and numerical simulations, 
we study the properties 
of vortices
and show they have a number of universal features: (1)
In three spatial dimensions, the generic defects
take the form of vortex rings.
(2) On average there is about
one vortex ring per de Broglie volume and (3) generically only single
winding ($\pm 1$) vortices are found in a realistic halo. (4) The density near a
vortex scales as $r^2$ while the velocity goes as $1/r$, where $r$ is
the distance to vortex. (5) A vortex segment moves at a velocity
inversely proportional to its curvature scale so that smaller
vortex rings move faster, allowing momentary motion exceeding escape
velocity. 
We discuss observational/experimental signatures from vortices and, more
broadly, wave interference. In the ultra-light regime, gravitational
lensing by interference substructures leads to flux anomalies 
of $5-10 \%$ in strongly lensed systems.
For QCD axions, vortices lead to a diminished signal in some detection
experiments but not in others. We advocate the measurement of
correlation functions by axion detection experiments as a way to
probe and capitalize on the expected interference substructures.
\end{abstract}
\newpage

\setcounter{tocdepth}{3}
\tableofcontents
\newpage
\renewcommand*{\thefootnote}{\arabic{footnote}}
\setcounter{footnote}{0}

\section{Introduction}
\label{sec:intro}

Though there is overwhelming evidence for the existence of dark matter
from a variety of astronomical probes, the constituents of dark matter
remain a mystery. Ideas range from a particle as light as $10^{-22}$~ 
eV to primordial black holes of tens of solar mass.
One useful demarcation in this immense spectrum is the division
between particle-like and wave-like dark matter. This boundary in phenomenology
occurs at masses around $30$ eV in a galactic halo like the Milky Way. 
A dark matter particle with mass below this $30$~eV threshold has a de Broglie wavelength that exceeds the
typical inter-particle separation. Such a particle is necessarily
bosonic, because of the Pauli exclusion principle for fermions.\footnote{This is the origin of the Tremaine--Gunn
  bound~\cite{Tremaine:1979we} for fermionic dark matter.}
In this case where the average particle number in a de Broglie volume is very large,
the corresponding quantum state has very high occupation number and its evolution is well approximated by a classical field~\cite{Guth:2014hsa},
much as a collection of many
photons is well described by electric and magnetic fields
satisfying Maxwell's equations.\footnote{\label{ft:hydrofootnote}For example, 
a coherent state with a large average
occupancy ({\it i.e.}, many particles in a de Broglie volume) 
behaves classically in the sense that quantum fluctuations are negligible. A collection of heavy particles, 
on the other hand,
is naturally described by a distribution function obeying the
Boltzmann equation. In this case there is also an effective description in terms of classical fields
after coarse-graining by taking
the hydrodynamic limit.  See {\it e.g.},~\cite{Ma:1995ey,Baumann:2010tm}. However, this situation is conceptually different and does not lead to wave-like phenomena on macroscopic scales.} In this
paper, we are interested in the {\it classical} wave-like behavior of dark matter in this regime.

There is a large literature on light dark matter of this sort, which is
commonly modeled by a scalar or pseudo-scalar
field~\cite{Baldeschi:1983mq,Press:1989id,Sin:1992bg,Peebles:2000yy,Goodman:2000tg,
Lesgourgues:2002hk,Chavanis:2011zi,Suarez:2011yf}.
A particularly appealing realization of the general framework is a
pseudo-Nambu--Goldstone boson, a prime example of which is the QCD
axion \cite{Peccei:1977hh,Kim:1979if,Weinberg:1977ma,Wilczek:1977pj,
Shifman:1979if,Zhitnitsky:1980tq,Dine:1981rt,Preskill:1982cy,Abbott:1982af,Dine:1982ah},
which could span a large range
of masses. Experimental efforts have focused in particular on masses
around $10^{-6}$ eV, with some reaching down to much lower values
(see reviews by \cite{Graham:2015ouw,Marsh:2015xka,Sikivie:2020zpn}).
More generally, axion-like-particles (ALPs) abound in string theory with a wide range of masses~\cite{Svrcek:2006yi,Arvanitaki:2009fg,Halverson:2017deq,Bachlechner:2018gew}.
An intriguing possibility is an ALP with mass around $10^{-22}$\,--\,$10^{-20}$~eV, which under fairly simple assumptions has a relic
abundance appropriate to be the dark matter.\footnote{More precisely---assuming an initial mis-alignment angle of
  order unity---the relic abundance of a particle in this mass range is 
$\Omega_{\rm ALP}\sim0.1\,(m/{10^{-22} {\,\rm eV}})^{1/2}\,(f/10^{17} {\,\rm
  GeV})^2$, where $m$ is the ALP mass and $f$ is the associated axion decay constant.
} In this extreme low mass limit, the de
Broglie wavelength is a significant fraction of the size of a galaxy, a possibility
often referred to as fuzzy dark matter~\cite{Hu:2000ke,Amendola:2005ad,Schive:2014dra,
Veltmaat:2016rxo,Schwabe:2016rze,Hui:2016ltb,Mocz:2017wlg,Nori:2018hud}. 
Beyond these concrete examples, the wave-like features  we study will be present in any model where the dark matter is light enough for the number of quanta in a de Broglie volume to be large, {\it e.g.},~\cite{Sikivie:2009qn,Berezhiani:2015bqa,Fan:2016rda,Alexander:2016glq,Alexander:2018fjp}.

For our purposes, the precise underlying model for dark matter
is not crucial, nor is its mass as long as it is under $\sim
30$ eV. What we do assume is that the dark matter is well-described by
a classical scalar field whose only non-negligible
interaction is gravitational.\footnote{For an axion, it can be shown
  self-interaction is subdominant compared to gravity or quantum
  pressure for a typical halo dark matter density. The relative
  importance of self-interaction, gravity and quantum pressure can be
  evaluated by comparing: $m^2 \phi^4/f^2$, $(m^2\phi^2)^2 r^2/M_{\rm
    Pl}^2$ and $\phi^2/r^2$. It can be shown the first term is smaller
  than the other two for all $r$ for a fixed $\rho = m^2\phi^2$ as
  long as $1 \, > \, 10^{-11}
  (10^{-6}{\,\rm eV\,}/m) (10^{12} {\,\rm GeV\,}/f)^2 (\rho / {\,\rm
    GeV/cm^3\,})^{1/2}$. The inequality also holds for fuzzy dark matter like values
  $m\sim 10^{-22} {\,\rm eV}$ and $f\sim 10^{17} {\,\rm GeV}$.
}
This is a reasonable approximation when the system is in a high-occupancy state, which in this context requires a sufficiently light dark matter mass.
Our starting point is to consider a real 
Klein--Gordon scalar $\phi$ minimally coupled to gravity, which is described by the following action
\be
S = \int\rd^4
  x\sqrt{-g}\left(-\frac{1}{2}(\partial\phi)^2-\frac{m^2}{2}\phi^2\right)
  \, .
\label{eq:scalaraction}
\ee
We are interested in
situations where the dark matter is non-relativistic. This applies
to much of the structure in the universe, the primary exception being 
dark matter close to black
holes ({\it e.g.},~\cite{Jacobson:1999vr,Arvanitaki:2010sy,Macedo:2013qea,Hui:2019aqm,Clough:2019jpm,Baumann:2019eav}). In the non-relativistic limit, it is convenient to
parameterize the field as\footnote{The real and imaginary parts of the complex field, $\Psi$,
  parameterize the two phase space degrees of freedom of the (real)
  scalar, $\phi$. For a discussion of the systematics of the
  non-relativistic limit, see~\cite{Mukaida:2016hwd,Namjoo:2017nia,Eby:2018ufi}.}
\be
\phi = \frac{1}{\sqrt{2m}}\left(\Psi e^{-imt}+\Psi^* e^{imt}\right)
       \, ,
\ee
where $\Psi$ is a complex scalar. 
Assuming $\Psi$
is slowly varying compared to the frequency $m$, {\it i.e.}, 
$\lvert\ddot\Psi\rvert \ll m\lvert \dot\Psi\rvert$, and dropping terms
that oscillate rapidly, 
the Klein--Gordon equation implies that
 $\Psi$ satisfies a Schr\"odinger equation coupled to gravity\footnote{Here and throughout this paper we normalize the wavefunction $\Psi$ so that $\lvert\Psi\rvert^2$ is a number density.}
\be
\label{eq:schrodinger}
i \partial_t \Psi = - {\frac{1}{2m}}\nabla^2 \Psi + m \Phi \Psi,
\ee
where $\Phi$ is the gravitational potential, which is itself fixed by
the Poisson equation sourced by the local mass density $\rho$
\begin{equation}
\label{eq:poisson}
\nabla^2 \Phi = \frac{\rho}{2M_{\rm Pl}^2} \quad {\rm where} \quad
\rho = m \lvert\Psi\rvert^2 \, .
\end{equation}
It is worth stressing that---despite the fact that the field $\Psi$ obeys a Schr\"odinger equation---this system of equations describes
the {\it classical} evolution of the collective dynamics of the underlying dark matter particles.
The wave-like nature of these classical field
configurations allows for a variety of behaviors that are analogous to
familiar quantum-mechanical phenomena, but it is important to keep in
mind that the physics is purely classical.\footnote{In particular, though we have set $\hbar = 1$, if we were to re-introduce these factors they would always appear in the combination $\hbar/m$, so that everything can be understood classically.}
In this paper we will be interested in the dynamics of this Schr\"odinger--Poisson system in gravitationally-bound objects like galaxies, so we neglect the background expansion of the universe, which is irrelevant on these scales.
For details of the cosmology of light dark matter see, {\it e.g.},~\cite{Marsh:2010wq,Hlozek:2014lca,Guth:2014hsa,Marsh:2015xka,Hui:2016ltb,Li:2018kyk}.

\begin{figure}[tb]
\centering
\hspace{-1.2cm}
\includegraphics[width=.42\textwidth]{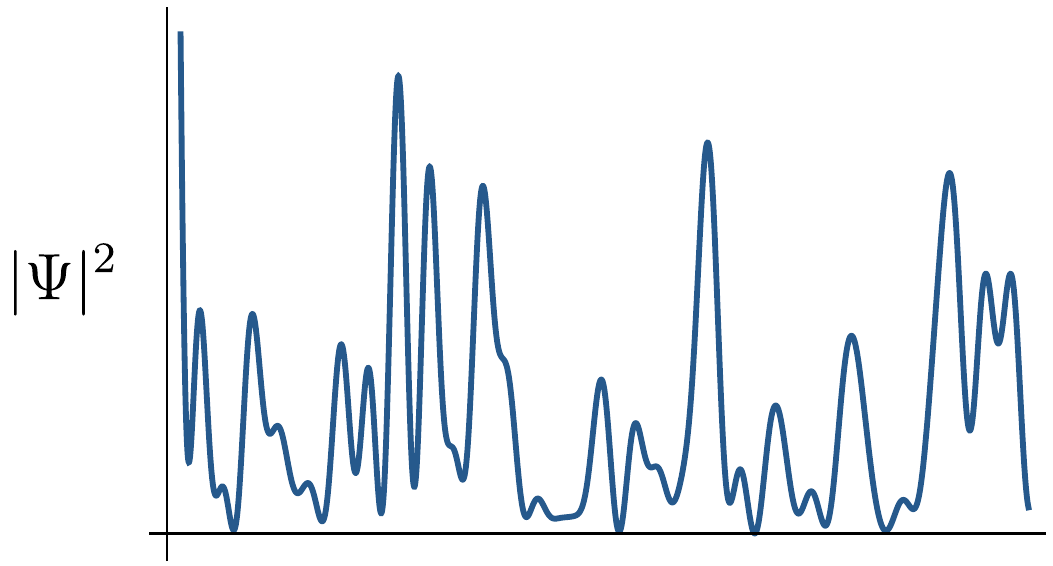}
\hspace{2cm}
\includegraphics[width=.3\textwidth]{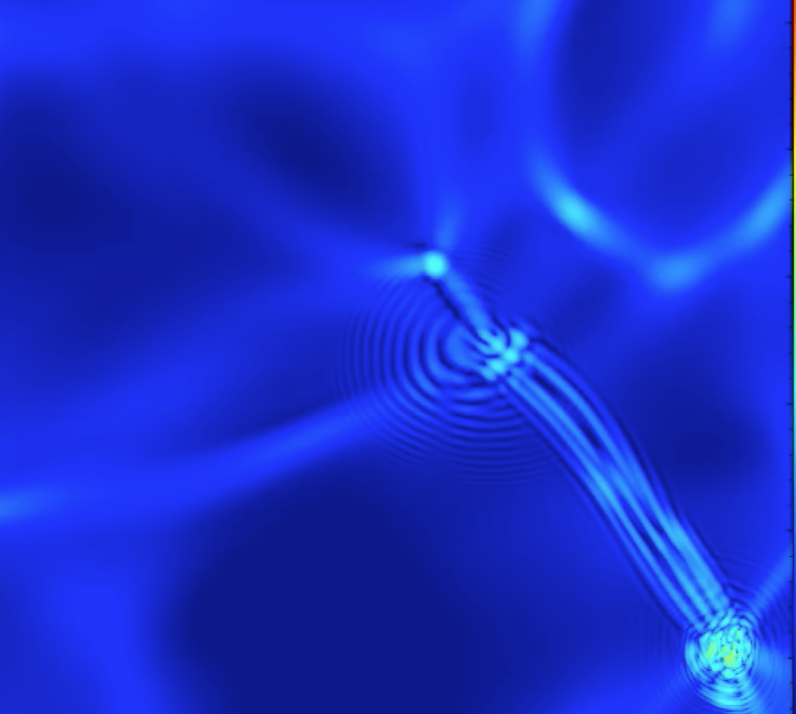}
\caption
{\small  In the outer regions of a light dark matter halo, the dynamics is well captured by a random superposition of waves. Chance destructive interference can take the local dark matter density all the way to zero. In the left panel, as an example we show a 1D plot of $\lvert\Psi\rvert^2$, where $\Psi$ is constructed from a sum of plane waves with random phases. Note that there are many places where the square of the wavefunction vanishes.
On the right, we show a more realistic scenario. This is a snapshot of
the dark matter density in a cosmological simulation of ultra-light
dark matter done in~\cite{Li:2018kyk}, showing interference fringes
from wave-like superposition in dark matter filaments and haloes
(darker shade means lower density). 
The locations where the dark matter density vanishes behave like
vortices (see \S\ref{sec:general}). 
At early times ({\it e.g.}, in the initial conditions) there is no complete destructive interference because fluctuations are too small. However, these fluctuations are amplified by gravitational collapse after which destructive interference occurs. It turns out that many of the features of the vanishing loci can be understood without gravity. In this paper we explore the formation and properties of these vortex features.
}
\label{fig:pictures}
\end{figure}

In the early universe, $\Psi$ is roughly homogeneous with small
perturbations, and linear perturbation theory suffices  to describe
the growth of structure. 
In the nonlinear regime, the 
fact that $\rho \propto |\Psi|^2$ leads to striking interference
patterns---for instance, a galactic halo typically consists of a
superposition of  $\Psi$ waves leading to order unity fluctuations in
the mass density $\rho$, as first demonstrated by Schive et
al.~\cite{Schive:2014dra}. 
This interference substructure has many interesting dynamical
consequences, some of which were explored in~\cite{Hui:2016ltb}.
In this paper, we are particularly interested in destructive
interference---we study locations where chance interference takes the
density all the way to {\it zero}. 
An illustration is shown in Figure
\ref{fig:pictures}, drawn from~\cite{Li:2018kyk}. 

What is so special about these sites of complete destructive
interference? These loci behave essentially like topological defects around which the phase of
$\Psi$ jumps or has a non-trivial winding, as will be explained in Section
\ref{sec:general}.\footnote{It is worth
  stressing that these are {\it not} the 
  axion strings that form as a result of the spontaneous breaking of the
  Peccei--Quinn $U(1)$ symmetry. Here, the relevant $U(1)$ symmetry is a more
  mundane one: that associated with particle number conservation. 
We discuss more how these defects should be thought of in the
underlying relativistic theory in Section~\ref{sec:conclude}.
}
They occur at a rate of about one per de Broglie volume in a
typical halo, and they have universal density profiles, leading to
interesting observational/experimental signatures. Our goal in this
paper is to demonstrate and explore this phenomenology with a number of numerical
and analytic computations. Gravity is important in the formation of
these objects---in the early universe fluctuations in $\Psi$ are
sufficiently small compared to the background value that nowhere 
is interference destructive enough to make $\rho$ vanish. 
The fluctuations of $\Psi$ are amplified by gravity until
complete destructive interference is possible and, as we will see, 
generically occurs. Interestingly, once order one fluctuations in $\Psi$ are
present, many of the features of these defects can be
understood without gravity.  We will present computations both with
and without gravity to illustrate these points. 

In many respects light dark mater resembles a superfluid, and
many phenomena familiar from the study of superfluids have avatars in
this setting. The topological defects of interest behave like
vortices from this point of view. In order to understand the correspondence, let us recall the fluid (Madelung \cite{Madelung1927})  formulation
of the Schr\"odinger equation.\footnote{An elementary discussion can be
  found in the Feynman lectures~\cite{Feynman:1963uxa}.}
We start by re-expressing the wavefunction in terms of its amplitude
and phase: 
\be
\Psi = \sqrt{\frac{\rho}{m}} e^{i\theta} \, ,
\label{eq:radialparam}
\ee
where $\rho$ and $\theta$ are interpreted as the fluid mass density
and the velocity potential, respectively. In other words, we have
\begin{equation}
\rho = m |\Psi|^2\, , \qquad   \qquad 
\vec v = \frac{1}{m} \vec \nabla \theta \,.
\label{eq:rhovdef}
\end{equation}

In these variables, eq.~\eqref{eq:schrodinger} takes the form of the familiar equations of hydrodynamics, albeit with a somewhat peculiar source of stress. Inserting the parametrization~\eqref{eq:radialparam} into the Schr\"odinger equation, the imaginary part gives a continuity equation for the density
\be
\partial_t \rho + \vec \nabla \cdot (\rho \vec v) = 0 \, ,
\label{eq:massconv}
\ee
which expresses the conservation of mass in the fluid system. This equation can of course equally well be interpreted as the continuity equation for the probability current. The real part of the Schr\"odinger equation gives an independent Euler-like equation
\be
\partial_t \vec v + (\vec v\cdot\vec\nabla)\,\vec v = - \vec \nabla \Phi + \frac{1}{
2 m^2} \vec \nabla \left( \frac{\nabla^2 \sqrt{\rho}}{ \sqrt{\rho}} \right).
\label{eq:euler}
\ee

The final term in~\eqref{eq:euler} is often interpreted as a ``quantum pressure," but this is somewhat misleading because it does not correspond to an isotropic stress (and in this context is not at all quantum).\footnote{The quantum pressure term can be re-written in terms of a stress tensor as
 $\frac{1}{2m^2}\partial_i\left(\frac{\nabla^2\sqrt\rho}{\sqrt\rho} \right) = \partial^j\sigma_{ij}$,
where $\sigma_{ij} = \frac{\rho}{4m^2}\partial_i\partial_j\log\rho$,
which typically has off-diagonal elements.
}
The change of variables between the Schr\"odinger formulation and the
Madelung equations is highly nonlinear, which makes it possible to
study solutions that are complicated in one set of variables in the
other formulation, where the solutions can be substantially simpler. It is worth noting that the fluid variables obtained in this fashion are not the same as the ones that would appear in the hydrodynamic description obtained from phase space coarse graining (see footnote~\ref{ft:hydrofootnote}), though they will agree at long distances.

In the fluid formulation, the velocity field is a gradient
flow, just as in a superfluid. At first sight, it therefore seems there is no room
for vorticity. Moreover, Kelvin's theorem \cite{thomson_1868}---the co-moving conservation of
circulation---follows from taking the curl of the Euler-like equation,
which tells us that if the initial vorticity vanishes everywhere,
it would continue to.\footnote{More explicitly the curl of the Euler equation is 
$\partial_t (\vec \nabla \times \vec v)^i + v^k \nabla_k (\vec \nabla
  \times \vec v)^i + (\vec \nabla \cdot \vec v)  (\vec \nabla
  \times \vec v)^i -  S^{ij} (\vec \nabla \times \vec v)^j = 0$
where $S^{ij} \equiv \partial^{(i} v^{j)}$. So, if the initial curl is zero, it will remain zero.
}
How then can vortices develop? The obvious loophole arises from places where the density
vanishes: there the phase $\theta$ and the velocity become ill-defined. 
As we will argue, the phase $\theta$ has a
non-trivial winding around these loci, leading effectively to a non-zero velocity
circulation $\oint \rd\vec\ell \cdot \vec v$, {\it i.e.}, vorticity $\vec \nabla \times \vec v$ is
non-vanishing at such locations.\footnote{Another way to say this is
  that Kelvin's theorem fails for the Madelung fluid because of the
  points where the density vanishes. Neither the fluid velocity nor the so
 -called quantum pressure are well defined at these points.
Vorticity (the velocity integrated along a closed loop) does not need to be conserved if
  the loop intersects a point where the fluid parameterization is
  ill-defined~\cite{Damski_2003,Nilsen7978}. 
  }

We can therefore interpret the loci where the wavefunction vanishes as
vortices. Note that this requires the vanishing of both the real and
imaginary parts of the wavefunction. Since the wavefunction is a
function of $d=D-1$ spatial variables, the vanishing of either its
real or imaginary part generically occurs along a co-dimension one
surface. These two surfaces will generically intersect along a
co-dimension two surface {\it i.e.}, a string in $d=3$. 
These strings are vortices from the fluid perspective. They do not
end, and thus generically take the form of vortex rings. 
In lower dimension---or in non-generic circumstances in three spatial
dimensions---the total wavefunction can vanish along a lower
co-dimension surface, leading to the appearance of a domain wall from
the fluid perspective. What we will show is that the phase and density
close to a defect (be it a wall or a vortex) behave in characteristic
ways, and that defects occur with a predictable frequency in a
galactic halo.\footnote{A possible point of confusion is that in the large $m$ limit, the Schr\"odinger--Poisson equations should match smoothly onto a particle description of the dark matter fluid~\cite{Widrow:1993qq}, which certainly can support vorticity without defects. The resolution of the apparent paradox is that the classical fluid description arises after coarse-graining. In particular, the Wigner function gives rise to a fluid phase space distribution function after smoothing on some scale, and vorticity can arise in the effective fluid as a byproduct of this coarse-graining procedure from structure smaller than the smoothing scale~\cite{Mocz:2018ium,Uhlemann:2018gzz}. 
}

Vortices are ubiquitous in physics. They appear, for example, as
defects in the Abelian Higgs model~\cite{Nielsen:1973cs}, as flux
tubes in confining gauge theories~\cite{Luscher:1980ac} and in both
superfluids~\cite{1949NCim....6S.279O} and in Bose--Einstein
condensates (BECs)~\cite{2008LaPhy..18....1F}. It is worth
emphasizing, however, that the vortices we consider in this paper
differ somewhat from the vortices that arise in these situations. In
particular, the fluid we consider behaves like a superfluid, but with
self-interactions turned off. In our context, the density is not a gapped degree of freedom, so there does not appear to be a well-defined effective field theory of these defects along the lines of~\cite{Luscher:1980ac,Polchinski:1991ax,LUND1991245,Dubovsky:2012sh,Aharony:2013ipa,Horn:2015zna}. Even within the context of wave-like dark matter vortices are well-studied~\cite{Silverman:2002qx,Brook:2009ku,Kain:2010rb,RindlerDaller:2011kx,Zinner:2011if,Banik:2013rxa}. However,
most of the previous studies have focused on the regime of BEC dark
matter where self-interactions dominate over the quantum pressure term
(sometimes called the Thomas--Fermi regime). Precisely the opposite
regime is of interest on galactic scales for weakly-coupled dark
matter with a large de Broglie length, and it is this regime that we
focus on in this paper.
In particular we combine simple analytical
estimates in the absence of gravity with numerical simulations to
ensure that our conclusions are robust to the inclusion of gravity and
to verify that the vortex configurations have the properties we
expect. Some properties of vortex solutions of the Schr\"odinger equation were also studied numerically in~\cite{2011JPhB...44k5101C}.

We find that vortex-like defects in light dark matter have a number of
universal properties which persist even in the presence of gravity. As
we justify below, on general grounds we expect that vortices will have
unit winding, and will have a characteristic density profile $\rho\sim
r^2$, and velocity profile $v \sim 1/r$, near the vortex location. All
of these expectations are borne out by numerical simulations. We
further expect based both on dimensional arguments and on simple
analytic estimates that the number density of defects should be
approximately one per de Broglie volume, in good agreement with
numerical results. We will investigate the implications of vortices, and
more broadly interference substructures, for astrophysical
observations (relevant for the ultra-light regime) and for axion
detection experiments (relevant for light, but not necessarily ultra-light, dark
matter). One feature we will use is the fact that vortices
can potentially move at high speed. Some of the most interesting
signatures have to do with general stochastic fluctuations associated
with the interference substructures.

The organization of this paper is as follows.
In \S\ref{sec:general}, we present a general Taylor-expansion argument to describe how $\Psi$
should behave close to a defect, from which we infer some general properties of defects.
In \S\ref{sec:exact}, we explore simple, exact solutions when
gravity is ignored, for walls \S\ref{sec:walldefect} and vortices
\S\ref{sec:vortices}. 
We show how defects can form and disappear.
In~\S\ref{sec:randomphase}, we introduce a simple statistical model for the fluctuations of the light dark matter field in a galactic halo and use this model to estimate the number density of defects.
In~\S\ref{sec:numerics}, we explore examples of defects in
numerical simulations, which incorporate the effect of gravity, and
show that their general characteristics are consistent with those of
the simple, exact solutions.
Possible observational/experimental signatures are discussed in
\S \ref{sec:observables}.
We conclude in \S\ref{sec:conclude} with a discussion of 
angular momentum and the role of vortices, along with future prospects
for observational/experimental signatures.
Several technical results are relegated to the Appendices.
As this manuscript was in preparation, a preprint by
\cite{2020arXiv200210473C} appeared on the subject
of gravitational lensing flux anomaly. It has some overlap with our
discussion in \S\ref{sec:observables}. 

\vskip4pt
\noindent
{\bf Conventions:} We employ the mostly-plus metric signature. We denote the spacetime dimension by $D$ and the spatial dimension by $d$, though we primarily work in four spacetime dimensions. Throughout the paper we have set Planck's constant, $\hbar$, and the speed of light, $c$, to one.

\section{A general existence argument for defects}
\label{sec:general}
Defects appear in fluid variables precisely at points where 
the square of the wavefunction vanishes, so that the
parameterization~\eqref{eq:rhovdef} is no longer well-defined.  This
corresponds to locations where both the real and imaginary parts of
the wavefunction are zero. We would like to understand how the system
behaves under generic circumstances in the vicinity of such a
defect. The system of
equations~\eqref{eq:schrodinger}--\eqref{eq:poisson} is
non-linear, so the general dynamics of defects is
complicated. However, very close to to the defect, it is possible to
characterize the behavior in a universal way. We therefore first focus
on the near-defect regime, before turning to numerical simulations to verify that defects behave as expected from our analytical arguments.

\subsection{A simple example: domain walls in $1+1$ dimensions}
\label{sec:walls}
\begin{figure}[tb]
\centering
\includegraphics[width=6cm]{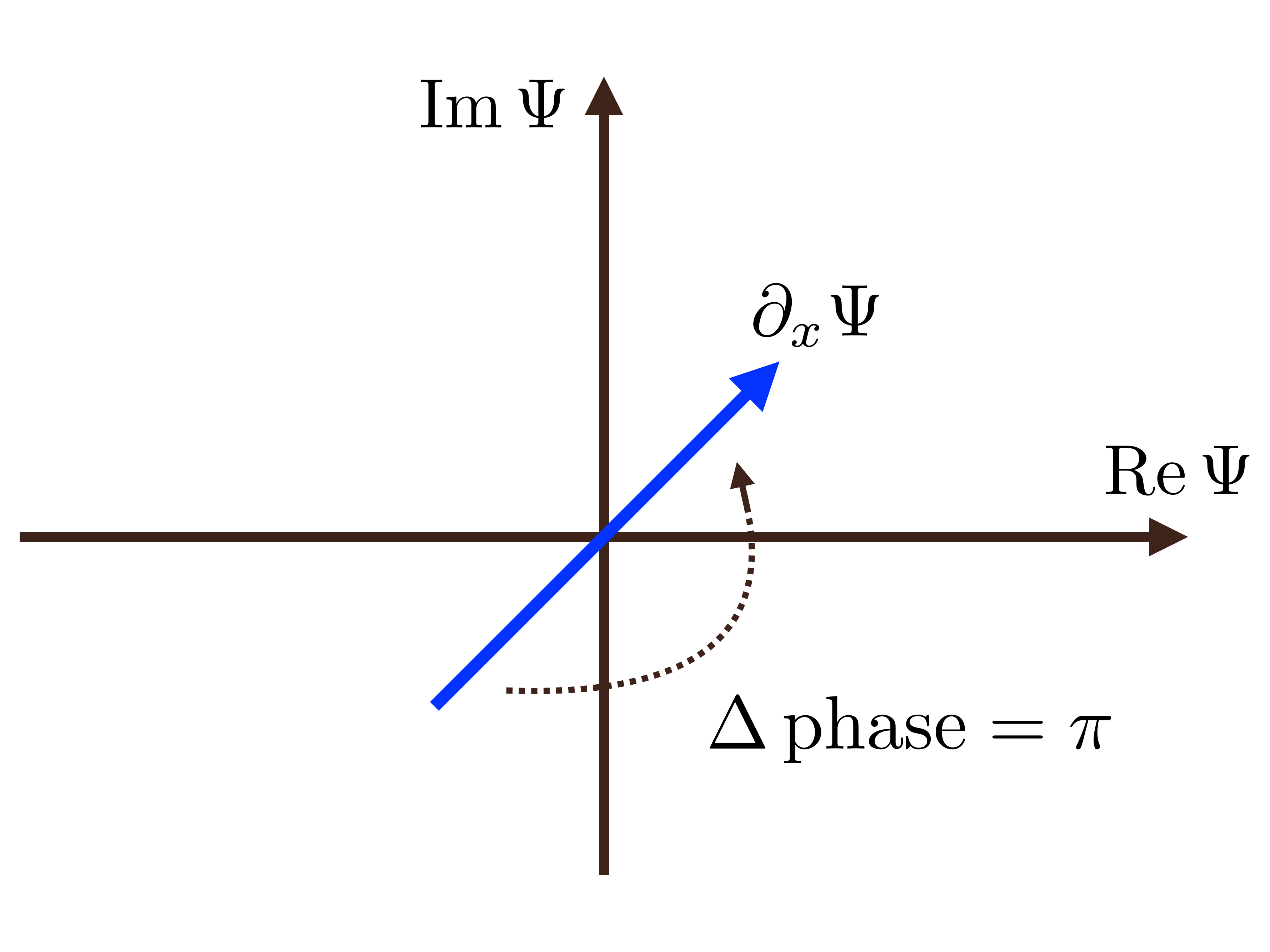}
\vspace{-.45cm}
\caption
{\small Behavior of the wavefunction in $1+1$ dimensions in the vicinity of a zero. As we move in the $x$ direction, the wavefunction varies along the (blue) arrow labeled by $\partial_x\Psi$. In order for the wavefunction to vanish at a point, but still have a finite first derivative the phase has to jump by $\pi$ across the origin, corresponding to the location of the domain wall.
}
\label{fig:wallphase}
\end{figure}

It is instructive to first consider the case of one spatial dimension. In this situation, the only possible defect has co-dimension one, corresponding to a wall. We wish to demonstrate that the behavior of the density and velocity in the neighborhood of the wall is universal.

Suppose there is a certain location, say
$x=0$, where the density happens to vanish (at some particular time
$t$), meaning both the real and
imaginary parts of $\Psi$ vanish. 
In the neighborhood of $x=0$, we can Taylor
expand the wavefunction as (for simplicity we restrict to static configurations)
\be
\Psi (x) \sim x\partial_x \Psi (0) +\cdots,
\ee
under the assumption that $\partial_x \Psi (0)$ does not vanish. In order for $\Psi$ to be well-defined as a complex function, its phase must jump by $\pi$ (or an odd integer multiple, $(2n+1)\pi$, $n\in{\mathbb Z}$)
as $x$ crosses zero. That is, the sign of $\Psi$ changes as $x$ passes through zero.
This indicates that there is a wall defect located at $x=0$. This situation is illustrated in Figure~\ref{fig:wallphase}. It is possible to tune things such that there is no phase jump at the location of the wavefunction's zero, by arranging for odd derivatives of the field to vanish but this is highly non-generic.

Notice that---given this profile---as we move away from the location where the domain wall is located, the square of the wavefunction scales as
\be
\rho \sim \lvert\Psi\rvert^2 \sim x^2,
\ee
so that the fluid density grows like $x^2$ in the vicinity of the defect.

\subsection{Vortices in higher dimensions}
\label{eq:genvortex}
In higher dimensions, the generic locus along which either the real or
imaginary part of the wavefunction vanishes will have co-dimension
one. The intersection of these two surfaces where the total
wavefunction vanishes will generally be a co-dimension two
surface. Therefore, in the particular case of three spatial
dimensions, our expectation is that the generic defects will be 
take the form of lines. Further, it is difficult to arrange for the
wavefunction to vanish along a line that extends all the way to
infinity, so we should expect that the typical defects will be closed
rings as opposed to infinite lines. 
As we will see in a moment, these lines/rings are vortices/vortex rings.
This is consistent with our expectation that dark matter haloes will be of finite size and therefore will only be able to support vortex loops of finite extent.

The vorticity of a flow is given by the line integral of the velocity field around a closed loop
\be
{\cal C} = \oint_{\partial A} \rd\vec\ell\cdot\vec v = \int_A \rd\vec A \cdot \vec\nabla\times\vec v.
\label{eq:circdef}
\ee
In the second equality we have used Stokes' theorem to write the circulation as an area integral. Since the fluid velocity is a gradient flow, this will always be zero unless the curl is singular somewhere, which happens exactly at the location of vortices.
Additionally,  if we do the line integral in~\eqref{eq:circdef}---possibly enclosing a vortex---the integral is related to the phase difference of the wavefunction transported around a closed circle, which must be an integer multiple of $2\pi/m$ for the wavefunction to be single-valued~\cite{Dirac:1931kp}
\be
{\cal C} = \frac{1}{m}\oint_{\partial A} \rd\theta = \frac{2\pi n}{m}\,,~~~{\rm with}~~~n\in{\mathbb Z}.
\label{eq:winding}
\ee
We therefore see that the vorticity carried by the Schr\"odinger vortices must be quantized, as in a superfluid~\cite{1949NCim....6S.279O}.

We can repeat the argument from Section~\ref{sec:walls} to see that in
the generic case the lines along which the wavefunction vanishes
indeed support vorticity. In the vicinity of a point (the origin, say)
where the wavefunction vanishes, we Taylor expand the wavefunction as
\be
\Psi(\vec x)\simeq \vec x\cdot\vec\nabla\Psi(0)+\cdots,
\ee
where the higher-order corrections are unimportant at small distances.
In the generic case, the wavefunction will vanish along a line (our
eventual vortex), so
that the derivative along the line direction will also
vanish. Without loss of generality we take the tangent to the line
to be pointing along the $z$ direction. It is then convenient to introduce complex coordinates in the transverse plane
\be
{\mathfrak z} \equiv x+i y\,,\qquad\qquad\qquad \bar {\mathfrak z} \equiv x-iy,
\label{eq:complexcoords}
\ee
In these coordinates, the wavefunction in the near-vortex regime takes the form\footnote{We define $\partial \equiv\frac{\partial}{\partial {\mathfrak z}}$ and $\bar\partial \equiv \frac{\partial}{\partial\bar {\mathfrak z}}$.}
\be
\Psi({\mathfrak z},\bar{\mathfrak z}) \simeq {\mathfrak z}\,\bar\partial\Psi(0)+\bar{\mathfrak z}\,\partial\Psi(0)+\cdots\simeq a\,  {\mathfrak z}+b \, \bar{\mathfrak z}+\cdots
\ee
The quantities $\partial\Psi(0)$ and $\bar\partial\Psi(0)$ are just (complex) constants, which we have written as $a$ and $b$ in the last equality. When $\lvert a\rvert > \lvert b\rvert$, the phase winds by $+2\pi$ as we traverse a circle enclosing the origin, while in the opposite case $\lvert a\rvert < \lvert b\rvert$ the phase winds by $-2\pi$ as we encircle the origin.\footnote{This can be understood as a consequence of the argument principle, generalized to harmonic complex functions~\cite{10.2307/2974933}.} So we see that both of these cases describe vortices, either of winding $1$ or winding $-1$. 

In the vicinity of the vortex, we can work out the density profile in the plane. It is of the form
\be
\rho\sim\lvert\Psi\rvert^2 \sim A x^2+B y^2+C xy,
\label{eq:perpdensity}
\ee
where the parameters $A, B, C$ can be solved for in terms of $a, b$, but their precise form is not important. What we learn from this expression is that the density
scales roughly as $\sim r_\perp^2$ (where $r_\perp$ is the radius in
cylindrical coordinates---the distance from the vortex) but is not in
general axially symmetric. Rather, lines of constant density are
ellipses. Also, the fact that we have a definite winding means 
the fluid velocity $v \sim 1/r_\perp$ around a vortex (see
eqs.~\eqref{eq:circdef} and~\eqref{eq:winding}). The
momentum density $\rho v$ vanishes towards the vortex, 
scaling as $r_\perp$.

There are two degenerate cases: if $\lvert a\rvert = \lvert b\rvert$ then the configuration is {\it not} a vortex, but is instead a domain wall, where the phase jumps by $\pi$ as we cross the wall. A special case of this situation is when both $a$ and $b$ vanish. In that case we have to keep the next order in the Taylor expansion and the defect is typically a vortex of higher winding, unless parameters are again tuned at this order. Our generic expectation is therefore that loci where the wavefunction vanishes in $3+1$ dimensions describe defects of winding $\pm1$.

We can also make some inferences about the dynamics of vortex lines
using these arguments. In this case, we imagine Taylor expanding 
in the vicinity of a defect, but additionally allow the configuration to evolve in time so that we have
\be
\Psi(\vec x,t)\simeq \frac{it}{2m}\nabla^2\Psi(0)+\vec x\cdot\vec\nabla\Psi(0)+\cdots,
\label{eq:vortexvel}
\ee
where we have used the Schr\"odinger equation to relate the
coefficient of the $t$ term to that of the ${\cal O}(x^2)$ terms. 
Here, $0$ denotes the origin in both time and space where the
wavefunction vanishes.
Near
the vortex, we can neglect the ${\cal O}(x^2)$ terms and again let us
choose the tangent to the vortex to point in the $z$ direction.
At some small $t$, we can infer the location of the vortex by setting
\eqref{eq:vortexvel} to zero, from which we read off the vortex velocity:
\be
\dot {\vec x}\cdot\vec\nabla\Psi(0)=- \frac{i}{2m}\nabla^2\Psi(0)\,.
\ee
By splitting the wavefunction into its real and imaginary parts as $\Psi = \Psi_1+i\Psi_2$
we can write this equation as a matrix equation for the instantaneous velocity in the plane perpendicular to the vortex:
\be
\label{velocitymatrix}
\left( \begin{array}{cc}
\partial_x \Psi_1(0) & \partial_y \Psi_1(0)\\
\partial_x \Psi_2(0) & \partial_y \Psi_2(0)
\end{array}
\right)
\left( \begin{array}{c}
\dot x \\ \dot y
\end{array}
\right)
=
\frac{1}{2m}\left(\begin{array}{c}
{\nabla^2 \Psi_2 (0) } \\
{-\nabla^2 \Psi_1(0) }
\end{array}
\right).
\ee
Inverting this matrix gives the local velocity,
it scales roughly like the ratio between the coefficient of the $t$ term divided by the coefficients of the $\vec x$ terms. Dimensionally this scales as 
\be
v_{\rm vortex} \sim \frac{1}{mR}\,,
\label{eq:vortexspeed}
\ee
where $R$ is the characteristic scale over which the configuration varies. Notice that for curved vortices, this characteristic scale is 
of order the curvature, so we expect that the velocity is inversely
proportional to the radius of curvature of the vortex, $R$. This is
consistent with the behavior of superfluid vortices, where smaller
vortex loops move faster.\footnote{From the point of view of the Taylor expansion argument
  presented in this section, one might reasonably ask: why not expand
  around some other (non-vanishing) values of the wavefunction? What
  is so special about a vanishing wavefunction? The winding of
  the phase, the $1/r_\perp$ velocity profile and
  the $r_\perp^2$ density profile are all special to the vortex where
  the wavefunction vanishes. On the other hand, the argument about 
  velocity in eq.~\eqref{velocitymatrix} works for any particular value
  of the wavefunction one chooses to focus on. This velocity is useful
  in the case of the vortex because it tells
  us how the center of this $r_\perp^2$ density profile moves. We will
  exploit this fact in discussing observational implications.
}

\section{Simple analytic solutions and their properties}
\label{sec:exact}

The arguments in Section~\ref{sec:general} are based purely on the  Taylor expansion of the wavefunction
near its zeros, and are therefore quite robust.
In particular, the fact that locations where
$\Psi$ vanishes support defects is essentially independent of the
details of the theory and is not affected by the presence of gravity. However, to this point we
have not actually solved the Schr\"odinger equation, so one
might wonder whether these vortices are actually present in
physical solutions, and whether vortices can form if they were 
initially absent. 
In this Section, we rectify this situation by verifying that there are
solutions to the Schr\"odinger equation that have the expected properties.\footnote{A small note about normalization: throughout this section we will be somewhat un-careful with the dimensions of the wavefunction, $\Psi$. The physical wavefunction should have dimensions such that $\lvert\Psi\rvert^2$ is a number density. However, since the solutions that we write down solve a linear equation, their overall normalization is arbitrary. We will therefore often set the dimension-ful scales in front of these solutions to $1$, but they can be restored by multiplying by a suitable power of any length scale desired.}

Deriving simple analytic solutions with vortices also allows us to get some intuition for how these objects behave.
Since many of the properties of defects are quite generic,
it is useful to first consider solutions in the absence of gravity.
In particular, this simplified setting allows us to understand how defects can form and disappear, and gives insight into the density profiles in the vicinity of vortices, which will turn out to be universal.
We expect that the coarse features of these solutions will continue to
persist even in the presence of gravitational interactions, at least
sufficiently close to the defect. We will also learn about the
fluid flow pattern around a vortex ring, and how it relates to the
ring's motion. We will study as well what happens when vortices encounter
each other.
In Section~\ref{sec:numerics} we
confirm many of the insights gained here by numerically simulating the formation and
evolution of defects, in the presence of gravity.

\subsection{Domain walls}
\label{sec:walldefect}
To begin, we look for solutions that describe domain walls. These are
the expected defects in $1+1$ dimensions, but should not occur
generically in $3+1$ dimensions. We will see later on that indeed such
defects do not form in the $3+1$ numerical simulations
of a generic gravitational collapse.

First, let us focus on a static defect in one spatial dimension.
With gravity turned off, the Schr\"odinger equation for time-independent wavefunctions takes the extremely simple form
\be
\partial_x^2 \Psi = 0,
\ee
so it is easy to write the most general solution to this equation as
\be
\Psi =  x + b,
\ee
where $b$ is a constant. Without loss of generality, we can
set $b=0$ by choosing the origin appropriately. Thus, 
$x=0$ is where $\Psi$, and therefore $\rho$, vanishes.
This is the location of the sought-after defect. The solution is so
trivial in appearance that one might miss its structure. To make things more apparent, rewrite $\Psi$ as an amplitude and a
phase: 
\be
\Psi = \left\lvert x\right\rvert e^{i\theta} \, .
\label{eq:wallsoln}
\ee
The phase, $\theta$, goes from $-\pi$ when $x <  0$ to $0$ for $x >
0$.
In other words, we learn a couple interesting things about this {\it
  wall} defect: the density goes like $x^2$ where $\left\lvert x\right\rvert$ is the distance
to the defect, and the phase $\theta$ experiences a jump.
One might complain that this solution is not very realistic: the
density diverges far away from the defect. 
Our claim is this: once
gravitational interactions are included, the divergence at infinity is regulated,
 but the description of the density and phase close to the
defect remains correct. This claim will be validated in the presence of gravity in Section~\ref{sec:numerics}.
Gravitational effects should become important parametrically around
the de Broglie wavelength $= 2\pi/(mv)$, where $v$ is the velocity dispersion in the gravitationally bound object.

\subsection{Vortices}
\label{sec:vortices}
Next we turn to consider vortex configurations. In Section~\ref{eq:genvortex}, we argued that these should be the generic defects in three spatial dimensions. In this Section we describe various simple analytic vortex configurations and study their properties in the absence of gravity.\footnote{As in the domain wall case, we expect that these analytic solutions should be reliable close to the defect, and we will later estimate where gravitational corrections become important.} These solutions, and many more complicated ones, can be efficiently constructed utilizing the solution-generating technique of~\cite{2000PhRvA..61c2110B}, which we review in Appendix~\ref{sec:solngenerating}. Some of the solutions that we describe have appeared in the literature before (for example in~\cite{2000PhRvA..61c2110B}), but we wish to review their properties in any case in order to understand the basic features that vortex lines have.

Since vortices are localized at the intersection where both the real and imaginary parts of the wavefunction vanish, we can visualize the evolution of vortex configurations by considering how these vanishing loci evolve in time. In Figure~\ref{fig:vortexsurfaces} we plot the vanishing surfaces for both real and imaginary parts of the wavefunction for some simple examples.

\subsubsection{Static vortex lines}
\label{eq:staticvortices}
Our goal is to find solutions to 
the free Schr\"odinger equation which describe a flow with vorticity. As mentioned before, this requires both the real and imaginary parts of the wavefunction to vanish along a line. We begin by considering the simplest possible vortex solution: an infinitely long straight vortex.

We first look for a time-independent axially-symmetric solution (independent of $t, z$). We therefore want to solve the Laplace equation in the $xy$ plane.
It is convenient to work in the complex coordinates~\eqref{eq:complexcoords}.
In these variables, the Laplace equation takes the simple form
\be
\partial\bar\partial \Psi({\mathfrak z},\bar {\mathfrak z}) = 0 .
\ee
As is well known, this equation is solved by an arbitrary (anti-)holomorphic function. First consider the holomorphic sector, 
$\Psi  \propto  {\mathfrak z}^n$.
It is straightforward to see that this solution describes a vortex located at $x= y=0$ with winding $n$. Similarly, the anti-holomorphic solutions, $\Psi  \propto \bar {\mathfrak z}^n$, describe anti-vortices of winding $-n$.

\begin{figure}[tb]
\centering
\subfigure[]{
\includegraphics[width=.3\textwidth]{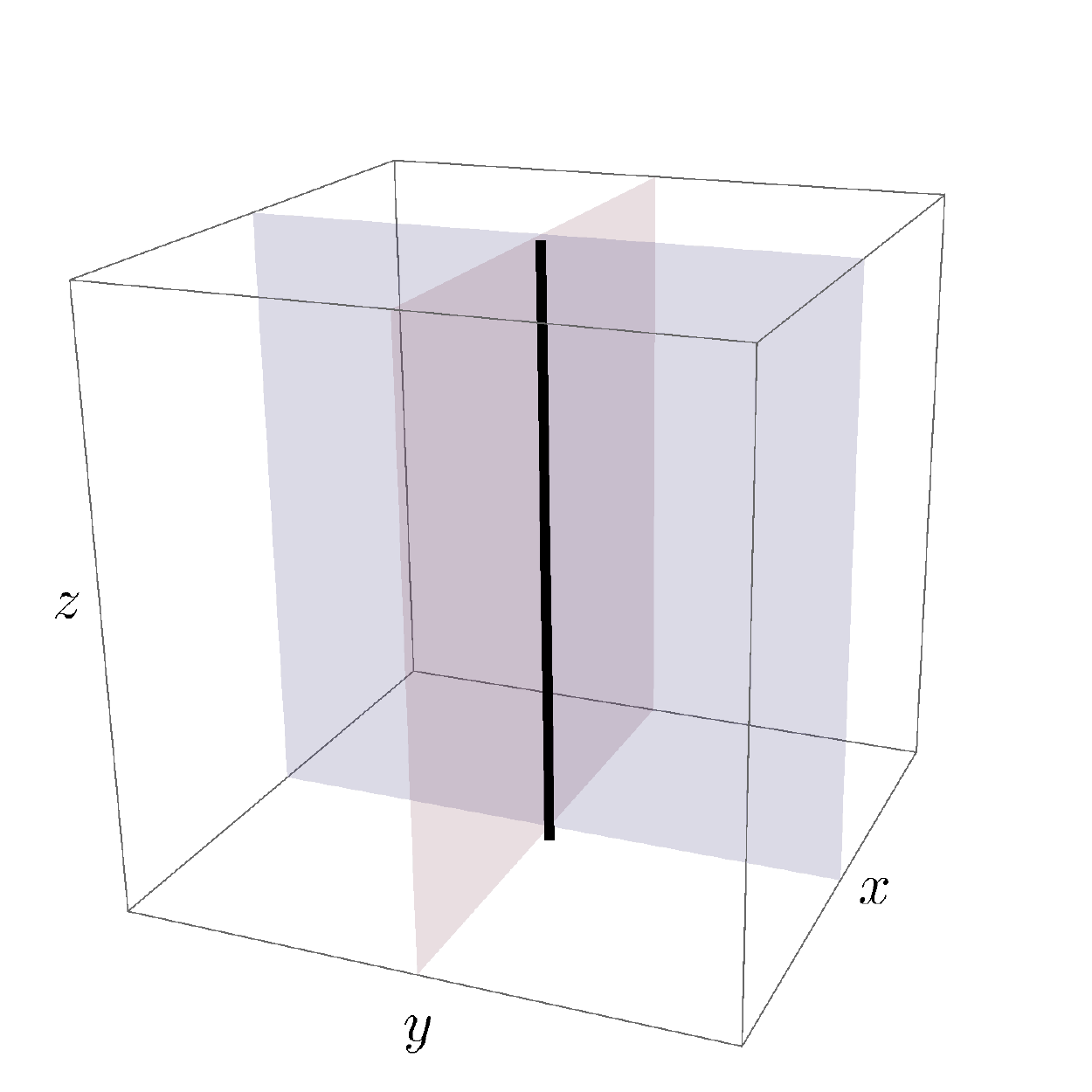}}
\hspace{1.5cm}
\subfigure[]{
\includegraphics[width=.32\textwidth]{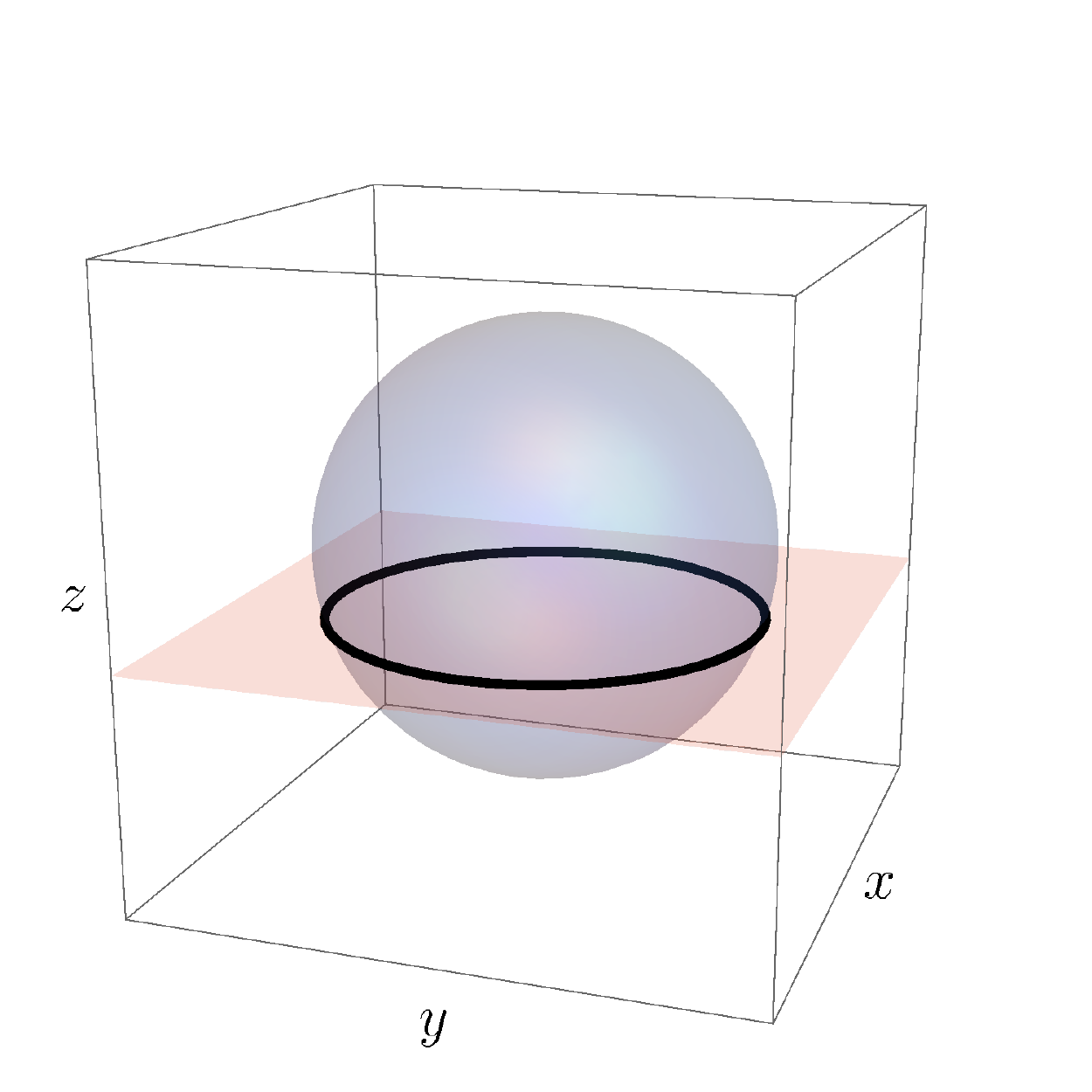}}
\caption
{
\small Loci where the real (blue) and imaginary (red) parts of the wavefunction $\Psi$ vanish for some examples. Vortices are localized on the co-dimension two strings where these surfaces intersect. {\bf a.} Static vortex line configuration where ${\rm Re} \,\Psi$ vanishes on the plane $x=0$ (blue surface) and ${\rm Im}\,\Psi$ vanishes along the plane $y=0$ (red surface). The vortex line lies at their intersection along the $z$ axis. {\bf b.} Dynamical nucleation of a vortex ring. The real part of the wavefunction ${\rm Re}\,\Psi$ vanishes along a sphere centered on the origin. ${\rm Im}\,\Psi$ vanishes in a plane parallel to the $xy$ plane which moves in the $z$ direction with velocity $v_z$. Shown is the intersection of these two surfaces at some intermediate time in the ring's evolution.
}
\label{fig:vortexsurfaces}
\end{figure}

The simplest vortex, of winding $1$, is described by the wavefunction
\be
\Psi = {\mathfrak z} = x+ i y,
\label{eq:winding1vortex}
\ee
The real part of the 
wavefunction vanishes along the $yz$ plane located at $x=0$ and the imaginary part vanishes along the $xz$ plane located at $y=0$. The intersection of these two planes is a line along the $z$ axis, where the total wavefunction vanishes and the vortex is located. 
In Figure~\ref{fig:vortexsurfaces} we visualize the intersection of the zero loci of the real and imaginary parts.

For purely holomorphic or anti-holomorphic vortices, the velocity flow is circular around the position of the vortex and it winds $+n$ or $-n$ times as we travel in a closed loop. However, we can consider an admixture of a vortex and anti-vortex solution
\be
\Psi \propto {\mathfrak z} + \alpha\, \bar {\mathfrak z}.
\ee
As was discussed in Section~\ref{eq:genvortex}, depending on the size of $\alpha$, the solution will behave differently. For $\alpha <1$, this describes a winding $n=1$ solution, but the velocity flow will now be elliptical (equivalently, the change in the argument is no longer constant with angle as we travel around in a circle). Similarly, the lines of constant density will be ellipses in the $xy$ plane. Once we have $\alpha >1$, the solution describes an anti-vortex also with an elliptical flow.
In the degenerate $\alpha=1$ case, the wavefunction happens to be purely real, and the solution now describes a domain wall in the $yz$ plane where the phase jumps by $\pi$ as we cross the wall.

\subsubsection*{Estimate where gravity becomes important} 

For the simplest vortex solutions of winding 1, as we move away from the vortex core, the density grows as 
\be
\rho(r)  = m\lvert\Psi\rvert^2= \rho_{\rm local} \left(\frac{r_\perp}{\lambda_{\rm dB}}\right)^2,
\label{eq:vortexdensityline}
\ee
where $r_\perp \equiv \sqrt{x^2+y^2}$. The proportionality constant is fixed by assuming that the dark matter density in the vicinity of the vortex will grow until it reaches the local background density, denoted by $\rho_{\rm local}$. The natural expectation is that this will occur around the de Broglie wavelength $\lambda_{\rm dB} = 2\pi/(mv)$.

This distance coincides with the characteristic length where we can no longer ignore gravitational effects. In the
vicinity of vortices, gravity is unimportant, but due to the
$r_\perp^2$ growth of the density, at some point it cannot be
neglected. Said differently, we want to estimate when the $m\Phi \Psi$
term in~\eqref{eq:schrodinger} becomes important relative to the
gradient (quantum pressure) term. 
If we assume that gradients in some region scale as the size of the region, $R$, the size of the quantum pressure term will decrease as we go to larger scales. When this term becomes of order the gravitational potential, we have
$\Phi \sim(mR)^{-2}$.
Inside a dark matter
halo, the virial theorem tells us that the gravitational potential is nearly constant $\Phi\sim v^2$ 
so we are able to solve for $R$.
We therefore find 
that gravitational effects become important at distance of order
the de Broglie wavelength $R \sim \lambda_{\rm dB}$. This is perhaps
expected because this is the characteristic length scale that governs
all aspects of the dynamics. Note that once we include the effects of
gravity, we expect the $r_\perp^2$ growth to stop and instead that the
wavefunction will eventually vanish as $r_\perp$ exceeds the size of
the halo, much like what happens for the Hydrogen atom.

\subsubsection*{Hydrogen atom-like wavefunctions}
Another broad class of static solutions can be constructed similarly straightforwardly. These are the wavefunctions that solve the Laplace equation in spherical coordinates $(r,\theta,\varphi)$. In fact, any static solution can be decomposed in this basis of functions. In spherical coordinates, the static Schr\"odinger equation reads
\be
\nabla^2\Psi = \frac{1}{r^2}\partial_r\left(r^2\partial_r \Psi\right)+\frac{1}{r^2}\Delta_{S^2}\Psi=0,
\ee
where $\Delta_{S^2}$ is the Laplacian on the sphere. This equation is easily solved via separation of variables, leading to solutions of the form
\be
\Psi_{l, m} =  \,r^{l}\,Y_l^m(\theta,\varphi).
\label{eq:hydrogenatomlikewfs}
\ee
Here $Y_l^m$ are the standard spherical harmonics.\footnote{We rely on context to differentiate $m$ as in $Y_l^m$
  versus $m$ as the mass of the dark matter particle.}
These solutions are analogous to the wavefunctions of the hydrogen atom, only in this case there is no Coulombic potential. Similar to the hydrogen atom problem, we have imposed regularity of the wavefunction at $r=0$ as a boundary condition.

Not all of the wavefunctions~\eqref{eq:hydrogenatomlikewfs} describe vortex configurations. Rather, only the ones with $m\neq 0$ do. As a special case the configurations with $m=\pm l$ reduce to the solutions considered above:
\be
r^l Y_l^{\pm l} \sim (x\pm iy)^l.
\ee
More generally, a solution with magnetic quantum number $m$ describes a vortex oriented in the $z$ direction with winding $m$. Aside from providing many examples of solutions with vortices, these solutions help illuminate the relation between angular momentum and the presence of vortices, which we discuss further in Section~\ref{sec:conclude}.

\subsubsection{Moving vortex solutions}
Once we have found a solution with a stationary vortex configuration, we can exploit the symmetries of the Schr\"odinger equation to obtain
solutions with vortices moving at constant velocity. A general Schr\"odinger group transformation acts quite nontrivially on the wavefunction~\cite{Niederer:1972zz}, but for our purposes, we only need the transformation of the wavefunction under 
galilean boosts:\footnote{The commutators of the Schr\"odinger algebra can be found, for example, in~\cite{Son:2008ye}.}
\be
\Psi(t,\vec x) \mapsto\exp\left(-im \vec x\cdot \vec v- \frac{im}{2} \vec v^2 t
\right)\Psi(t,\vec x+\vec v t),
\label{eq:wavefboost}
\ee
where $\vec v$ is the boost parameter. 
Using this we can transform a static configuration into one moving at velocity $\vec v$. 
For example, an infinitely long vortex line moving with velocity $-v$ in the $x$ direction is given by
\be
\Psi =e^{-im xv - \frac{imv^2 t}{2}}(x+v t+iy).
\ee
It is also possible to generate boosted versions of more complicated vortex configurations. Note that these boosted configurations are somewhat peculiar in that the entire fluid is also boosted, so they do not represent solutions where only the vortex is moving through a fluid at rest.

\subsubsection{Vortex rings}
The infinitely long vortex lines discussed so far are the simplest configurations, and they capture some important aspects of the physics, but we do not expect that these precise configurations will arise in dark matter haloes, which are of finite extent. In more realistic situations, the generic defects will consist of closed vortex loops (consistent with the overall conservation of vorticity at infinity). We therefore would like to explore the properties of simplified vortex ring solutions. The simplest example is given by
\be
\Psi_{\rm ring}(\vec x, t) = r_\perp^2 -\frac{\ell^2}{4} +2i\left(-az+\frac{ t}{m}\right)
\ee
where $r_\perp = x^2 + y^2$, and $a$ is a parameter with units of length. The real part of the wavefunction vanishes on a cylinder oriented in the $z$ direction with diameter $\ell$, while the imaginary part vanishes in the $x y$ plane with a $z$ coordinate that increases at a constant rate with time. This configuration therefore describes a circular ring of radius $\ell/2$ moving at $v_{\rm ring} = (am)^{-1}$ in the $+z$ direction. Note that this velocity is the speed of the ring as measured relative to the rest frame of the fluid very far from the vortex.
 
We can gain some intuition for the properties of this solution by expanding $r_\perp = \ell/2 +\delta r_\perp$ so that the near-ring solution takes the form (at $t=0$)
\be
\Psi_{\rm near-ring}(\vec x, 0) = \ell\,\delta r_\perp - 2aiz.
\ee
In the $(\delta r_\perp, z)$ plane, we can write this in terms of the complex coordinate $u\equiv \delta r_\perp + iz$ as
\be
\Psi_{\rm near-ring}(u,\bar u) = \left(\ell/2-a\right) u+ (\ell/2+a) \bar u.
\label{eq:nearring}
\ee
From this, we see that for a generic value of $a$, the fluid flow around this ring is non-circular, corresponding to a mix of a vortex and anti-vortex, though it will always have winding $-1$. For $a=\ell/2$, the near-ring geometry is precisely that of a pure vortex of winding $-1$ and in this case, 
there is a relation between the size of the ring and its velocity, $v_{\rm ring} \sim (m\ell)^{-1}$, which is consistent with the behavior of superfluid vortex rings. 
We do not expect realistic vortex rings to be circular, but even in
that case we generically expect the characteristic scale that governs the dynamics, $a$, to continue to be $a\sim\ell$, so that the relation between size and velocity persists.
\begin{figure}[tb]
\centering
\includegraphics[width=.45\textwidth]{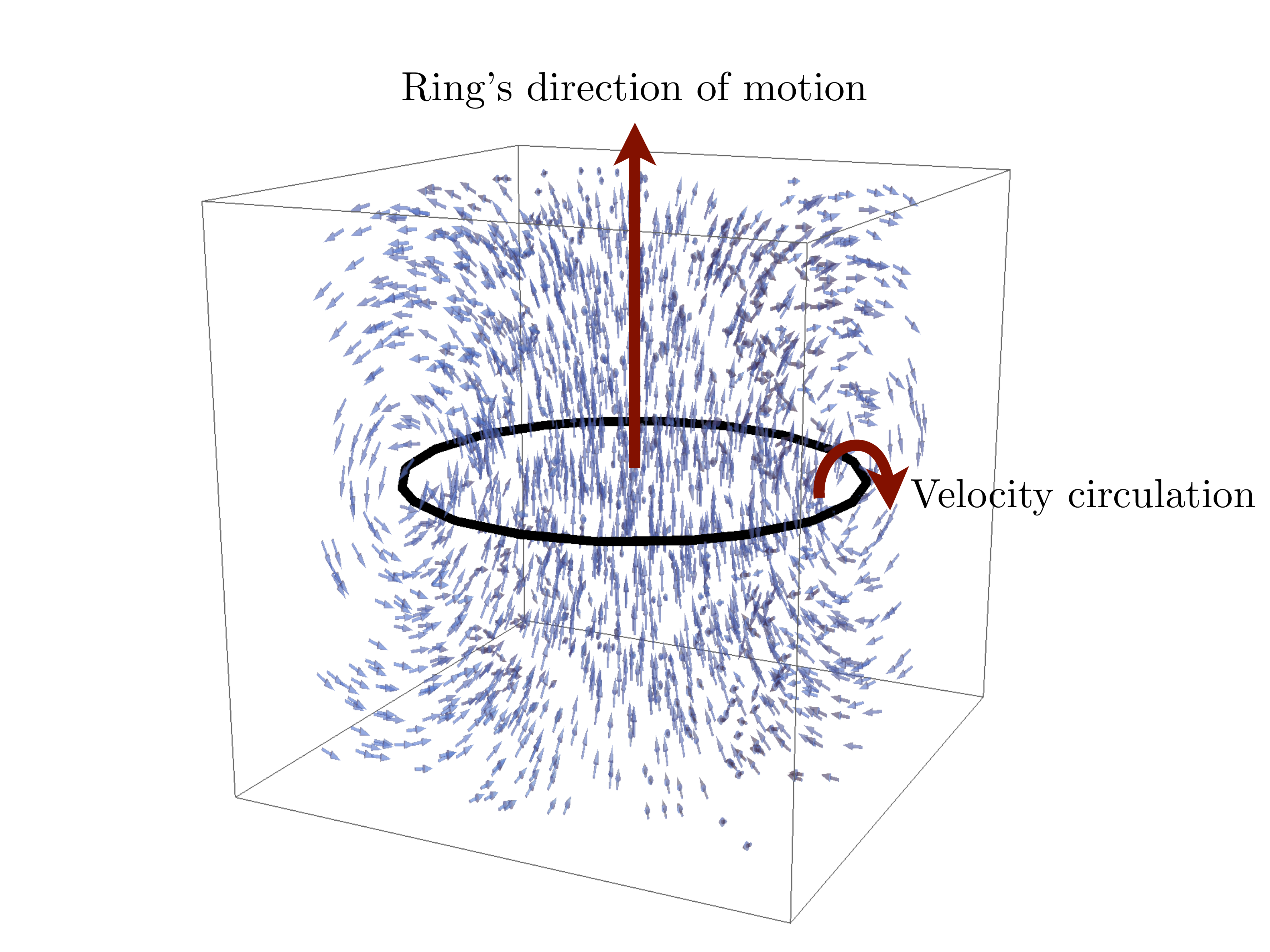}
\vspace{-.1cm}
\caption
{
\small Cartoon of the velocity flow around a vortex ring moving in the $z$ direction. Note that the fluid velocity is such that it flows through the ring in the same direction that the ring is moving. The flow then circulates around the ring as shown.
}
\label{fig:vortexringflow}
\end{figure}

In Figure~\ref{fig:vortexringflow} we show the velocity flow around a vortex ring with $a=\ell/2$. Note in particular that the fluid velocity flows through the center of the ring in the same direction that the vortex moves. We also depict the velocity circulation around the ring. An interesting feature of the fluid flow is that the ring moves with a different velocity than the fluid around it. For example, at the very center of the ring, the fluid velocity is $v_{\rm fluid} = 2v_{\rm ring}$.

We can also investigate the density profile in the vicinity of a vortex ring. Very close to the ring, where the near-ring approximation~\eqref{eq:nearring} is reliable, the density grows as $r_\perp^2$ as expected, where $r_\perp$ is the distance from the vortex ring. Note however that very far from the ring, the density profile is anisotropic. In the plane of the ring, the density grows like $r_\perp^4$. However, in the transverse direction, the density only grows like $z^2$. These precise scalings depend on the fact that the ring in the solution lies in a plane,
so we do not expect them to persist in more realistic scenarios, where the only truly robust signature will be $r_\perp^2$ growth of the density in the near-ring region.

\subsubsection*{Nucleation of a vortex ring}
An important feature of light dark matter vortices is that they can appear and disappear dynamically. We can understand this phenomenon in a simple analytic setting by considering the nucleation of a vortex ring.
We take the real part of the wavefunction to vanish along a 2-sphere and have the imaginary part again vanish along a plane moving in the $z$ direction. The wavefunction in this case is given by:
\be
\Psi_{R}(\vec x, t) = r^2 -\frac{\ell^2}{4} + i\left(-\ell
  z+\frac{3t}{m}\right) \, ,
\label{eq:vortexringnucl}
\ee
where $r^2 = x^2 + y^2 + z^2$.
This describes the nucleation of a vortex ring at $x = y = 0$, $z = -\ell/2$ at $t = -\frac{m\ell^2}{6}$ which moves in the $z$ direction with velocity, $v = \frac{3}{m\ell}$. The radius of the ring at height $z$ is given by $R = \sqrt{\ell^2/4 - z^2}$. The vortex ring re-combines and annihilates at $t = \frac{m\ell^2}{6}$. This is visualized in Figure~\ref{fig:vortexsurfaces}.

Note that the vortex ring moves at constant velocity in the $z$ direction, but the $v_x$ and $v_y$ components of points on the ring change as a function of its size, so that the magnitude of the ring's velocity increases as the size of the ring decreases. This behavior is consistent with our expectation from Taylor expansion~\eqref{eq:vortexspeed}, which suggests that smaller vortex rings move with faster linear velocity. In fact, as the ring shrinks to zero size, the velocity in the $x$ and $y$ directions becomes large, which may lead to interesting observational signatures, as we explore more in Section~\ref{sec:observables}.\footnote{The velocity approaches the speed of
  light as the ring size approaches the Compton scale. By this scale,
  the non-relativistic wavefunction description also breaks down.
}

\subsection{Configurations of vortices}

It is also illuminating to 
 consider slightly more complicated vortex configurations where vortex-anti vortex pairs spontaneously appear and disappear, as well as solutions where vortices intersect.

\subsubsection*{Nucleation of vortex--anti-vortex pair}
If we choose the imaginary plane in eq.~\eqref{eq:vortexringnucl} to instead be oriented in the $yz$ direction, this will describe the intersection of a plane and a cylinder along two lines, which has the interpretation of the creation of a vortex--anti-vortex pair. We have a solution of the form\footnote{Note that this is again not the most general solution, we have chosen the factors so that the velocity flow around the vortices is circular.}
\be
\Psi_{\rm v\bar v}(\vec x, t) = x^2 + y^2 -\frac{\ell^2}{4} +2i\left(-\frac{\ell x}{2}+\frac{ t}{m}\right).
\ee
For $t < -m\ell^2/4$, this solution has no vortices. At $t= -m\ell^2/4$ a vortex--anti-vortex pair nucleate at $x = -\ell/2, y=0$, they move around the ring $x^2+y^2 = \ell^2/4$ in opposite directions, being separated by $\ell$ at $t=0$ and recombine at $t =m\ell^2/4$, after which they annihilate.

\subsubsection*{Intersection and re-connection of vortices}
\label{sec:reconnect}
A natural question to ask about vortex lines is: what happens when they intersect? The generic behavior of the analytic solutions we have considered is for the vortices to re-connect whenever they come into contact, as opposed to becoming frustrated and forming a defect network. However, even arranging for vortex lines to intersect is somewhat non-trivial. The generic behavior seems instead to be for them to develop cusps that then meet at a point. As an example, we can consider the simplest solution that describes the collision of two vortex lines:
\be
\Psi = (x+iy)(y+iz)-\frac{t}{m}.
\ee
At $t=0$ this describes two intersecting vortex lines oriented along
the $z$ and $x$ axes. However, as a function of time, this solution
describes two cusped vortices coming together at a point at $t=0$,
which is indistinguishable at that moment from two infinitely long
vortices intersecting. It is possible to construct more complicated
configurations with intersecting vortices---for example solutions
involving linked vortex rings---the fact that they intersect at a point and pinch off appears to be a generic feature.

\section{The random phase model and the number density of vortices}
\label{sec:randomphase}

Now that we have analyzed some generic features and convinced
ourselves that vortices can appear, 
we would like to estimate the frequency of their occurrence.
One might think with a random superposition of waves, the
probability that chance interference takes $\Psi$ all the way to zero
must be negligibly small. 
In fact---as
we will verify using numerical
simulations---such vortices (or defects) are ubiquitous: they exist with a
number density of about one per de Broglie volume. 
In this section, we wish to gain some insight into this phenomenon by
studying a simple analytic model---the random phase model. As a
by-product we will work out statistical properties of
this model, facts which are useful beyond the immediate application to calculating
defect number density---for instance, we will refer to these
properties when we explore
observational and experimental implications in Section
\ref{sec:observables}.
The random phase model is used in some of the axion detection
literature, for
example~\cite{Derevianko:2016vpm,Foster:2017hbq,Centers:2019dyn}; 
here we spell it out in detail and derive several statistical consequences.

\subsection{The random phase model}
\label{rpm}

In order to make progress, we need a stochastic model for $\Psi$. 
This model can be formulated in any number of dimensions.
Suppose that $\Psi$ consists of a superposition of waves of different
momenta, $\vec k$:
\be
\label{PsiExpand}
\Psi (\vec x, t) = \sum_{\vec k} \left(A_{\vec k} \, e^{i B_{\vec k}}\right)e^{  i \vec k
  \cdot \vec x - i \omega_k t} \, ,
\ee
where we have split the coefficients of the plane waves into an amplitude and a phase so that
$A_{\vec k}$ and $B_{\vec k}$ are real. We have represented each
mode as a plane wave; in a realistic halo, the mode functions might be
more complicated and our model can be adapted accordingly (see
footnote~\ref{modefunctions} below). 
In the same vein, $\omega_k$ is assumed to be $k^2/(2m)$ for
simplicity, but the dispersion relation can also be easily modified. 

It is helpful to split $\Psi$ into its real and imaginary parts as
$\Psi = \Psi_1 + i \Psi_2$, where the explicit expressions for its components are:
\begin{align}
\Psi_1 (\vec x, t) &= {1 \over 2} \sum_{\vec k} \left(A_{\vec k} \, e^{i B_{\vec k}}\right) e^{  i \vec k
  \cdot \vec x - i \omega_k t} + \left(A_{\vec k} \, e^{-i B_{\vec k}}\right) e^{ - i \vec k
  \cdot \vec x + i \omega_k t} \, , \\
\Psi_2 (\vec x, t) &= {1 \over 2i} \sum_{\vec k} \left(A_{\vec k} \, e^{i B_{\vec k} }\right)e^{ i \vec k
  \cdot \vec x - i \omega_k t} - \left(A_{\vec k} \, e^{-i B_{\vec k}}\right) e^{- i \vec k
  \cdot \vec x + i \omega_k t}  \, .
\end{align}
In order to streamline expressions, it is 
helpful to introduce a more compact
notation:
\be
\Psi_i (\vec x, t) = {1\over 2} \sum_{\vec k} \left(A_{\vec k} \, e^{i
  b_{\vec k}^i }\right) e^{ i \vec k
  \cdot \vec x - i \omega_k t} + \left(A_{\vec k} \, e^{-i b^i_{\vec k}}\right) e^{ - i \vec k
  \cdot \vec x + i \omega_k t} \,,
\ee
where the index $i$ (to be distinguished from the imaginary number
$i$) equals $1$ or $2$, and 
\be
b_{\vec k}^i =
 \begin{cases}
   \displaystyle B_{\vec k} & {\rm if}~~ i=1  , \\[5pt]
   \displaystyle B_{\vec k} - {\pi \over 2} & {\rm if} ~~ i=2\, .\\
  \end{cases}
\ee

We are interested in the two-point function between all possible combinations of real and imaginary parts
$\langle \Psi_i (\vec x_1, t_1) \Psi_j (\vec x_2, t_2) \rangle$.
The statistical average can be done in two steps, first averaging over
the phase $b^i_{\vec k}$, and then subsequently averaging over the amplitude
$A_{\vec k}$. Note that $A_{\vec k}$ actually does not need to be stochastic at all. A
simple, reasonable, model would be to have the a {\it fixed}  $A_{\vec k}$. For
instance, it can be a Gaussian, representing the distribution of
momenta in an isothermal halo. In what follows, we will take
this point of view, though it is a simple matter to take our
expressions and average over $A_{\vec k}$, if one wishes $A_{\vec k}$ to be
stochastically drawn from some distribution.

The key assumption of the random phase model is about the statistical properties of the phases of the coefficients of the Fourier modes. In particular, we have that
\be
\langle e^{i b^i_{\vec k}} \, \, e^{-i b^j_{\vec k'}} \rangle = \delta_{\vec
  k, \vec k'} \sigma_{ij} \qquad {\rm where} \qquad \sigma_{11} =
  \sigma_{22} = 1 \,\, , \,\, \sigma_{12} = -\sigma_{21} = i \, .
  \label{eq:randomphasestatistics}
\ee
This holds if the phase, $B_{\vec k}$, for each Fourier mode is independently
randomly drawn from a uniform distribution between $0$ and $2\pi$. 
One can think of this as the analog of populating a halo with stars,
where the stellar orbits have random phases.
With this understanding, the one-point function vanishes (as do all
correlation functions involving an odd number of points). 
For the two-point function, it can be shown using~\eqref{eq:randomphasestatistics} that
\be
\label{2pointPsi}
\langle \Psi_i (\vec x_1, t_1) \Psi _j(\vec x_2, t_2) \rangle
= \frac{1}{ 4} \sum_{\vec k}  A_{\vec k}^2\left( \sigma_{ij} \, e^{i \vec k \cdot
  (\vec x_1 - \vec x_2) - i \omega_k (t_1 - t_2)} + \,
  \sigma_{ji} \, e^{-i \vec k \cdot
  (\vec x_1 - \vec x_2) + i \omega_k (t_1 - t_2)}\right)  \, .
\ee
It is worth noting that $\Psi_1$ and $\Psi_2$ are in general
correlated. However, at equal time $t_1 = t_2$, and
assuming isotropy so that $A_{\vec k} = A_{-{\vec k}}$, they are
uncorrelated. 

Using the two-point function~\eqref{eq:randomphasestatistics}, we can compute higher-point functions by Wick contraction. For example we have
\be
\langle e^{i b_{\vec k_1}^{ i_1}} \, \, e^{i b_{\vec k_2}^{ i_2}} \,\,
  e^{- i b_{\vec k_3}^{i_3}} \, \, e^{- i b_{\vec k_4}^{ i_4}} \rangle
  = 
\delta_{\vec k_1 , \vec k_3} \delta_{\vec k_2 , \vec k_4} \sigma_{i_1
  i_3} \sigma_{i_2 i_4} + \delta_{\vec k_1 , \vec k_4} \delta_{\vec k_2 , \vec k_3} \sigma_{i_1
  i_4} \sigma_{i_2 i_3}  \, .
\ee
Using this formula, we can likewise compute the four-point function of the real and imaginary parts of $\Psi$:
\be
\begin{aligned}
\label{WickThm}
\langle \Psi_{i_1} (\vec x_1, t_1) \Psi_{i_2} (\vec x_2, t_2)
  \Psi_{i_3} (\vec x_3, t_3) \Psi_{i_4} (\vec x_4, t_4) \rangle
=&~ \langle \Psi_{i_1} (\vec x_1, t_1) \Psi_{i_2} (\vec x_2, t_2)
  \rangle \langle \Psi_{i_3} (\vec x_3, t_3) \Psi_{i_4} (\vec x_4,
  t_4) \rangle
  \\ &+ (2 \leftrightarrow 3) + (2
                                \leftrightarrow 4) \, .
\end{aligned}
\ee
This Wick contraction pattern continues to higher point correlation functions
so that $\Psi_1$ and $\Psi_2$ are {\it correlated} Gaussian random
fields.\footnote{\label{GaussRandomNote}It is worth emphasizing that the random phase assumption alone
is sufficient to guarantee $\Psi_1$ and $\Psi_2$ are Gaussian random
fields. That is, only $B_{\vec k}$ needs to be stochastic.
This might at first sight seem surprising: a real Gaussian
random field in general requires Fourier coefficients that have random
phases, {\it and} that have amplitudes which are stochastically distributed in a
Gaussian fashion. In our model, the {\it complex} field $\Psi$ has
Fourier amplitudes $A_{\vec k}$ which are not (or at least not
necessarily) stochastic. It can be easily checked that its real and
imaginary parts $\Psi_1$ and $\Psi_2$, do have stochastically
distributed Fourier amplitudes, simply from the randomness of the
phases $B_{\vec k}$. 
}

We can differentiate eq.~\eqref{2pointPsi} to obtain correlators
between $\Psi$ and its derivatives. For our application below---computing the average number
density of vortices or defects---we will be interested in correlations
at coincident points, {\it i.e.}, setting $\vec x_1 = \vec x_2$ and $t_1 =
t_2$ (after taking derivatives) which leads to the correlations
\be
\begin{aligned}
\label{eq:PsiPsi}
& \langle \Psi_i \Psi_j \rangle = \delta_{ij} \,\, {1\over 2} \sum_{\vec k}
  A_{\vec k}^2 \, , \\
& \langle \Psi_i \partial_\mu \Psi_j \rangle = {1\over 4} \sum_{\vec k}
  A_{\vec k}^2 \, i k_\mu \left(- \sigma_{ij} + \sigma_{ji} \right) \, ,
 \\
& \langle \partial_\nu \Psi_i \partial_\mu \Psi_j \rangle =
   \delta_{ij} {1\over 2} \sum_{\vec k}
  A_{\vec k}^2 k_\nu k_\mu ,
\end{aligned}
\ee
where we have defined $k_\mu$ so that  $k_0=-\omega_{k}$  and where its spatial components are $\vec k$.\footnote{\label{modefunctions}If we had used some mode
  functions other than the plane waves {\it i.e.}, replacing
  eq.~\eqref{PsiExpand} by $\Psi(\vec x, t) = \sum_{\vec k} A_{\vec k}
  e^{i B_{\vec k}} f_{\vec k} (\vec x, t)$ where $f_{\vec k}$'s are the
  appropriate mode functions for the halo in question and ${\vec k}$ is
  merely a label for them, essentially all the results go through with
  minimal modifications. For instance, the factor of
$e^{i\vec k \cdot (\vec x_1 - \vec x_2) - i \omega_k (t_1-t_2)}$ in
  eq.~\eqref{2pointPsi} is replaced by $f_{\vec k} (x_1) f^*_{\vec k}
  (x_2)$. For eq.~\eqref{eq:PsiPsi}, the right hand side for each line reads:
$\delta_{ij} \sum_{\vec k} A^2_{\vec k} \left\lvert f_{\vec k}\right\rvert^2 /2$, 
$\sum_{\vec k} A^2_{\vec k} (\sigma_{ij} f_{\vec k} \partial_\mu
f_{\vec k}^* + \sigma_{ji} f^*_{\vec k} \partial_\mu f_{\vec k} )/4$,
and $\sum_{\vec k} A^2_{\vec k} (\sigma_{ij} \partial_\nu f_{\vec
  k} \partial_\mu f_{\vec k}^* + \sigma_{ji} \partial_\nu f^*_{\vec
  k} \partial_\mu f_{\vec k})/4$, respectively. Note that in general,
these expressions depend on space and time ($\vec x_1$, $t_1$). Even in this more general context, eq.~\eqref{WickThm}
remains correct.}
If the distribution of momenta is isotropic so that $A_{\vec k} =
A_{-\vec k} = A_k$, then~\eqref{eq:PsiPsi} simplifies to:
\be
\begin{aligned}
& \langle \Psi_i \Psi_j \rangle = \delta_{ij} \,\, {1\over 2} \sum_{\vec k}
  A_k^2 \, , \\
& \langle \Psi_1 \partial_t \Psi_2 \rangle = - \langle
   \Psi_2 \partial_t \Psi_1 \rangle = {1\over 2} \sum_{\vec k}
  A_{k}^2 \, \omega_k \, , \\
& \langle \partial_t \Psi_i \partial_t \Psi_j \rangle = \delta_{ij}
   {1\over 2} \sum_{\vec k} A_{k}^2 \, \omega_k^2 \, , \\
& \langle \partial_m \Psi_i \partial_n \Psi_j \rangle = \delta_{ij}
   \delta_{mn} 
   {1\over 2d} \sum_{\vec k} A_{k}^2 \,k^2 \, .
\end{aligned}
\ee
Here the simplifications occur because the positive and negative $\vec k$ terms in many of the sums cancel out.

As an example of how these quantities might be used, suppose we are
interested in the joint probability distribution of $\Psi_1$,
$\Psi_2$ and their spatial derivatives at a single point. Recalling that $\Psi_1$ and
$\Psi_2$ are Gaussian random fields, the distribution is given by---assuming 2 spatial dimensions:
\be
\label{P6D}
P\left(\Psi_i, \partial_j \Psi_k\right) = {1\over (2\pi)^3 \Gamma^2
  \tilde\Gamma^4}\exp \left(- {\Psi_1^2 +
  \Psi_2^2 \over 2\Gamma^2} - { (\partial_1 \Psi_1)^2 + (\partial_2
  \Psi_1)^2 + (\partial_1 \Psi_2)^2 + (\partial_2 \Psi_2)^2 \over
  2 \tilde\Gamma^2} \right) \, ,
\ee
where the quantities $\Gamma$ and $\tilde\Gamma$ are defined by:
\be
\label{Gammadef}
\Gamma^2 \equiv {1\over 2} \sum_{\vec k} A_k^2\, , \qquad \qquad 
\tilde\Gamma^2 \equiv {1\over 2d} \sum_{\vec k} A_k^2\, k^2 \, .
\ee
Here $P\left(\Psi_i, \partial_j \Psi_k\right)$ is a shorthand for the 6-dimensional probability distribution
\be
P\left(\Psi_i, \partial_j \Psi_k\right)\equiv P(\Psi_1, \Psi_2, \partial_1 \Psi_1, \partial_2 \Psi_1 , \partial_1
  \Psi_2 ,\partial_2 \Psi_2).
\ee
Note that this probability has
has been properly normalized so that
integrating it over its 6 arguments yields unity.

An even simpler application is to consider the probability distribution of $\Psi_1$ and $\Psi_2$ at a single point.  That is, consider\footnote{We use the same symbol $P$ to denote probability
distribution in different contexts, so for example $P(\Psi_1, \Psi_2)$ should
not be confused with $P(\Psi_1, \Psi_2, \partial_1\Psi_1, \partial_2
\Psi_1, \partial_1 \Psi_2, \partial_2\Psi_2)$, nor with $P(|\Psi|)$ below.}
\be
\rd\Psi_1 \rd\Psi_2 P(\Psi_1, \Psi_2) = \rd\Psi_1 \rd\Psi_2 {1\over 2\pi
  \Gamma^2} {\,\rm exp\,} \left(- {\Psi_1^2 +
  \Psi_2^2 \over 2\Gamma^2}\right)
\ee
If we rotate this to ``polar coordinates'' by defining $\Psi_1 =
\left\lvert\Psi\right\rvert  \cos\theta$ and $\Psi_2 = \left\lvert\Psi\right\rvert \sin\theta$, we
see that the probability distribution is independent of $\theta$, and
the probability distribution for $\left\lvert\Psi\right\rvert = \sqrt{\Psi_1^2 + \Psi_2^2}$
is 
\be
\label{RayleighP}
\rd \left\lvert\Psi\right\rvert P\left(\left\lvert\Psi\right\rvert\right) = \rd\left\lvert\Psi\right\rvert {\left\lvert\Psi\right\rvert \over \Gamma^2} {\,\rm exp\,}
  \left(- {\left\lvert\Psi\right\rvert^2\over 2\Gamma^2}\right) \, .
\ee
The above reproduces
the Rayleigh distribution for the field amplitude, discussed in~\cite{Centers:2019dyn}.
In Section~\ref{sec:numerics}, we compare this distribution against
the distribution measured from numerical simulations of haloes from
gravitational collapse. We will see that this distribution captures
the low density, or low $\lvert\Psi\rvert$, part well, but not the high density
tail.

It is worth noting that with the distribution given by
eq.~\eqref{RayleighP}, the expectation value of $\left\lvert\Psi\right\rvert$ is $\langle \left\lvert\Psi\right\rvert \rangle =
\sqrt{\pi} \Gamma/2$, while its variance is given by $\langle \left\lvert\Psi\right\rvert^2 \rangle = 2 \Gamma^2$. 
Recalling that the dark matter density is given by $\rho = m\left\lvert\Psi\right\rvert^2$, we see that the
probability distribution for density is exponential:
\be
\label{expP}
\rd\rho \, P(\rho) = {\rd\rho\over \bar\rho} \, e^{-\rho/\bar\rho} \, ,
\ee
where $\bar\rho$ is the mean density $\bar\rho=\langle \rho \rangle = 2 m
\Gamma^2$. More generally, the higher moments of the distribution are given by $\langle \rho^n \rangle = \Gamma (n+1)
\, \bar\rho \, {}^n$, where $\Gamma(n+1)$ is the Gamma function, not to be
confused with $\Gamma$ above.\footnote{For instance: $\Gamma (n+1) = n!$ for integer $n$,
$\Gamma (3/2) = \sqrt{\pi}/2$ and so on.}
The fact that the
density distribution
peaks at $\rho = 0$ is a distinctive feature of destructive
interference. The locations where $\rho$ vanishes are precisely the
sites of the vortices we have been focusing on.

Direct detection experiments are typically sensitive to the real scalar field, $\phi$. It is straightforward to deduce the 
statistical properties of $\phi$ from the relation to $\Psi$:
\be
\label{phiFromPsi}
\phi (\vec x, t) = {1\over \sqrt{2m}} \Big( 2 \Psi_1 (\vec x , t)
  {\,\rm cos\,}(mt) + 2 \Psi_2 (\vec x , t) 
  {\,\rm sin\,}(mt) \Big) \, ,
\ee
along with the facts that in the random phase model $\Psi_1$ and $\Psi_2$ are correlated Gaussian random fields
with the correlation functions given by eq.~\eqref{2pointPsi}. The
amplitude of $\phi$ oscillations is essentially $\lvert\Psi\rvert = \sqrt{\Psi_1^2 + \Psi_2^2}$.
We will use this random phase model to explore implications for axion detection
experiments in~\S\ref{sec:observables}.

\subsection{The number density of vortices}
\label{sec:numberdensityvortices}

Before carrying out the computation of the average number of vortices
or defects, let us discuss what this means in different number of spatial dimensions.
Recall in one dimension (where the defects take the form of walls), 
it is not generic to have both the real and imaginary parts of $\Psi$
vanish at the same point; thus it is not useful to compute the
average number density of defects in that case.
In three spatial dimensions, the vanishing of both the real and
imaginary parts of $\Psi$ typically occurs along a line, and since such a line
should not end, generically one expects a vortex ring defect. It
is therefore meaningful to ask what the number density of vortex rings is. 
However, working with an extended object like a ring is not so straightforward:
computing the expected number density analytically involves considering the
multiple-point probability distribution for $\Psi$, which is somewhat complex.
The case of two spatial dimensions is
special: here, the vortex or defect takes the form of a point, and it is
possible to work out the average number density of such points
by considering the one-point probability
distribution for $\Psi$ and its derivatives---a problem we can solve exactly.

%

The number density of vortex defects can be expressed as
\be
\label{ndefect}
n_{\rm vortex} (\vec x) = \sum_a \delta (\vec x - \vec x_a) \, ,
\ee
where $\vec x_a$'s are the locations where $\Psi$ vanishes in some volume, and
$\delta$ represents the Dirac delta function. If we are interested in the average number of vortex defects averaged over many realizations we can compute it schematically as~\cite{Bardeen:1985tr}
\be
\langle n_{\rm vortex}\rangle  = \int {\cal D}\Psi(\vec x)  \sum_a \delta (\vec x - \vec x_a) P\left(\Psi(\vec x)\right)\,,
\ee
where $P\left(\Psi(\vec x)\right)$ is the probability density for general field configurations. In order to actually evaluate this integral we have to make some simplifying assumptions, we 
Taylor expand $\Psi$ around its vanishing points $\vec x_a$, as
\be
\Psi (\vec x) \simeq \vec \nabla \Psi \big|_{\vec x_a} \cdot (\vec x - \vec x_a) \, .
\ee
We are interested in two spatial dimensions: let us use $x_1$ and $x_2$ to denote the
two spatial coordinates (with $\partial_1$ and $\partial_2$ the
respective derivatives).\footnote{Note we use the index in two distinct ways: the index $a$ of
  $\vec x_a$ (or $1$ as in $\vec x_1$ in \S \ref{rpm}) labels which
  particular point we are interested in; the index $i$ of $x_i$ ($i =
  1, 2$ for two spatial dimensions) labels
  the spatial component.
}
It is convenient to use $\Psi_1$ and $\Psi_2$ to denote the real and imaginary
parts of $\Psi$ as we did earlier, so that $\Psi = \Psi_1 + i \Psi_2$. Thus, $\vec \nabla
\Psi$ can be thought of as a $2 \times 2$ matrix which can be inverted
to express $\vec x - \vec x_a$ in terms of $\Psi$ and its
derivatives. We can then change variables in the delta function to obtain
\be
\langle n_{\rm vortex} \rangle = \int \rd^4 (\partial \Psi)\, \rd \Psi
 \, \left\lvert\det\vec\nabla\Psi\right\rvert \delta(\Psi)\,P(\Psi, \partial \Psi) \, .
\ee

We can write the average number density of point vortices in two dimensions in a slightly more explicit form as
\be
\label{averagen}
\langle n_{\rm vortex} \rangle = \int \rd^4 (\partial \Psi) 
 \, \left\lvert\partial_1 \Psi_1 \partial_2 \Psi_2 - \partial_2 \Psi_1 \partial_1
  \Psi_2\right\rvert \, P(\Psi_1 = 0, \Psi_2 = 0, \partial \Psi) \, ,
\ee
where 
we use $\partial \Psi$ as a shorthand for
the 4 combinations of derivatives of $\Psi_1$ and $\Psi_2$.
The probability density $P$ is the 6-dimensional joint
distribution for $\Psi_1$, $\Psi_2$ and $\partial\Psi$. 
Note that $P \, \rd\Psi_1\, \rd\Psi_2\, \rd^4 (\partial\Psi)$ gives the probability
for these 6 quantities taking the prescribed values
at a single point in space.

\begin{figure}[tb]
\centering
\subfigure[]
{\label{fig:contourphase}
\includegraphics[width=0.48\textwidth]{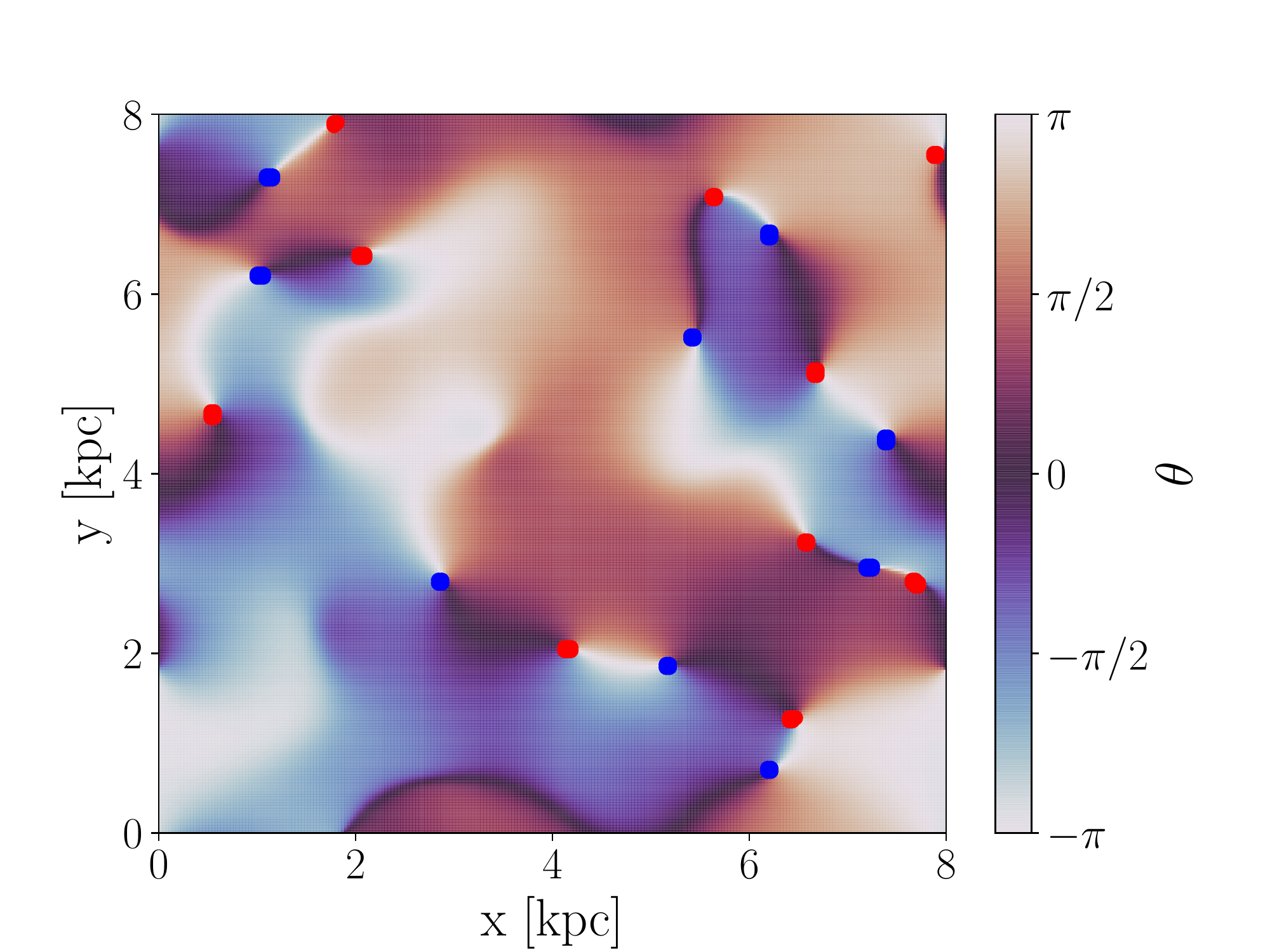}
}
\subfigure[]
{\label{fig:contourdensity}
\includegraphics[width=0.48\textwidth]{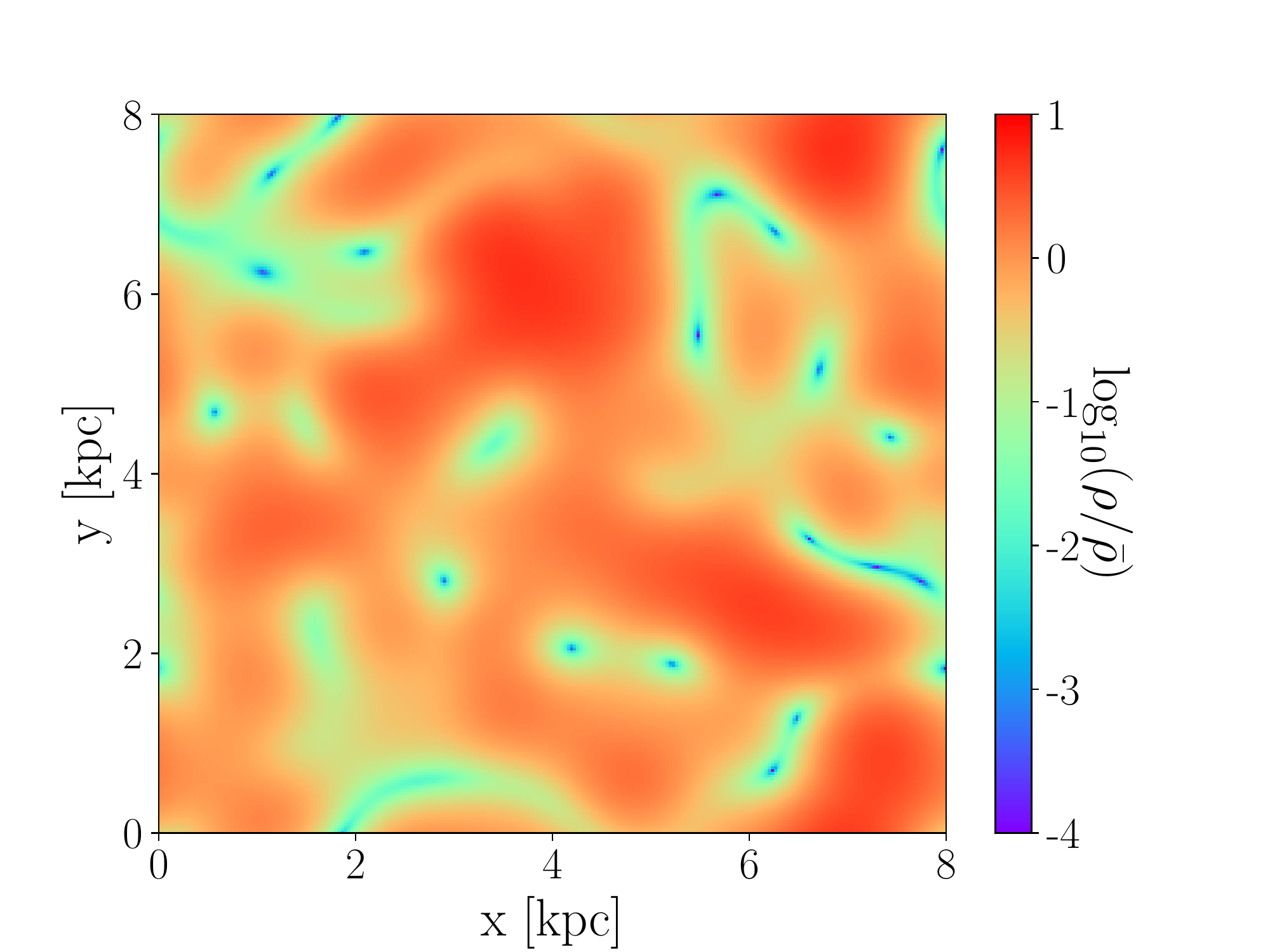}
}
\caption
{\small A snapshot of the phase (left panel) and density (right panel)
  from a numerical realization of the random phase model in two
  spatial dimensions. Here, $A_{\vec k}$ (see eq.~\eqref{PsiExpand}) is chosen to be $\propto
  e^{-k^2/k_0^2}$ with $k_0 = 2.63 {\,\rm kpc}^{-1}$. 
  The corresponding 2D RMS momentum (weighted by $A_{\vec k}^2$---{\it i.e.}, mass
  weighting) is $k_0/\sqrt{2}$, and thus the de Broglie wavelength
  $\lambda_{\rm dB} = \sqrt{2} \,2\pi/k_0 = 3.38 {\,\rm kpc}$. 
  On the left panel, the red and blue dots indicate vortices of
  opposite winding, that is, locations around which the phase winds
  by $2\pi$ (clockwise for red, counter-clockwise for blue).
  On the right panel, the density
  normalized by the mean density $\bar\rho$ is shown on a logarithmic
  scale. It can be seen that the locations of vortices are also sites
  of very low density (consistent with zero), as expected.
}
\label{fig:contour.1.paper}
\end{figure}

The equation~\eqref{averagen} is general, in the sense that it holds for any
probability distribution $P$. To make progress, we need a statistical
model for $\Psi$, and this is where we employ the random phase model. 
Plugging the expression for $P$ from eq.~\eqref{P6D} into
eq.~\eqref{averagen}, we find:\footnote{Evaluating this expectation
  value takes some work. The integral~\eqref{averagen} can be done by rotating to polar coordinates in field space and using the Gaussian probability distribution for $\partial\Psi$~\eqref{P6D}.}
\be
\langle n_{\rm vortex} \rangle = {1\over 2\pi} {\tilde \Gamma^2 \over \Gamma^2}\,,
\ee
where $\Gamma$ and $\tilde\Gamma$ are defined in eq.~\eqref{Gammadef}.
Suppose $A_k$, which enters into the definitions for $\Gamma$ and
$\tilde\Gamma$, takes the form $A_k\propto e^{-k^2/k_0^2}$. 
Replacing the summation over $\vec
k$ by an integral, we arrive at the average density of vortices
({\it i.e.}, number of zero crossings) in two dimensions:\footnote{In other words, we replace $\sum_{\vec k} \mapsto L^2/(2\pi)^2
  \int {\rd^2 k}$ in two spatial dimensions, where $L$ is the infrared
  length scale so that $2\pi/L$ is the fundamental mode. For three spatial
  dimensions, insisting $\langle \rho \rangle = m (\langle \Psi_1^2
  \rangle + \langle \Psi_2^2 \rangle) = m \sum_{\vec k} A_k^2$ would
  mean $A_k = \sqrt{\langle\rho\rangle/m} (\sqrt{8\pi}/L k_0)^{3/2}
  {\,\rm exp\,}(-k^2/k_0^2)$. 
}
\begin{equation}
\label{ndefectresult}
\langle n_{\rm vortex} \rangle = {k_0^2 \over 8 \pi} \, .
\end{equation}
Note that $k_0^2/2$ is the momentum variance in 2D.\footnote{In other words, $\int \rd^2 k \, k^2 \, {\,\rm
    exp\,}(-2 k^2/k_0^2) / \int \rd^2 k  {\,\rm
    exp\,}(-2 k^2/k_0^2) = k_0^2/2$. This gives the momentum variance
  in two-dimensions, and so the momentum variance in say the
  $x$-direction is $k_0^2/4$. We square $A_k$ in computing this average
because it is $\left\lvert\Psi\right\rvert^2$ that contributes to the mass density. 
}
The associated de Broglie wavelength is $\lambda_{\rm dB} = \sqrt{2} \, 2\pi/k_0$. 

The average vortex density can be rewritten as 
$\langle n_{\rm vortex} \rangle = \pi/\lambda_{\rm dB}^2$, which is roughly 
3 per de Broglie area.
As a check, we numerically generate realizations of the random phase model. A
two-dimensional example is shown in Figure \ref{fig:contour.1.paper},
where the red and blue dots highlight the vortices of opposite
winding. All vortices have $+1$ or $-1$ winding---higher winding does
not appear, consistent with our arguments in Section~\ref{sec:general}.
The analytic prediction for the expected number
density of vortices is confirmed by these simulations.

  \begin{figure}[tb]
\centering
\subfigure[]
{\label{fig:waves2}
\includegraphics[width=.4\textwidth]{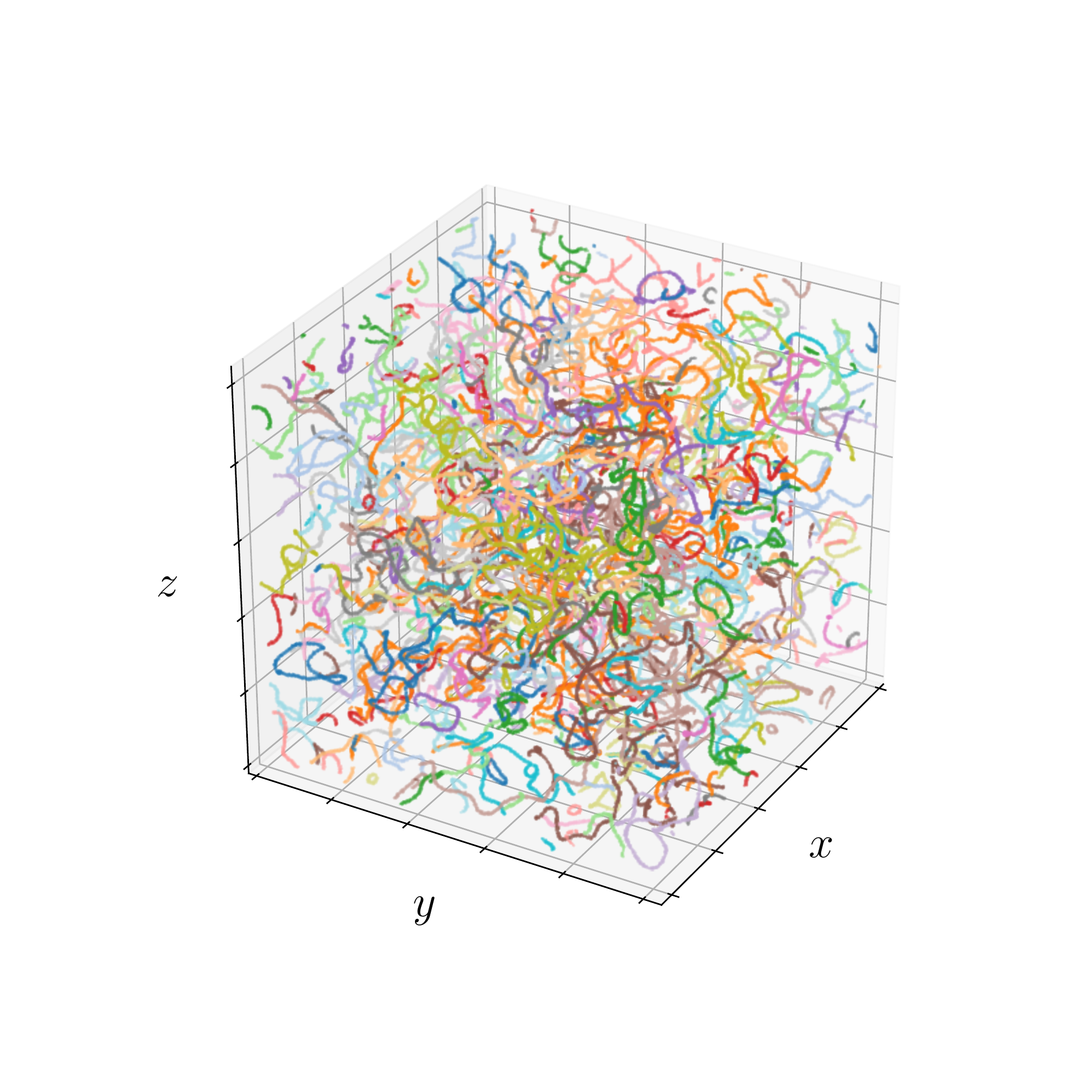}
}
\subfigure[]
{\label{fig:sizehist2}
\includegraphics[width=.48\textwidth]{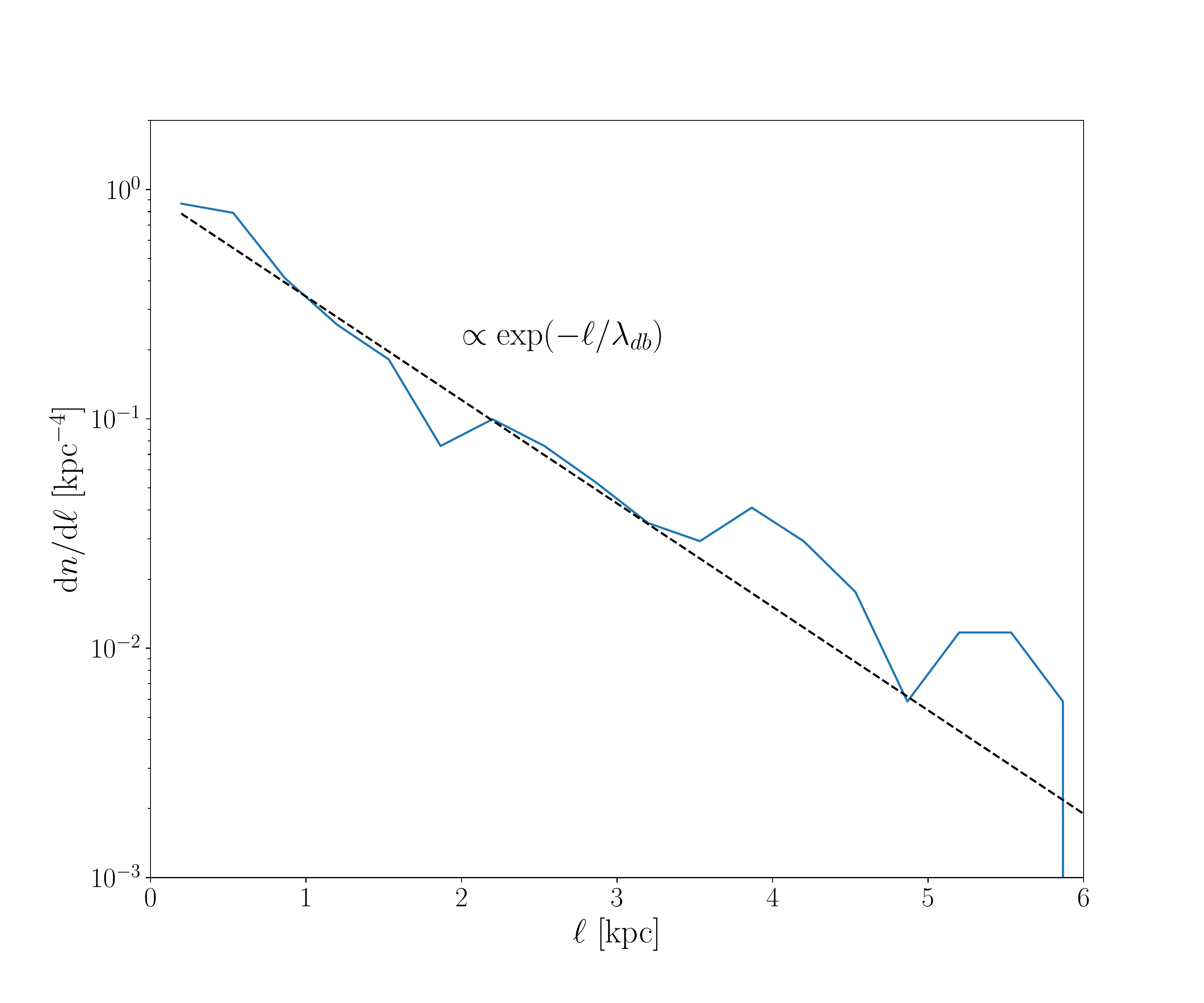}
}
\vspace{-0.25cm}
\caption
{\small Numerical realizations of the random phase model in three dimensions. {\bf a.} The lines highlight vortices in a three-dimensional
  simulation of the random phase model. All vortices take the form of
  vortex rings---note that cases that appear not to form a closed loop
  actually do once accounting for the periodic boundary condition.
{\bf b.} Size distribution of vortex rings in a
  three-dimensional random phase simulation. The size of a ring is
  defined to be the maximum distance between two points on the ring. 
Here, the de Broglie wavelength $\lambda_{\rm dB} = 1.15$ kpc. The size distribution is approximately exponential.
}
\label{fig:randomphases}
\end{figure}

We also generated three-dimensional realizations of the random phase
model. Here, the vortices take the form of rings, and as discussed
earlier, an analytic prediction of the average number density of rings
is more challenging to work out, but we can measure it
straightforwardly from the
numerical realizations. The average number density of vortex rings
comes out to be about $1.5/\lambda_{\rm
  dB}^3$, where $\lambda_{\rm dB}$ is $2\pi$ divided by the square
root of the 3D momentum variance.
An example is shown in Figure \ref{fig:waves2}. The
vortex identification algorithm is discussed in \S \ref{sec:numerics}.

It is also possible to measure the size distribution of rings
(defined as the maximum distance between two points on a ring)---this
is shown in Figure \ref{fig:sizehist2}. The distribution is roughly
exponential:
\be
{\rd n_{\rm vortex} \over \rd\ell} \sim {1.5 \over \lambda_{\rm dB}^4}
  e^{-\ell/\lambda_{\rm dB}} \, ,
\ee
where $\ell$ is the size of a ring, 
and $n_{\rm vortex}$ is the average 
number density of vortex rings. The fact that large rings
are rare is not surprising: the real space correlation function of
$\Psi$ is suppressed beyond the de Broglie scale. The fact that the
distribution is roughly flat for small ring sizes might come as a
surprise, given that the momentum distribution is suppressed at high
$k$'s. It should be kept in mind that ring size (or more precisely,
the local ring curvature) is determined by 
a nonlinear combination of derivatives of $\Psi$. 

\section{Vortex configurations in simulated haloes with gravity}
\label{sec:numerics}
So far, our discussions of vortices have mostly been based on analytic
solutions/models where the effects of gravity have been ignored.
It is helpful to know to what extent our expectations about
the behaviors of vortices hold up even when gravity is included. In this Section, we address this
question using numerical simulations with gravity---essentially the
analog of $N$-body simulations but for waves~\cite{Schive:2014dra,Schive:2014hza,Mocz:2015sda,Veltmaat:2016rxo,Schwabe:2016rze,Veltmaat:2018dfz,Nori:2018hud,Edwards:2018ccc,Li:2018kyk}. 
Let us begin by summarizing a number of expectations based on the
above discussion:

\begin{itemize}
\vskip5pt
\item  We expect to find vortices in a collapsed halo which consists
  of a superposition of waves with different momenta. Such vortices
  form as a result of chance interference, and should exist even if
  the halo has no net angular momentum. The vortices
 should generically have winding $\pm 1$, where the winding is given by~\eqref{eq:winding}.

\vskip5pt
\item We expect vortices to come in the form of rings in three spatial
  dimensions, with a density of roughly one vortex ring per de Broglie
  volume. The de Broglie scale is $\lambda_{\rm dB}= 2\pi/(mv)$ where
$v$ is the (three-dimensional) velocity dispersion of the halo.

\vskip5pt
\item In the vicinity of vortices, the mass density should scale as $\rho \propto
  r_\perp^2$ where $r_\perp$ is the distance to the vortex as in~\eqref{eq:vortexdensityline}, and 
the proportionality constant is roughly $\rho_{\rm local}
  /\lambda_{\rm dB}^2$ with $\rho_{\rm local}$ being the mean mass 
  density of the halo in the neighborhood of the vortex, the idea being that the density near a vortex grows like $r_\perp^2$ until it reaches the local mean density, which happens approximately at the de Broglie wavelength. Further, we expect that the lines of constant density will be elliptical in the plane transverse to the vortex line as in~\eqref{eq:perpdensity}.

\vskip5pt
\item Near the vortices, the tangential (circulation) velocity is expected to scale
  inversely with distance to the vortex {\it i.e.}, $v_\parallel \sim 1/(m r_\perp)$. 

\vskip5pt
\item The vortex rings are dynamical objects: they can form and
  expand, or shrink and disappear, on a time scale of $\lambda_{\rm
    dB}/v$ or $2\pi/(mv^2)$, where $v$ is the velocity dispersion of
  the halo. When vortices intersect, they reconnect as
  opposed to getting frustrated. Vortex rings are expected to move
  with a size-dependent velocity $v_{\rm ring} \sim 1/(m\ell)$ where $\ell$ is the size of
the ring. More generally, a segment of a vortex is expected to move
with the same scaling, with $\ell$ being the local curvature scale.
\end{itemize}
\vskip3pt
We confirm all of these expectations with our simulations, except the
last bullet point, which requires examining simulation outputs at
multiple times and will be left for future work. (See~\cite{2011JPhB...44k5101C}, for a study of some aspects of vortex dynamics.)

\subsection{Numerical algorithm}

Our numerical simulations are generated using the \texttt{SPoS} code presented
and tested in \citep{Li:2018kyk}. The code solves the combined Schr\"odinger--Poisson
system of equations in three spatial dimensions. 
Details of the initial conditions and parameters for
different runs will be given below, but let us make a few general
comments about them here.
Cosmological simulations---meaning simulations that start from
Gaussian random initial conditions, with a suitably large box (larger
than $10$ Mpc co-moving)---do not quite have haloes that are
sufficiently well-resolved for our purpose. We do observe
vortex rings in such cosmological simulations~\citep{Li:2018kyk}, but in order to study
the vortex rings in detail, we will instead focus on simulations with a small box
size and somewhat artificial initial conditions chosen to result in a collapsed halo. 
In all cases we study, there are no vortices in the initial conditions. Yet,
vortices generically appear after gravitational collapse brings
about a superposition of waves in a virialized halo. This is the scenario we
expect in the cosmological setting: the early universe is fairly homogeneous
with small fluctuations (in particular $\Psi$ does not vanish anywhere in the early
universe). Gravity amplifies the fluctuations and creates nonlinear
objects in which a superposition of waves with suitably large
amplitudes exist, making destructive interference, and therefore
vortices,
possible.

Given a simulation output, we identify vortex lines by
searching for locations of high vorticity
$\vec\omega=\vec\nabla\times \vec v$.\footnote{Note that even though the fluid vorticity is expected to be
  zero away from vortices, there is always numerical noise (such as
  from taking gradients) that generates vorticity. Vortices are
  identified as places where the vorticity is exceptionally high.}
In order to do this, we first compute the normalized wavefunction $\tilde{\Psi}\equiv  \Psi/\sqrt{\rho/m} = e^{i\theta}$.
We then take the numerical gradient to compute the velocity $\vec v =
\tilde{\Psi}^* \vec \nabla \tilde{\Psi}/im$, and from that we can
compute the 
vorticity field $\vec \omega$.
Our numerical results show that $\vec\omega$ is close to zero
in most regions, but has large values at the locations of vortex lines.
One might imagine alternatively identifying vortices by searching for
locations of zero density. Of course, numerically, the density is
never strictly zero and we find that regions with low density sometimes
also have small $\vec\omega$, indicating that small
density alone is not a sufficiently robust indicator of vortex lines, and it is therefore better to search for regions of high vorticity.

After a vortex line is identified, we also check that the phase does have
the required characteristic winding around it. This is important: given limited
resolution, it might be difficult to ascertain that a particular point
has low density or high vorticity, but the winding phase (or,
equivalently, the non-zero velocity circulation) is something that can
be checked in the neighborhood of that point and is therefore robust.
The discrete grid points with large $\vec\omega$ are grouped into vortex lines using the DBSCAN (Density-Based Spatial Clustering of Applications with Noise) algorithm~\citep{Ester96adensity-based}.

\subsection{Numerical solutions including gravity}
\label{sec:simulations}

\begin{figure}[tb]
\centering
\subfigure[]
{\label{dens_profile}
\includegraphics[width=.48\textwidth]{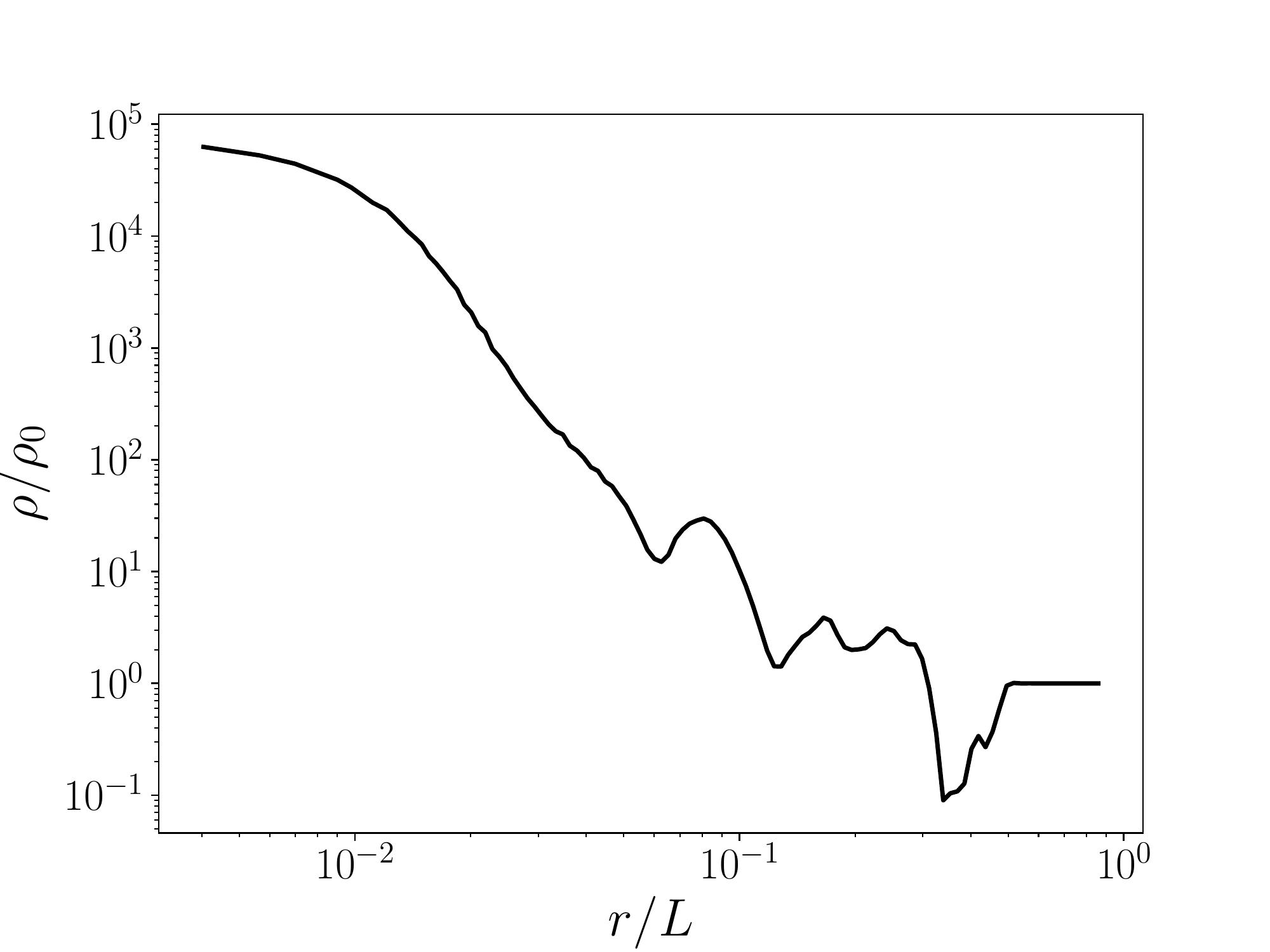}
}
\subfigure[]
{\label{vortexline}
\includegraphics[width=.4\textwidth]{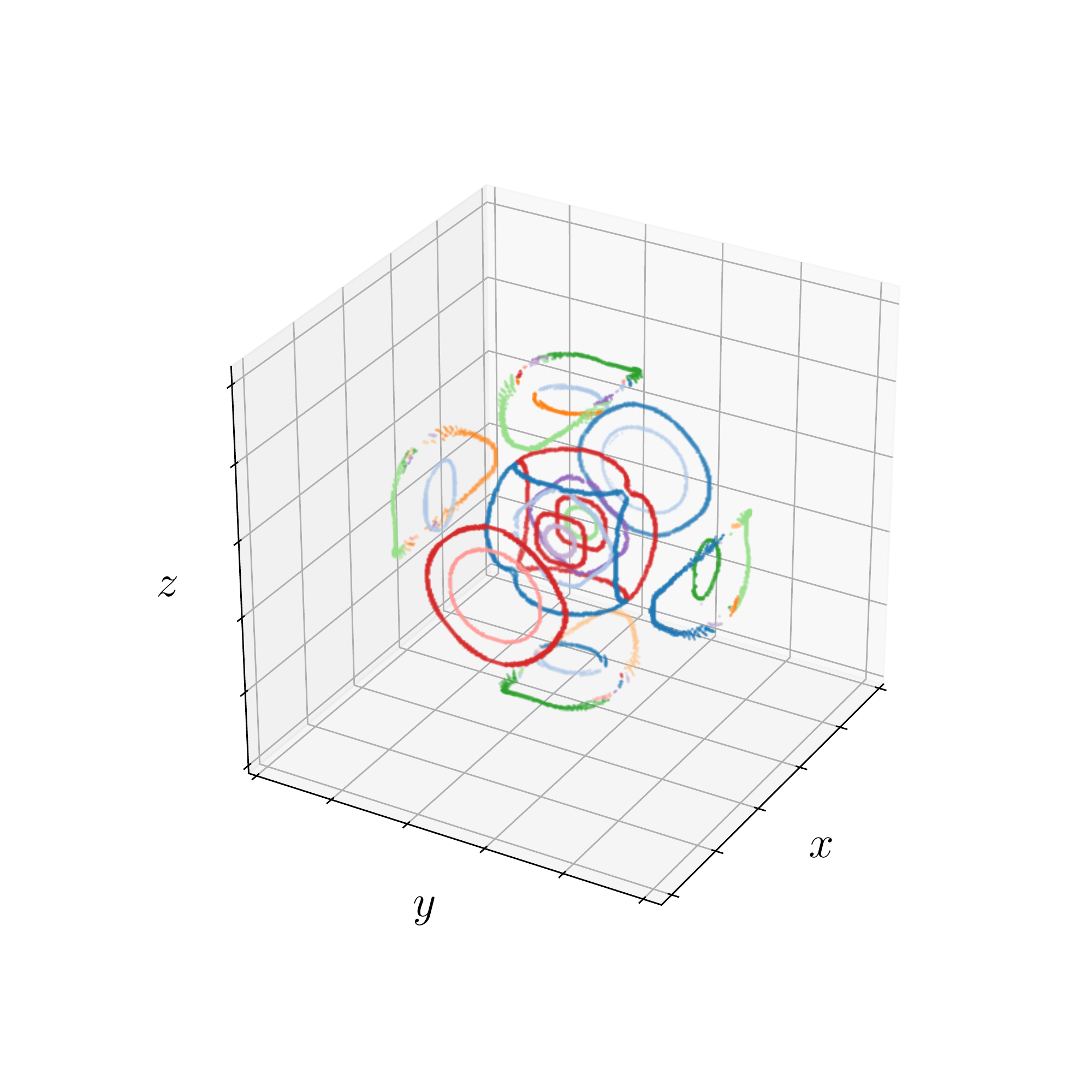}
}
\caption
{\small Numerical simulation of a halo formed from the collapse and
  merger of a symmetric configuration of four Gaussian peaks (see text for details). {\bf a.} Spherically averaged density profile (as a function of radius) 
of the virialized halo that forms from the
  coalescence of the peaks.
{\bf b.} Vortex rings formed
in the same halo.
}
\label{dens_profile_0}
\end{figure}

As a first example, we investigate a fairly symmetric configuration: a
halo that forms from the coalescence of four evenly spaced density
peaks. The three-dimensional computational box has a length of $L=0.5$
Mpc on each side. (In figures shown below, the origin is at the center
of the box.) Note that our eventual conclusions---for instance about
the abundance of vortices---can be scaled to whatever dark matter mass
one is interested in, but for concreteness, we choose a rather low mass, 
$m=6 \times 10^{-24}$ eV. The relevant de Broglie scale is well-resolved with a simulation grid of $256^3$ points. 
We have also performed convergence tests with a higher resolution grid of $512^3$ points and found that our conclusions
are not sensitive to the precise resolution.

The simulation is initialized from a symmetric configuration where four
identical peaks are present on top of a background density of 
$\rho_0=1.88\times 10^{-29}$~g/cm$^3$. 
The background density is chosen to resemble
the cosmological one, even though there is no cosmological
background expansion in the simulation. In fact, the precise value of
$\rho_0$ does not have much impact on the eventual structure of the
collapsed halo---its presence is primarily to ensure the density does not
vanish anywhere in the initial condition.
Each of the density peaks has the Gaussian profile
\be
\label{peakprofile}
    \rho = 250\rho_0 \exp \left[-\left(\frac{\vec r-\vec r_c}{r_0}\right)^2 \right]
\ee
where $r_0=0.05 L$ is the radius of the Gaussian core and $\vec r_c$
denotes the center of the peak. In the initial state,
the four peaks are placed on the four vertices of a square chosen to lie
in the $x=0$ plane, and which is centered at the origin with a side length
of $L/10$. 
The total mass in the computational box is $M=5.8\times 10^{10}M_\odot$.
The initial wavefunction is taken to be real $\Psi = \sqrt{\rho/m}$,
so there is neither initial velocity nor angular momentum.
We use an absorbing boundary condition to allow waves to leave the
box.\footnote{This is accomplished by adding
an extra term $-C(\Psi-\sqrt{\rho_0/m})$ to the Schr\"odinger equation
with $C=\arctan (10^3 d/L)/\Delta t$ with $d=\max(0,r-0.48L)$, where
$r$ is the distance from the center of the box.
This extra term is appreciable only in the region $r>0.48L$ near the
boundaries, removing fluctuations of $\Psi$ around the constant background.
For the most of the box $r<0.48L$, the evolution of $\Psi$ is
essentially unaltered.}
Figure \ref{dens_profile} shows the density profile of the
quasi-stationary dark matter
halo that forms after the initial four peaks merge under the influence
of gravity. 
A soliton core is formed in the inner region of size $r\sim 0.01L$, and
the density falls roughly as $r^{-3}$ 
in the outer region.

\begin{figure}[t]
\centering
\subfigure[]
{\label{vline_dens}
\includegraphics[width=.48\textwidth]{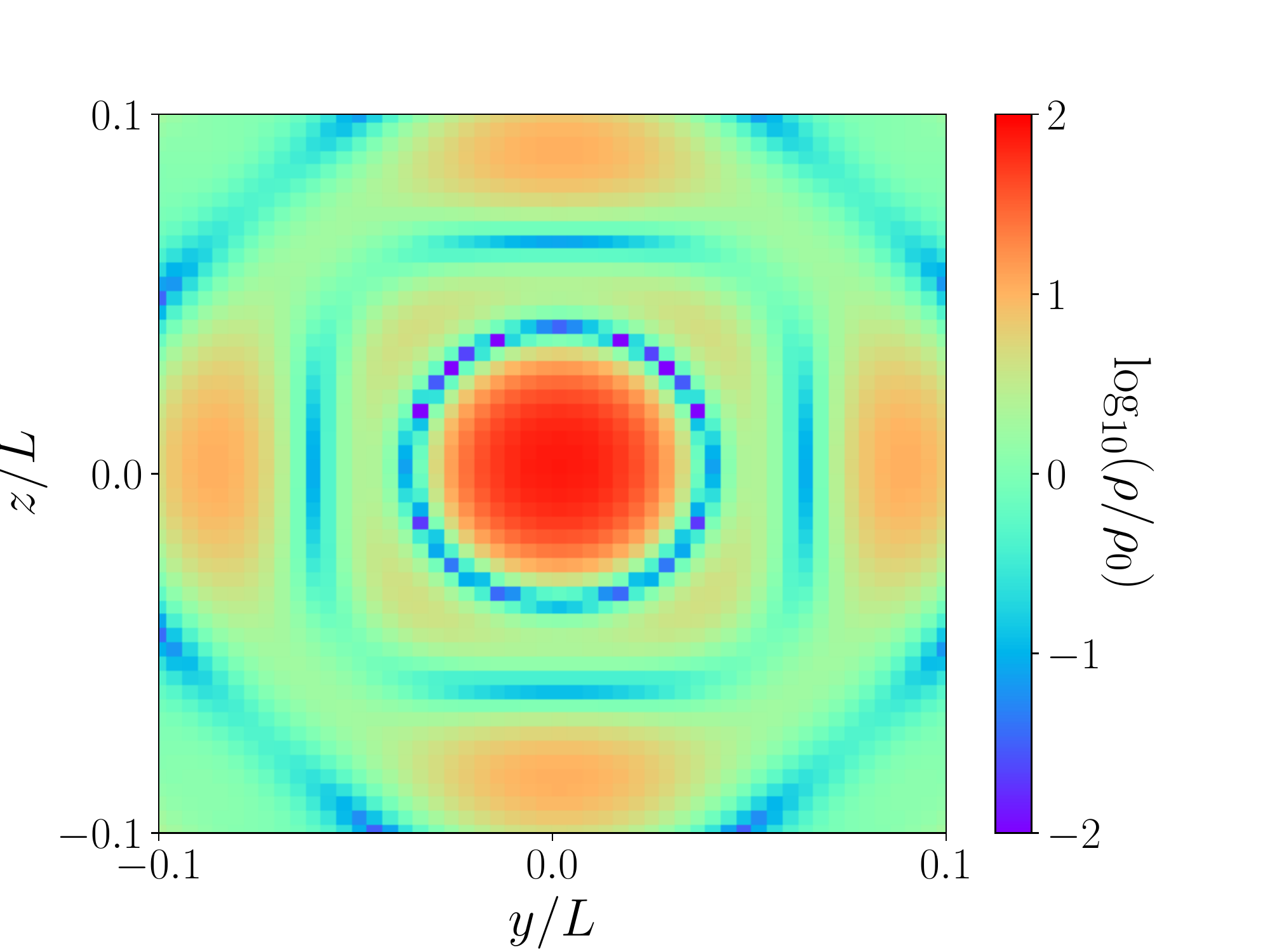}
}
\subfigure[]
{\label{vline_phase}
\includegraphics[width=.48\textwidth]{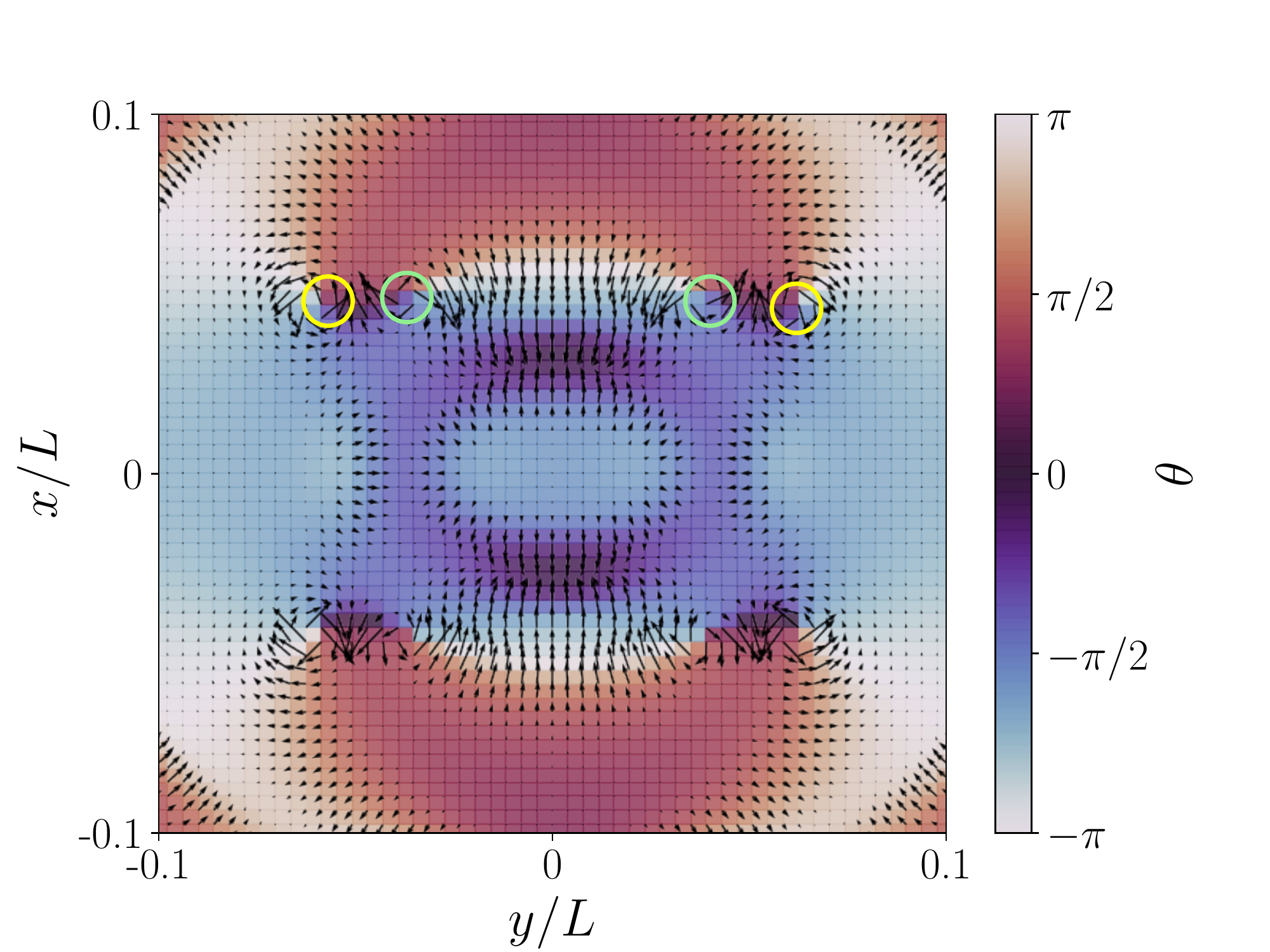}
}
\caption
{\small
More detailed views of a halo depicted
in Figure \ref{dens_profile_0}, formed from collapse of a symmetric configuration of four Gaussian peaks. {\bf a.} A zoomed-in slice of the density field in the $y-z$ plane at
$x=0.05L$, showing vortex rings (where the density vanishes) observed near halo center.
{\bf b.} A zoomed-in slice of the phase $\theta$ in the $yx$ plane at $z=0$.
The colors show the phase value and the arrows trace the velocity
field in the plane. The green and yellow circles indicate where
the circular and square vortex rings in the left panel intersect the $z=0$
plane. Note that there is a pair of such rings at $x=0.05 \, L$, and
a pair at $x= -0.05 \, L$, due to the symmetric nature of the initial condition.}
\label{vortexline2}
\end{figure}

Figure \ref{vortexline} visualizes the vortex lines identified in the collapsed
halo. The soliton core is rather smooth at the center, and vortices
can be found outside the core. Around each vortex line the winding number is $\pm 1$. 
All the vortex lines found are in the form of vortex rings.
It is worth stressing that these vortex loops emerge 
from an initial configuration which does not have vorticity or angular
momentum. By measuring the 3D velocity dispersion of the outer halo (away
from the soliton core), we infer
a de Broglie scale of $\lambda_{\rm dB} \equiv 2\pi/mv\approx
0.12 L$.
Note that the soliton core size of $\sim 0.01 L$ is in fact closer to
$\lambda_{\rm dB}/(2\pi)$ rather than $\lambda_{\rm dB}$. 
With a halo radius of about $0.2 L$, our one vortex ring per $\lambda_{\rm
  dB}^3$ estimate corresponds to an expected total of $\sim19$ rings, in reasonable
agreement with the 20 seen in the simulated halo.

Figure \ref{vortexline2} shows two zoomed-in slices for the density and phase
corresponding to the snapshot in Figure \ref{dens_profile_0}. 
The density slice plot on the left shows the $yz$ plane at 
$x=0.05\, L$. 
(The symmetric initial condition is such that the $x=-0.05\, L$ plane
looks more or less the same.)
Near the center, there is a circle and a square loop of low density
corresponding to the two vortex rings.
The phase slice plot on the right shows the $yx$ plane at $z=0$, 
with arrows indicating the velocity field on that plane.
The green and yellow circles show the position of the circular and
square vortex rings as they intersect the $z=0$ plane (one pair at
$x=0.05 \, L$ and one pair at $x=-0.05 \, L$). 
The phase $\theta$ around each vortex smoothly varies
from $-\pi$ to $\pi$, with winding number $\pm 1$.
Correspondingly, the velocity field wraps around each vortex line, as expected.

\begin{figure}[tb]
\centering
\subfigure[]
{\label{den_r2_1}
\includegraphics[width=.48\textwidth]{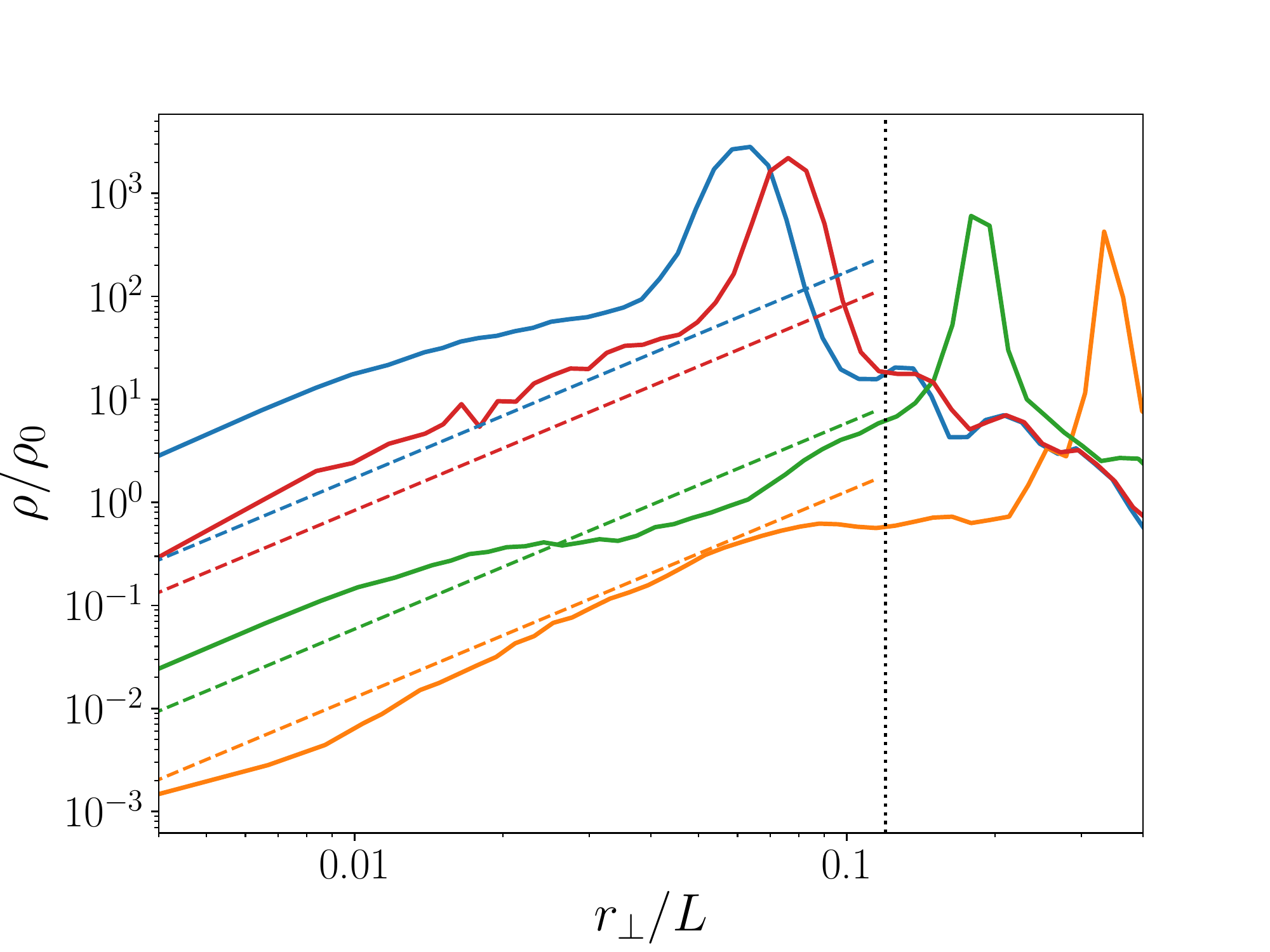}
}
\subfigure[]
{\label{den_r2_2}
\includegraphics[width=.48\textwidth]{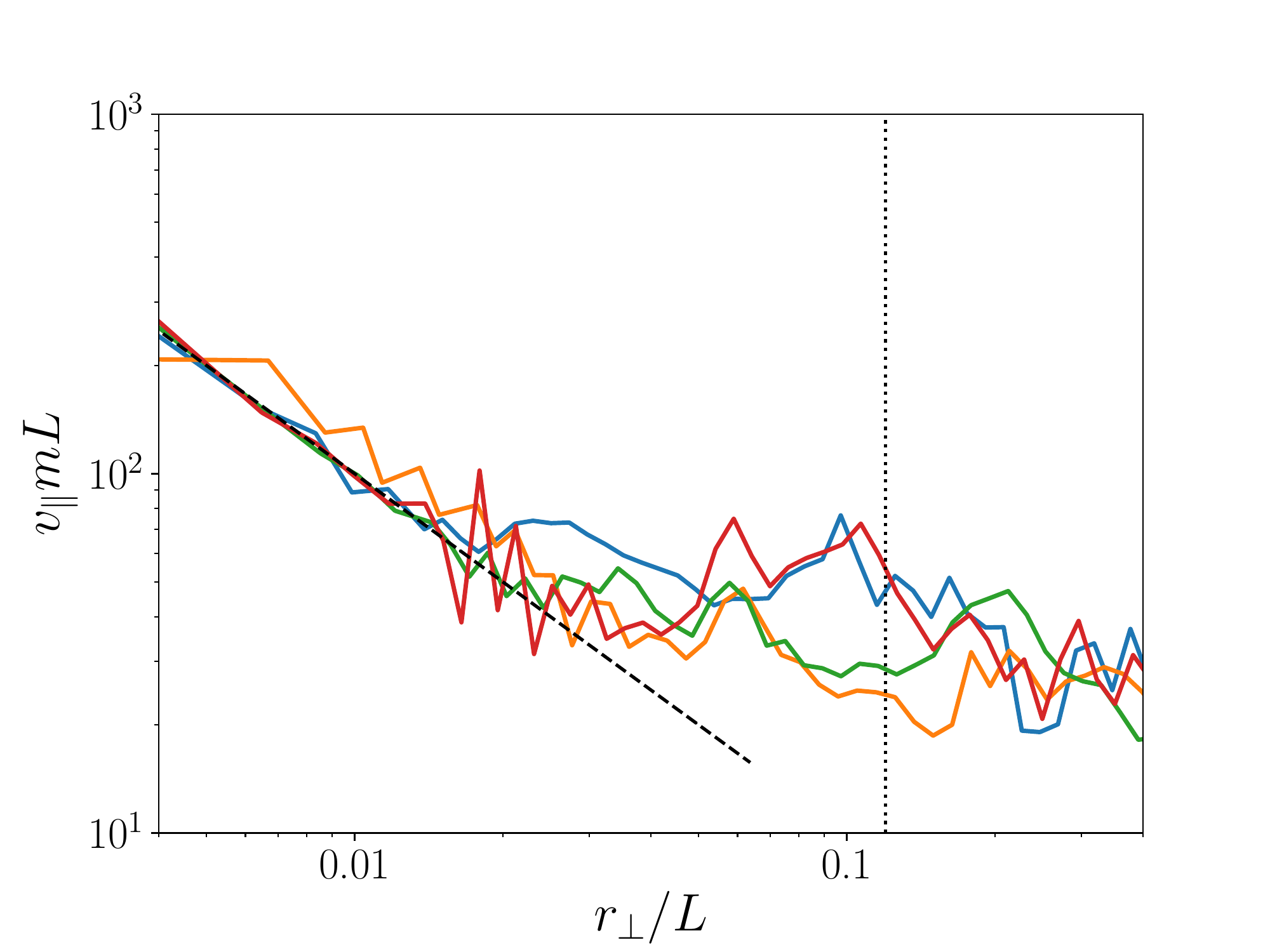}
}
\caption
{\small Density and velocity profiles around four vortices in Figure
  \ref{vortexline}. The profiles are cross-sectional, in the sense that they depict
  (circularly averaged) profiles on a plane perpendicular to the
  vortex, intersecting the vortex at some point. {\bf a.}
Density profiles (solid lines) in the radial direction;
the dashed lines indicate $\rho_{\rm local} \,
(r_\perp/\lambda_{\rm dB})^2$, where $\rho_{\rm local}$ is the mean
density interior to the location of the vortex (see text), and $r_\perp$ is the distance to
the vortex line. This shows that the near-vortex density profiles obey the expected $r_\perp^2$ scaling. {\bf b.} Velocity profiles (solid lines);
the black dashed line shows the tangential (circulation) velocity $v_\parallel = 1/(m\, r_\perp)$. 
The vertical dotted line in both panels shows the scale of $\lambda_{\rm dB}$. 
Lines of the same color on the left and right correspond to the same
vortex.
}
\label{dens_r2}
\end{figure}

Figure~\ref{dens_r2} shows the density and velocity profiles close to
four different vortices seen in Figure~\ref{vortexline}. 
Lines of the same color on the left and right correspond to the same
vortex.
Both the density and velocity profiles are cross-sectional, in the
sense that they are profiles on a plane intersecting the
vortex at some point, oriented perpendicular to the local vortex line direction.
The (circularly averaged) density profile as a function of the distance from the
vortex $r_\perp$ is shown on the left.
As a comparison, we show in the same plot four dashed
lines (color coded, one for each vortex) indicating $\rho = \rho_{\rm local} (r_\perp/\lambda_{\rm
  dB})^2$, where $\rho_{\rm local}$ is the mean mass density at radii
smaller than the location of the corresponding vortex. The
$r_\perp^2$ profile is a reasonable approximation close to the
vortex, but generally breaks down before $r_\perp$ reaches the de
Broglie scale. The normalization by $\rho_{\rm local}$ captures the
qualitative trend that vortex rings closer to the halo center have
a higher normalization, but can deviate from the actual normalization by an order of magnitude. The density bumps in the curves correspond to
locations where one starts encountering the central region of the halo.
The velocity profile is shown in Figure~\ref{den_r2_2}. 
It depicts the (circularly averaged) tangential velocity $v_\parallel$
as a function of $r_\perp$. The black dashed line indicates $v_\parallel = 1/(m
r_\perp)$ which describes the velocity profiles reasonably well.

\begin{figure}[tb]
\centering
\subfigure[]
{\label{vline_dens2}
\includegraphics[width=.48\textwidth]{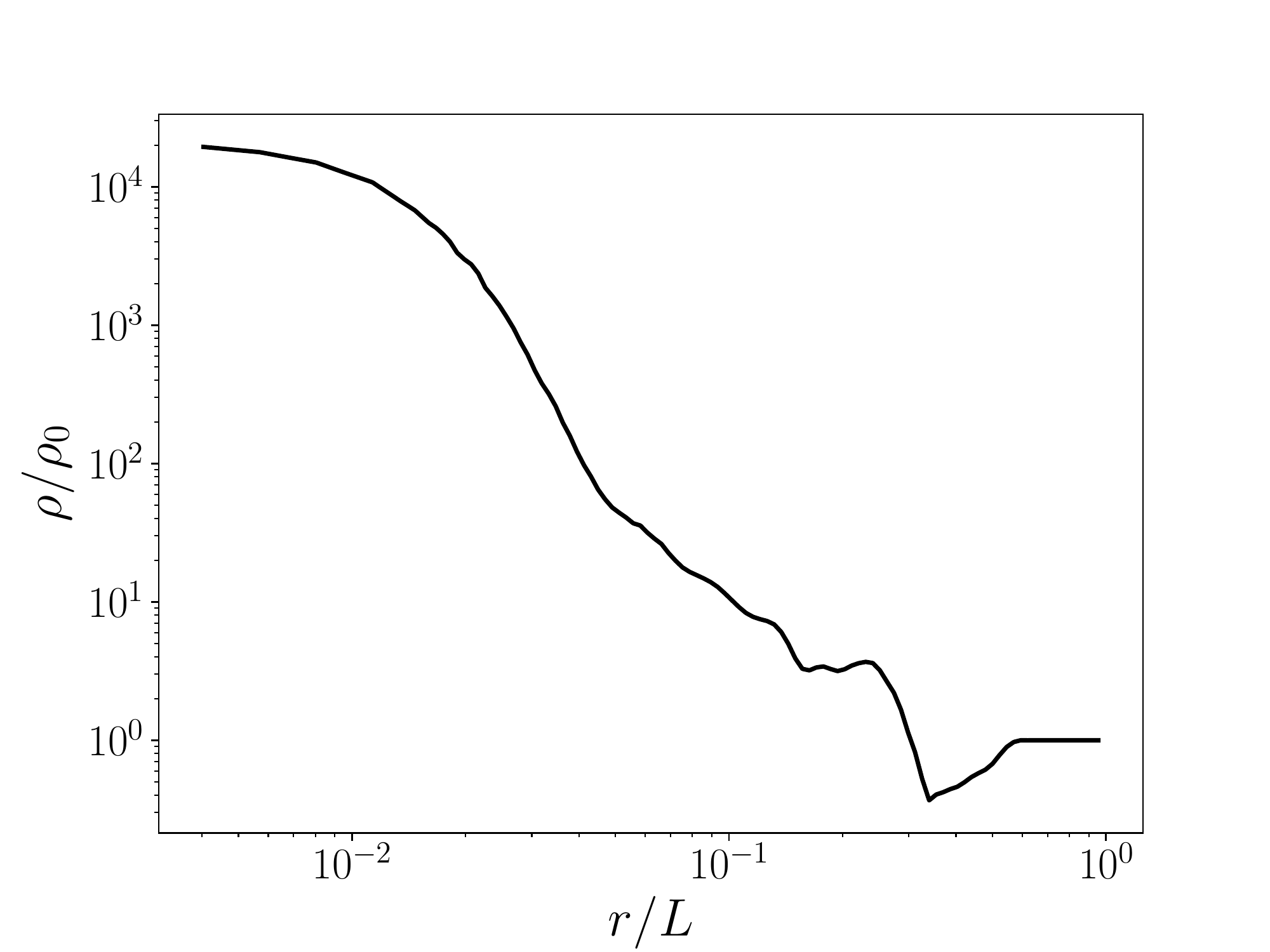}
}
\subfigure[]
{\label{vline_phase2}
\includegraphics[width=.4\textwidth]{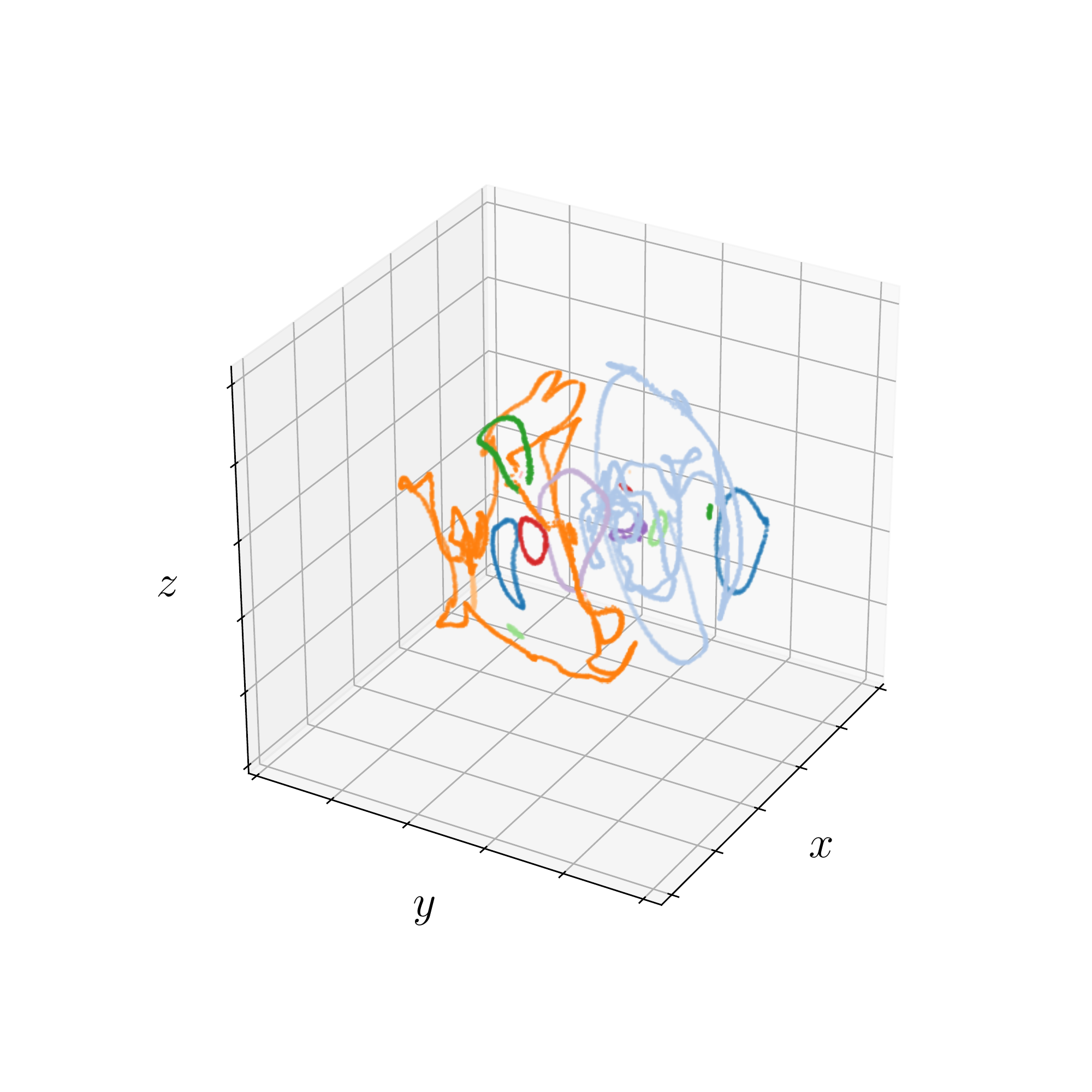}
}
\caption
{\small Features of a dark matter halo formed from collapse from asymmetric initial conditions  (see text).
{\bf a.}~Density profile of the halo.
{\bf b.}~Visualization of vortex lines found in the halo.
}
\label{vortexline_random}
\end{figure}

The simulated halo discussed above arises from rather symmetric
initial conditions, so one might be concerned that the formation of vortices and their properties are artifacts of these special initial conditions. As a second example, we therefore performed a simulation with more
asymmetric initial conditions: there are initially 10 identical peaks
(with profiles given by eq.~\eqref{peakprofile}) randomly placed close to the center of the
box (they are within $0.2\, L$ of each other).
The total mass in the computational box is kept the same.
Figure~\ref{vortexline_random} shows the density profile of the
resulting halo from the coalescence of these peaks under gravity, with the vortex
lines highlighted on the right.
The vortex lines look less symmetric, but all qualitative features
remain the same: (1) they all close in loops; (2)
the phase winds in the cross-section of each line with winding number
$\pm  1$; (3) the density profile scales as $r_\perp^2$ and velocity
profile scales as $1/r_\perp$ close to the vortex; (4) the number
density of vortex rings remains roughly one per de Broglie volume.

\begin{figure}[tb]
\centering
\raisebox{0.2\height}{
\includegraphics[width=.45\textwidth]{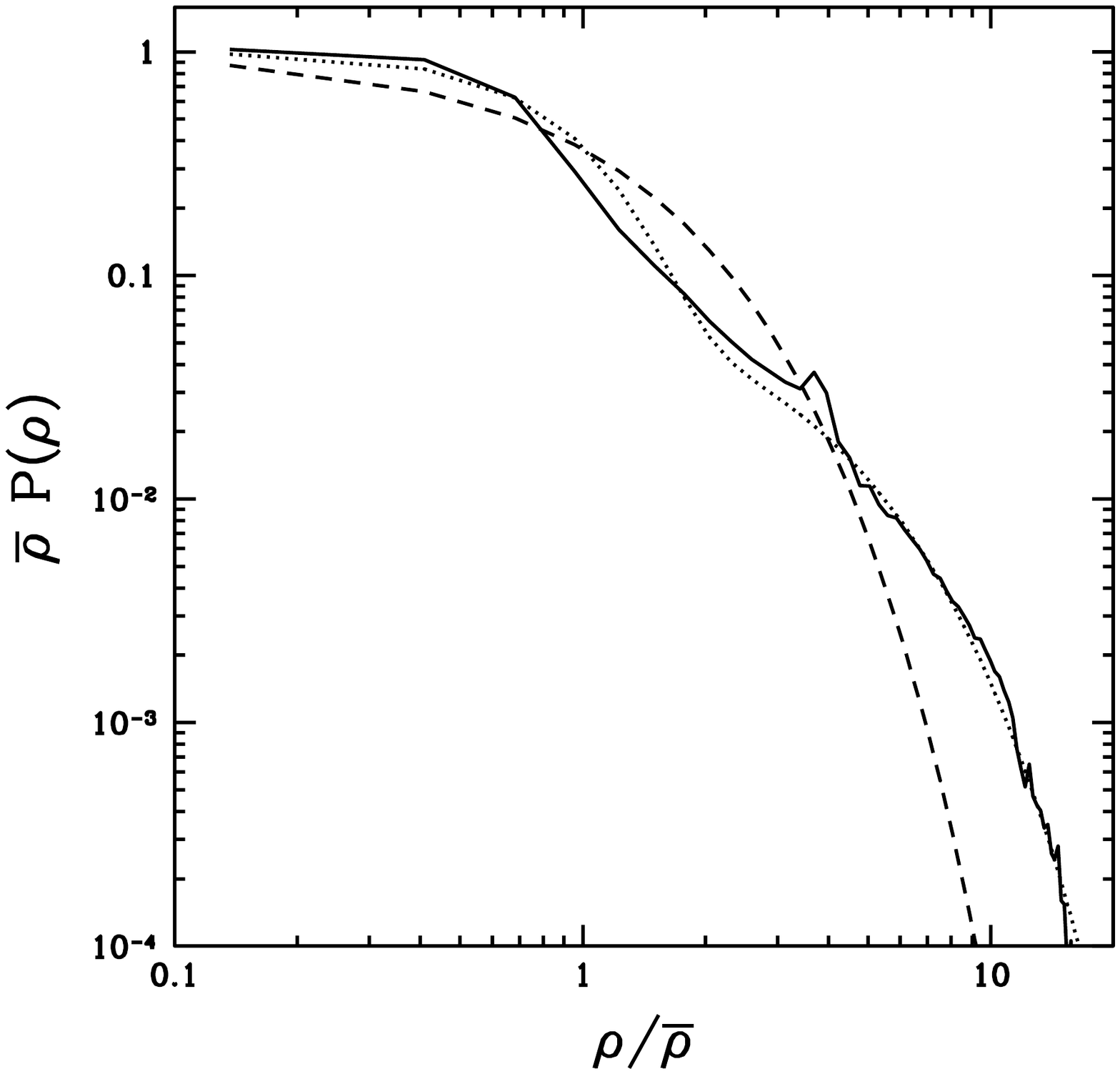}
}
\raisebox{0.2\height}{
\includegraphics[width=.45\textwidth]{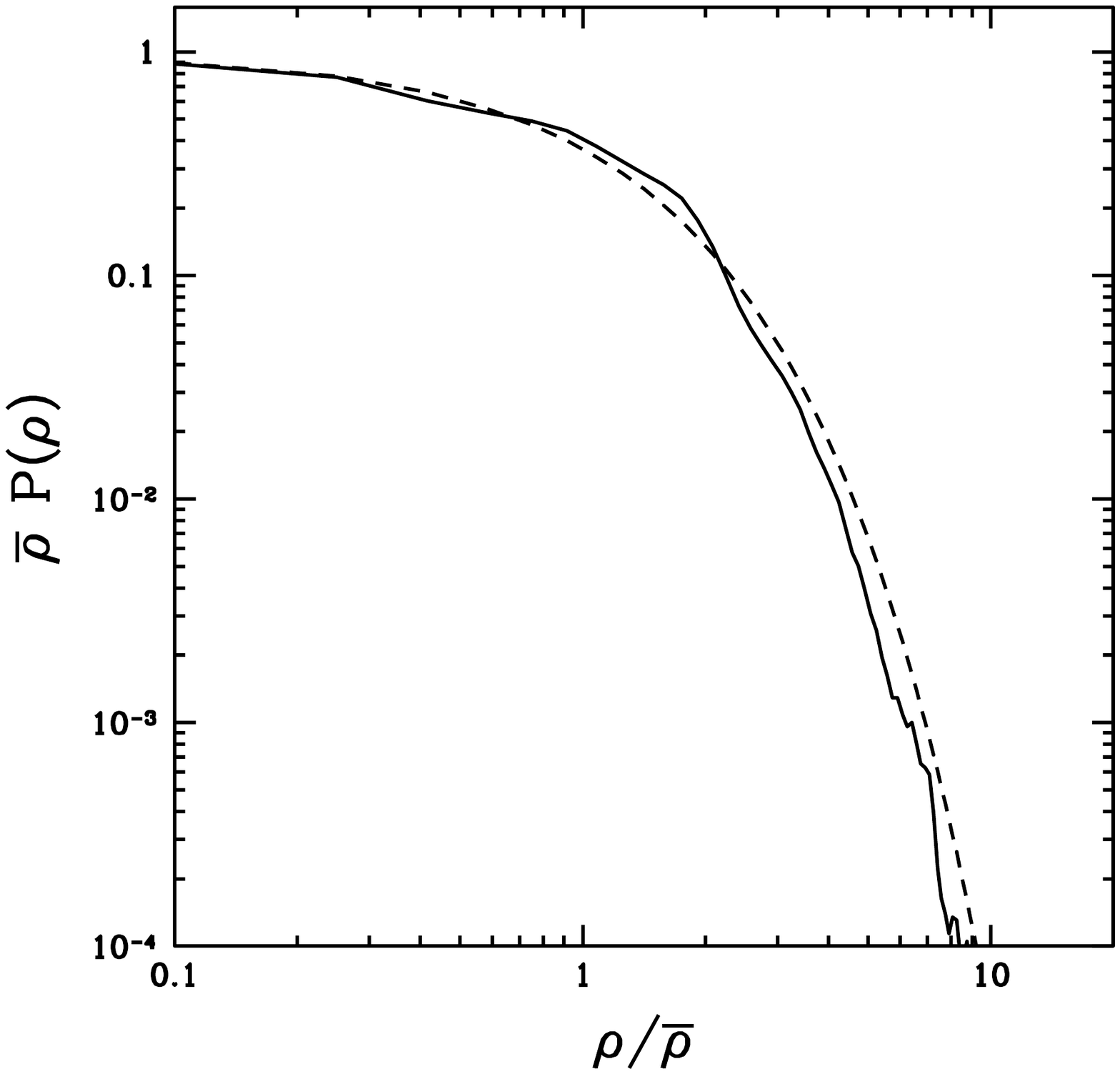}
}
\vspace{-4.0cm}
\caption
{\small
One-point probability distribution of density (solid lines)
for haloes formed from
gravitational collapse with symmetric
(left panel) and asymmetric (right panel) initial conditions
(see text). Here $P(\rho) \, \rd\rho$ gives the probability
that $\rho$ falls within the interval $\rd\rho$; $\bar\rho$ is 
the mean density. The dashed line in both panels denotes
the random phase model prediction: $\bar\rho P(\rho) =
e^{-\rho/\bar\rho}$ (eq.~\eqref{expP}). 
The dotted line in the left panel is an analytic
approximation to the numerical data: 
$\bar\rho P(\rho) = 0.9 \, e^{-1.06(\rho/\bar\rho)^2} + 0.1 \, e^{-0.42(\rho/\bar\rho)}$.
}
\label{psipdf}
\end{figure}

It's not surprising that these qualitative features should hold---they
are expected on fairly general grounds with or without gravity
(recall the Taylor expansion argument articulated in Section~\ref{sec:general}). The average number density of vortex rings is a
quantity which in principle could be significantly modified when gravity is
included, though it does not appear to be.
Our computation in Section~\ref{sec:randomphase} is based on
the random phase model, which raises the question: how good a description
is this of a halo that forms from gravitational collapse? 
As an indication, we show in Figure~\ref{psipdf} the one-point
probability distribution of density $\rho$ for the haloes depicted in 
Figure~\ref{dens_profile_0} and Figure~\ref{vortexline_random},
respectively.\footnote{The density $\rho$ of course has a systematic trend with radius i.e., overall,
lower density at larger radii. We remove this systematic trend by
considering $\rho / \bar\rho$, where $\bar\rho$ is the local mean
density, estimated by averaging $\rho$ over spherical shells at the 
radii of interest. The probability distribution is then obtained from a histogram of $\rho/\bar\rho$.}
The solitonic core is excluded in measuring the probability distribution.
In both cases, the distribution is reasonably approximated by the analytic
prediction of the random phase model (eq.~\eqref{expP}),
especially for small or moderate values of $\rho$
(see also eq.~\eqref{RayleighP} for the corresponding distribution for
$|\Psi|$). The fact that
the probability distribution $P(\rho)$ is exponential, and
thus flat at small $\rho$, is a telltale sign
of destructive interference. The random phase model prediction matches
well the simulation results for the halo depicted
in Figure \ref{vortexline_random}, but less well for the halo
in Figure \ref{dens_profile_0}. 
In the latter case, there appears to be some memory of
the initial conditions, which are more symmetric; there is a 
prominent high $\rho$ tail, which 
is not captured by the random phase model.
Some fraction of the halo substructure is in the form of sub-haloes---much like what one expects in conventional cold dark matter---and they
contribute to this tail. We provide an approximate analytic fit to
$P(\rho)$ in this case.

\section{Observational and experimental implications}
\label{sec:observables}

We have seen that dynamic small-scale structure is a generic feature of light dark matter. 
In this section, we discuss the observational and experimental
implications of these interference substructures including, but not
limited to, vortices. 
In order to organize our thinking, it is helpful to remind ourselves of certain characteristic length
and time scales:
\be
\begin{aligned}
 \lambda_{\rm c} &\equiv {\pi \over {m v}} = 0.24 {\,\rm kpc}  \left(
  {10^{-22} {\,\rm eV} \over m} \right) \left( {250 {\,\rm km/s} \over
  v} \right) = 7.4 \times 10^4 {\,\rm cm} \left(
  {10^{-6} {\,\rm eV} \over m} \right) \left( {250 {\,\rm km/s} \over
  v} \right) \, , \\
 t_{\rm c} &\equiv {\lambda_{\rm c} \over v} = 9.4 \times 10^5 {\,\rm yr.} 
\left( {10^{-22} {\,\rm eV} \over m} \right) \left( {250 {\,\rm km/s} \over
  v} \right)^2 = 3 \times 10^{-3} {\,\rm s} \left(
  {10^{-6} {\,\rm eV} \over m} \right) \left( {250 {\,\rm km/s} \over
  v} \right)^2 \, , \\
\lambda_{\rm osc.} &\equiv {\pi \over mc} = 0.2 {\,\rm pc} \left(
  {10^{-22} {\,\rm eV} \over m} \right) = 62 {\,\rm cm} \left(
  {10^{-6} {\,\rm eV} \over m} \right) \, , \\
 t_{\rm osc.} &\equiv {\lambda_{\rm osc.} \over c} = 0.6 {\,\rm yr.} \left(
  {10^{-22} {\,\rm eV} \over m} \right) = 2 \times 10^{-9} {\,\rm s} \left(
  {10^{-6} {\,\rm eV} \over m} \right) \, ,
\end{aligned}
\label{eq:characteristicscales}
\ee
where we give baseline values both for the ultra-light extreme ($10^{-22}$ eV)
and the merely light, characteristic of QCD axion experiments
($10^{-6}$ eV). The general features we discuss apply more broadly to
a wide range of possible masses. We have restored the speed of light $c$
here for clarity, though $\hbar$ remains $1$. 
The velocity dispersion $v$ is chosen to be close to that of the Milky Way.

There is some freedom in the definition of the coherence length
$\lambda_{\rm c}$ and time $t_{\rm c}$---for instance, we could have chosen to
equate $\lambda_{\rm c}$ with the de Broglie scale $\lambda_{\rm dB}$, which is factor of $2$
larger. 
The choice of $\lambda_{\rm c}$ is motivated by the random phase
model discussed in Section~\ref{sec:randomphase}: there the two-point function is given by $\langle \Psi_1 (\vec x, t) \Psi_1
(\vec y, t)\rangle /\langle \Psi_1 {}^2 \rangle = {\,\rm exp} \left(- k_0^2 \lvert\vec x - \vec
  y\rvert^2/8\right)$, and the coherence length $\lambda_{\rm c} = \pi/(\sqrt{3} k_0/2)$ is the separation
at which the two-point function is suppressed by a factor of $0.2$, while
at a separation of $\lambda_{\rm dB}$ the two-point function drops to 
$10^{-3}$.\footnote{In the random phase model, the  3D momentum variance is 
$\int \rd^3 k \,k^2 \exp (-2k^2/k_0^2) / \int \rd^3 k \exp
(-2k^2/k_0^2) = 3 k_0^2/4$.}
Note that 
the oscillation length and time scales, $\lambda_{\rm osc.}$ and
$t_{\rm osc.}$, are related to the Compton (as opposed to de Broglie)
scale---they are mostly relevant for axion direct detection
experiments, whereas the characteristic scales $\lambda_{\rm c}$ and
$t_{\rm c}$ are relevant for both astrophysical observations and
detection experiments.

\subsection{Observational implications}
\label{sec:observe}

If the dark matter mass $m$ is in the ultra-light regime ($\lsim \,
10^{-20}$ eV, {\it i.e.}, fuzzy dark matter), there are many
astrophysically observable effects, some of which are detailed in 
\cite{Hui:2016ltb}. Here, we wish to focus in particular on effects
of the inevitable halo substructure due to wave interference.
We concentrate mainly on gravitational lensing effects, that is the
scattering of photons, and will briefly remark on the
scattering of stars in Section \ref{sec:conclude}.

\subsubsection{Lensing magnification}
\label{sec:magnify}

We first consider the effects of light dark matter on lensing magnification in the context of strong lensing. 
We will show that the small-scale structure associated to interference generically leads to flux anomalies between images of order $10\%$, making lensing a promising observational avenue to search for this substructure, including vortices.

At a basic level, gravitational lensing can be thought of as a mapping of the 2D plane to itself.
The fundamental lensing equation
takes the form (see {\it e.g.}, the review~\cite{Bartelmann:2010fz}):
%
\be
\label{lenseqt}
\vec \theta_S = \vec \theta_I - 2 \int_0^{r_S}\rd r_L {r_S - r_L \over r_S} \,
  \vec\nabla_\perp \Phi \,  \, ,
\ee
where $\vec \theta_S$ and $\vec \theta_I$  are
the 2D angular positions of the source and of the image in the
sky, $\Phi$ is the (lens) gravitational potential, 
$r_S$ is the distance (from observer) to the source, 
$r_L$ is the distance (from observer) to the lens.
The derivative $\vec\nabla_\perp$ is the spatial
gradient perpendicular to the line of sight.\footnote{The expression~\eqref{lenseqt} holds even in an expanding universe with
the metric $\rd s^2 = - \rd t^2 + a(t)^2 (\rd r^2 + r^2 \rd\theta^2 +
r^2 {\,\rm sin}^2\,\theta \rd\varphi^2)$, with the understanding that 
$r_S$ and $r_L$ are co-moving distances and the gradient $\vec
\nabla_\perp$ is also co-moving. In our discussion, we ignore the
expanding background, and treat all distances as physical or proper.
Note also our sign convention: the deflection angle for a point mass
lens of mass $M$ and (physical) impact parameter $b$ is:
$\vec\alpha \equiv -2\int \rd r_L\vec \nabla_\perp \Phi \, = -(4GM/b)\,\hat b$, where $\hat b$ is the unit vector pointing from
lens to the point of closest approach of the photon.
}

We decompose $\Phi$ into two parts:
\be
\label{Phidecompose}
\Phi = \Phi_{\rm smooth} + \Phi_{\rm pert.}  \, ,
\ee
where $\Phi_{\rm smooth}$ represents the gravitational potential from
a smooth halo or lens, and $\Phi_{\rm pert.}$ is the perturbation from
substructure, such as that due to wave interference.

Let us first study the effect of  substructure on
lensing magnification. 
The $2 \times 2$ lensing distortion matrix is
\be
\label{lensmatrix}
{\cal A}_{ij} \equiv {\partial \theta_S^i \over \partial \theta_I^j} \equiv
\left(
\begin{array}{cc} 1-\kappa - \gamma_+ & -\gamma_\times \\
-\gamma_\times & 1 - \kappa +\gamma_+
\end{array}
\right) =
\delta_{ij} - 2 \int \rd r_L {(r_S - r_L) r_L \over r_S} \,
  \nabla_\perp {}_i \nabla_\perp {}_j \Phi \, ,
\ee
where the quantity $\kappa$ is known as convergence, and $\gamma_+$ and $\gamma_\times$
are the two components of the lensing shear~\cite{Bartelmann:2010fz}. It is the inverse of this matrix that dictates the
mapping from source to image (in particular it is the Jacobian for this mapping). Note that the determinant 
$\lvert\det{\cal A}^{-1}\rvert$ gives the lensing magnification. 

In most directions in the sky, $\kappa$ and $\gamma_{+,\times}$ are small---at
the percent level---and substructure introduces small corrections
on top of that, which are likely difficult to observe.
Occasionally, strong lensing occurs with multiple highly
magnified images: in other words the determinant of ${\cal A}_{ij}$ is
nearly zero. This provides an opportunity for the small contributions
from substructure to make an observable difference.
The idea that substructure introduces observable corrections to
lensing magnification has been discussed by many authors ({\it e.g},~\cite{Mao:1997ek,Chiba:2001wk,Metcalf:2001ap,Dalal:2001fq,Kochanek:2003zc,Mao:2004iw,Amara:2004dr,Hezaveh:2016ltk,Alexander:2019qsh,Alexander:2019puy,Maleki:2019xya}). Most have
focused on substructure in the form of sub-haloes in the parent lensing
halo; here we are interested in substructure from wave interference.

Consider a location in the image plane where ${\cal A}_{ij}$ has a
zero eigenvalue, let us call it the origin. 
Let us choose the orientation of axes such that
$\gamma_\times = 0$ at that point, and the zero eigenvalue is $1 - \kappa -
\gamma_+ = 0$. At the origin the lensing matrix therefore takes the form
\be
\label{Aij0}
{\cal A}_{ij} (0) = 
\left(\begin{array}{cc} 0 & 0 \\
0 & 2(1 - \kappa_0)
\end{array}\right) \,,
\ee
where $\kappa_0$ is the value of $\kappa$ at the origin. 
Generically, at the origin where $1 - \kappa - \gamma_+$ vanishes,
both $\kappa$ and $\gamma_+$ are of order unity so that $2(1 - \kappa_0)$
is expected to be of order unity.

We are interested in images close to but not at the origin.
Assuming for the moment that $\Phi$ arises entirely from $\Phi_{\rm smooth}$, we can Taylor
expand ${\cal A}_{ij}$ for $\vec \theta_I$ close to the origin:
\be
\label{Aijclose}
{\cal A}_{ij} (\vec \theta_I) = \left( 
\begin{array}{cc} -\vec \nabla (\kappa +
  \gamma_+) \big|_0 \cdot \vec \theta_I &{\cal O}(\theta_I) \\
{\cal O}(\theta_I) & 
2(1 - \kappa_0) + {\cal O}(\theta_I) 
\end{array} \right) \, .
\ee
There is thus a line, the critical curve in the image
plane\footnote{Its counterpart in the source plane is
 known as the caustic. What we are describing here is the neighborhood of a fold caustic. See for example~\cite{Schneider}.}  defined by
$\vec \nabla (\kappa + \gamma_+) \big\rvert_0 \cdot \vec \theta_I = 0$,
where the magnification becomes very large.
Suppose the origin in the image plane maps
to/from some point in the source plane, say $\vec \theta_{S0}$. 
It can be shown the structure of ${\cal A}_{ij}$ given in
eq.~\eqref{Aijclose} implies, for a source close to~$\vec \theta_{S0}$:
%
\begin{align}
\label{thetax}
\theta_S^x - \theta_{S0}^x &= -{1\over 2} \partial_x (\kappa +
  \gamma_+) \Big|_0 \theta_I^x {}^2 + {\cal O}(\theta_I^x \theta_I^y , \theta_I^y
  {}^2) \, , \\
\label{thetay}
\theta_S^y - \theta_{S0}^y &= 2(1-\kappa_0) \theta_I^y + {\cal O}(\theta_I^x
  {}^2 , \theta_I ^y {}^2 , \theta_I^x \theta_I^y) \, ,
\end{align}
where we have introduced $x$ and $y$ as coordinates to represent the components $i=1$ and $i=2$, for
clarity. Assuming $\partial_x (\kappa + \gamma_+) |_0 \ne 0$, we see
that $\theta_I^x \sim \pm \sqrt{-2(\theta_S^x - \theta_{S0}^x)/\partial_x
  (\kappa+\gamma_+) \rvert_0}$
and
$\theta_I^y \sim (\theta_S^y - \theta_{S0}^y)/[2(1-\kappa_0)]$.  
Therefore, there are {\it two} images and they
are equidistant from the origin.
We can also see that if $(\theta^x_S -
\theta^x_{S0})$ and $(\theta^y_S - \theta^y_{S0})$ are of the same
order (and small),
then $\theta_I^y$ is small in magnitude compared to
$\theta_I^x$.\footnote{It can be checked this is a self-consistent solution, 
in that 
terms ignored in eq.~\eqref{thetax} are ${\cal O}([\theta_S -
\theta_{S0}]^{3/2})$ or higher.
The fact that
$\theta_I^x \sim \pm \sqrt{-2(\theta_S^x - \theta_{S0}^x)/\partial_x
  (\kappa+\gamma_+) |_0}$ also implies that having two images requires
$\theta_S^x$ to be on the correct side of the caustic so the quantity
under the square root is positive.
}
Therefore, if the source is on the correct side of the caustic, there
are two images with magnification $\lvert\mu\rvert$, where the inverse magnification is given by the determinant of~\eqref{Aijclose}
\be
\label{musmooth}
\mu^{-1} \sim -2(1-\kappa_0) \partial_x (\kappa + \gamma_+) \rvert_0 \,
  \theta_I^x \sim \pm 2(1-\kappa_0) \sqrt{2 \big\lvert\partial_x (\kappa + \gamma_+)
  \rvert_0 \, (\theta_S^x - \theta_{S0}^x) \big\rvert } \equiv \mu^{-1}_{\rm smooth}\, ,
\ee
where we define $\mu^{-1}_{\rm smooth}$ for later convenience.
This is the standard result that the lensing magnification $\lvert\mu\rvert$ 
scales inversely with distance from the critical curve in the image plane, or
the square root of distance from the caustic in the source
plane. Most relevant for our purpose: the two images have equal magnification~$\lvert\mu\rvert$. 

It is interesting that the above argument is similar in spirit to
the argument for the generic properties of vortices in
Section~\ref{sec:general}: both are based on a Taylor series
expansion.\footnote{This type of argument is employed in
  catastrophe theory.}
The lensing argument
is predicated upon the assumption that the gravitational potential
$\Phi$ is smooth on the scale of the image separation. 
Adopting the split $\Phi = \Phi_{\rm smooth} + \Phi_{\rm pert.}$, 
perturbations on the scale of the image separation
lead to a differential magnification between the two
images. 
This allows us to use anomalous flux ratios to study halo
substructure, as advocated
by~\cite{Mao:1997ek,Chiba:2001wk,Metcalf:2001ap,Dalal:2001fq,Kochanek:2003zc,Mao:2004iw,Amara:2004dr,Hezaveh:2016ltk}.\footnote{It is important to distinguish between two different
sources of differential magnification. The Taylor expansion argument
tells us that two images close to a critical curve 
should have equal 
magnification {\it to the lowest order.} To the next order, assuming a
smooth potential,
the two images do differ slightly in
magnification, fractionally at the level of $1/|\mu_{\rm smooth}|$.
We are interested here in differential magnification caused by
substructures {\it i.e.}, perturbations to the smooth potential that introduce
new scales, capable of enhancing the differential magnification
beyond the minimal floor of $1/|\mu_{\rm smooth}|$.
}

For us, the relevant substructure is the
order unity fluctuations in density due to wave interference.
The angular scale subtended by the corresponding coherence scale is:
\be
\theta_{\rm c} \equiv {\lambda_{\rm c} \over r_L} = 0.006\, {\,\rm arcsec} \,
\left({10^{-22} {\,\rm eV} \over m}\right) 
\left({1000 {\,\rm km/s} \over v}\right) 
\left({2 {\,\rm Gpc} \over r_L}\right) \, ,
\ee
which is smaller than the typical image separation in 
systems with observable multiple images. We choose a velocity
dispersion of $v \sim 1000$ km/s in order to correspond to a cluster scale lens.
A galactic lens with $v$ of a few hundred km/s would have a larger 
$\theta_{\rm c}$ but still smaller than the typical image separation.
In order to model the effect of substructure from wave interference, suppose that
$\Phi = \Phi_{\rm smooth} + \Phi_{\rm pert.}$ in the expression
for ${\cal A}_{ij}$, eq.~\eqref{lensmatrix}. The determinant of this matrix can then be written as
\be
\label{mudecompose}
\mu^{-1} = \mu^{-1}_{\rm smooth} + \Delta \mu_{\rm pert.}^{-1} \, ,
\ee
where $\mu^{-1}_{\rm smooth}$ is defined in eq.~\eqref{musmooth},
and 
\be
\Delta \mu_{\rm pert.}^{-1} \equiv -4 (1-\kappa_0) 
\int\rd r_L {(r_S - r_L) r_L \over r_S} \,
  \nabla_\perp {}_x \nabla_\perp {}_x \Phi_{\rm pert.} \,  ,
\ee
where recall that $2(1-\kappa_0)$ is order unity.
A quick way to estimate $\Delta \mu^{-1}_{\rm pert.}$, which is likely an
overestimate, is to note that wave interference leads to order unity
density fluctuations on scale of $\lambda_{\rm c}$. Adding up these
fluctuations in a random walk fashion along the line of sight through
the lensing halo produces
%
\be
\Delta \mu_{\rm pert.}^{-1} ~\lsim ~ \sqrt{\lambda_{\rm c} \over R_{\rm halo}}\sim
\begin{cases}
  ~0.007 
\left({10^{-22} {\,\rm eV} \over m}\right)^{1/2}
\left({1000 {\,\rm km/s} \over v}\right)^{1/2}
\left({1 {\,\rm Mpc} \over R_{\rm halo}}\right)^{1/2} ~~{\rm cluster}\,, \\[5pt]
~0.06 \left({10^{-22} {\,\rm eV} \over m}\right)^{1/2}
\left({250 {\,\rm km/s} \over v}\right)^{1/2}
\left({50 {\,\rm kpc} \over R_{\rm halo}}\right)^{1/2} ~~\,~~{\rm galaxy}\,,
\end{cases}
\label{eq:overestimate}
\ee
where $R_{\rm halo}$ is the size of the lensing halo (so that there are $N = R_{\rm halo}/\lambda_{\rm c}$ steps to the random walk, and we are estimating the average)  and we give
values for both a cluster and for a galaxy scale lens.
This is an overestimate because most of the (smooth halo) contribution
to the order unity $\kappa_0$ comes from the
central region, where the supposed soliton resides, whereas the
interference substructures are located in the outer halo where
the density is lower. 

As an alternative estimate, 
consider a halo of mean density $\bar\rho$ with
$\nabla^2 \Phi_{\rm pert.} = 4\pi G\bar\rho \delta$ where $\delta$ is
the density fluctuation. The root mean square (RMS) fluctuation for $\Delta \mu_{\rm
  pert.}^{-1}$ can be computed using the Limber approximation:
\be
(\Delta \mu_{\rm pert.}^{-1})_{\rm rms}  = 16(1-\kappa_0) \pi G\bar\rho
{(r_S - r_L) r_L \over r_S} \sqrt{{3R_{\rm halo}\over 4} 
\int {\rd^2 k_\perp \over (2\pi)^2} P_\delta (k_\perp) } \, ,
\ee
where $P_\delta$ is the 3D power spectrum of $\delta$.\footnote{We have used the fact that $r_L$, the distance to the mean location
  of the lensing halo, is much larger than the radius of the halo
  $R_{\rm halo}$. The integral is assumed to give $\int \rd r_L = 2R_{\rm
    halo}$. The Limber approximation is: $\int \rd r_L' \int \rd r_L \langle \nabla_\perp {}_x
  \nabla_\perp {}_x \Phi_{\rm pert.} (r_L) \nabla_\perp {}_x
  \nabla_\perp {}_x \Phi_{\rm pert.} (r_L') \rangle 
= 2R_{\rm halo} \int {\rd^2 k_\perp \over (2\pi)^2} k^4_\perp {}_x 
  P_{\Phi_{\rm pert.}} (k_\perp)$, where the power spectrum for 
$\Phi_{\rm pert.}$ is $P_{\Phi_{\rm pert.}} (k) =
  k^{-4} (4\pi G \bar\rho)^2 P_\delta (k)$. The notation $P_{\Phi_{\rm
      pert.}} (k_\perp)$ denotes $P_{\Phi_{\rm pert.}}
  (k=k_\perp)$. In other words, the essence the Limber approximation
  is to use the line of sight integral to set the line of sight
  momentum $k_\parallel$ to $0$. 
}
The momentum integral is convergent because fluctuations are smooth
for $k_\perp \, \gsim \, \pi/\lambda_{\rm c}$. A reasonable estimate is that
the integral is dominated by fluctuations on the characteristic scale
$k_\perp \sim \pi/\lambda_{\rm c}$, so the integral is roughly
$\lambda_{\rm c}$ (in other words, we assume $4\pi k_\perp^3
P_\delta(k_\perp)/(2\pi)^3$ is order unity at the characteristic scale). Putting this together, we deduce
\be
(\Delta \mu_{\rm pert.}^{-1})_{\rm rms}  = 8(1-\kappa_0) \pi G\bar\rho
{(r_S - r_L) r_L \over r_S} \sqrt{3R_{\rm halo}\lambda_{\rm c}} \, 
= 2(1-\kappa_0) 3^{3/2} 
v^2 {(r_S - r_L) r_L \over r_S R_{\rm halo} }\sqrt{\lambda_{\rm c} \over
  R_{\rm halo}} \, ,
\ee
where $v^2 = GM_{\rm halo}/R_{\rm halo} = 4\pi G\bar\rho R_{\rm
  halo}^2/3$. A rough numerical estimate is therefore
%
\be
( \Delta \mu_{\rm pert.}^{-1})_{\rm rms}\sim
\begin{cases}
  ~4 \times 10^{-4} 
\left({10^{-22} {\,\rm eV} \over m}\right)^{1/2}
\left({v \over 1000 {\,\rm km/s}}\right)^{3/2}
\left({1 {\,\rm Mpc} \over R_{\rm halo}}\right)^{3/2} ~~~~~~{\rm cluster}\,, \\[5pt]
~4.5 \times 10^{-3} \left({10^{-22} {\,\rm eV} \over m}\right)^{1/2}
\left({v \over 250 {\,\rm km/s}}\right)^{3/2}
\left({50 {\,\rm kpc} \over R_{\rm halo}}\right)^{3/2} ~~\,~~{\rm galaxy}\,,
\end{cases}
\ee
where we have assumed $2(1-\kappa_0) \sim 1$ and $(r_S - r_L)r_L/r_S
\sim 1$ Gpc. Note that this estimate is about a factor of 10 smaller than~\eqref{eq:overestimate}.

In order to understand the impact of this substructure component, recall
eq.~\eqref{mudecompose}:
$\mu^{-1} = \mu^{-1}_{\rm smooth} + \Delta\mu_{\rm pert.}^{-1}$. 
If substructure were absent, the two images close to the critical
curve would have the same magnification $\lvert\mu_{\rm smooth}\rvert$. 
Substructure makes $\lvert\mu\rvert \sim \lvert\mu_{\rm smooth}\rvert (1 - \mu_{\rm
  smooth} \Delta\mu_{\rm pert.}^{-1})$, where $\Delta\mu_{\rm pert.}^{-1}$ differs
between the two images by an amount indicated by the RMS value
estimated above. 
Thus, the flux anomaly (the departure from
unity of the flux ratio
between two images close to a critical curve) is
\be
\lvert\mu_{\rm smooth} \rvert  \Delta\mu^{-1}_{\rm pert.}  \sim
\begin{cases}
  ~0.06 \left({\lvert\mu_{\rm smooth}\rvert \over
  150}\right) \left({\Delta\mu^{-1}_{\rm pert.} \over 4 \times 10^{-4}}\right) ~~~~~~~~{\rm cluster}\,, \\[5pt]
~0.45 \left({\lvert\mu_{\rm smooth}\rvert \over
  100}\right) \left({\Delta\mu^{-1}_{\rm pert.} \over 4.5 \times
  10^{-3}}\right) ~~\,~~~~{\rm galaxy}\,,
\end{cases}
\ee
where the values for $|\mu_{\rm smooth}|$ correspond to those found
for some cluster and galaxy lenses~\cite{Dalal:2001fq,Dai:2020rio}.
 If we had used the quick
(over-)estimate for $\Delta\mu^{-1}_{\rm pert.}$ derived earlier, the flux
anomaly would be significantly bigger, e.g.
$\sim 1$ for the cluster example.
A recent paper based on simulations finds similar results to
our conservative estimate~\cite{2020arXiv200210473C}. It is also worth noting that a recent measurement of the flux
anomaly in a cluster lens gives an anomaly at the
level of tens of percent~\cite{Dai:2020rio}. 

Two more comments before we leave this subject.
(1) With more data, it would be possible to measure the flux anomaly
as a function of image separation, in which case, one has effectively
a means to measure the substructure power spectrum~\cite{Hezaveh:2014aoa,Rivero:2017mao}. The
interference substructure has a characteristic scale below which the
power spectrum should be suppressed, which might provide a way to distinguish
this kind of substructure from more conventional substructure due to sub-haloes.
(2) In the random
phase model, while the wavefunction $\Psi$ is Gaussian random, the
density fluctuation is highly non-Gaussian. Thus
computing the RMS density fluctuation does not quite
capture the full picture of what is going on. In the language of vortices, we know
from Fig.~\ref{fig:sizehist2}
there are vortices significantly smaller than the
coherence scale $\lambda_{\rm c}$. The associated steep spatial gradient
could give enhanced lensing effects. This is worth further investigation.

\subsubsection{Lensing deflection}
\label{sec:deflect}

A natural question to ask is whether there is an
observable signal that probes specifically vortices or
vortex rings (as opposed to generally probing small-scale
interference). 
One distinctive feature of vortices is that they are associated with 
under-dense fluctuations, leading to de-magnification compared to the
smooth background.
Another interesting feature is that vortices can 
in principle move very fast (see eq.~\eqref{eq:vortexspeed}): $v_{\rm
  vortex} \sim 1/(mR)$ where $R$ is either the size of a ring, or the
curvature scale of a portion of a ring. Normally, in a given halo with
velocity dispersion $v$, we would not consider objects moving at a
much greater speed than $v$, because they would have escaped the halo.
Vortices are an exception: $v_{\rm vortex} \gg v$ is possible, because
vortices can change and disappear before they could escape. Indeed, they are
locations of zero density, so the contrast with their surrounding
could give rise to interesting lensing signals. 
To exploit these features, we consider possible time-dependent
observables.

The time varying flux of a distant source, due to gravitational
lensing by a transiting vortex, is in principle an interesting
signal---the flux would drop instead of increase during the transit.
But the time scale of variation is very long. 
It is roughly $t_{\rm c} \sim 9 \times 10^5~{\rm yr.}$ for $v = 250$
km/s, or $\sim 5.6 \times 10^4~{\rm yr.}$ for $v = 1000$ km/s
(assuming $m = 10^{-22}~{\rm eV}$ )
\eqref{eq:characteristicscales}.
Consider for instance the flux anomaly deduced in the previous section
(which is caused not by a single vortex but a random superposition of
coherent patches): dividing the $10\%$ anomaly by $t_{\rm c}$ gives an un-observably small
variation per year.

Angular positions ({\it i.e.}, astrometry) can be measured
to much higher precision than fluxes. Is it possible to detect time
varying lensing deflection? This was advocated by~\cite{VanTilburg:2018ykj,Mondino:2020rkn} as a way to
search for sub-haloes inside the Milky Way halo. Sub-haloes can have
much higher density than the average density of the parent halo,
whereas the interference substructure gives rise to only order unity
fluctuations in density, making it more challenging to detect. 
It is natural to ask if a high vortex 
velocity can make up for the shortfall. 

Recall the setup: consider the lensing equation~\eqref{lenseqt} with
the gravitational potential split into $\Phi_{\rm smooth}$
and $\Phi_{\rm pert.}$. We are interested in $\Phi_{\rm pert.}$ being
sourced by the under-dense fluctuation associated with a moving vortex
that happens to be crossing the line of sight to a distant source.
The vortex resides in some galaxy along the line of sight---it could
even be our own galaxy.
Taking the time derivative of the lensing equation, and assuming
$\dot \theta_S$ is negligible, such as for a distant source 
(an assumption we will revisit below), we obtain
\be
\dot \theta_I \sim 2 v_{\rm vortex} \int\rd r_L {r_S - r_L \over r_S} \,
  \nabla_\perp^2 \Phi_{\rm pert.} \,   \, .
\ee
where we have made the estimate that $\dot\Phi_{\rm pert.}$ is of order $v_{\rm vortex}$ times a
spatial derivative, and we are not keeping track of the vectorial
nature of $\theta_I$ for simplicity. 
The gravitational potential $\Phi_{\rm pert.}$
obeys the Poisson equation $\nabla^2 \Phi_{\rm pert.} = 4\pi G \delta\rho$, where
$\delta\rho$ is the density fluctuation which we can model as
\be
\delta\rho = \begin{cases}
  ~\bar\rho \left( {r_\perp^2 \over r_0^2} - 1 \right)
  {\quad \rm for \quad} r_\perp < r_0 \\[5pt]
~0 ~\qquad\qquad\quad\,{\rm otherwise}\
\end{cases}
.
\ee
The deflection from such an
object can be worked out exactly
(Appendix \ref{sec:lensvortex})
 but for our
purpose, dimensional analysis suffices (we ignore effects associated with relative
orientation between vortex and line of sight). The time-variation of the deflection angle scales as
\be
\dot\theta_I \sim 8\pi G\bar\rho r_0\,v_{\rm vortex} {r_S - r_L \over
  r_S} \, ,
\ee
where we assume $r_0 \ll r_L$ ($r_L$ being the location of the
vortex lens). 

There are two scenarios we can consider. 
A conservative scenario is to assume $v_{\rm vortex} = 1/(m r_0)$,
with $r_0$ signifying the size of the vortex ring in question. In this
scenario, we see that the advantage from the high velocity of a small
ring is completely washed out by the diminished effect from a small
ring:
\be
\label{smallthetadot}
\dot\theta_I \sim 4 \times 10^{-11} {\,\rm arcsec}/{\,\rm yr.} 
\left({\bar\rho \over 0.3 {\,\rm GeV\,cm^{-3}}}\right) 
\left({10^{-22} {\,\rm eV} \over m} \right) \, ,
\ee
assuming $(r_s - r_L)/ r_s$ is order unity.\footnote{Note that the location of the vortex lens does not matter much: 
it can be in a galaxy
somewhere halfway between the source and us, or it can be located very close to
us, including in our own galaxy. The factor of $r_S - r_L / r_S$
remains order unity, unless the lens is located very close to the source.}
In this scenario---since small and large rings give similar
contributions---there is no reason to single out the deflection from a
single vortex ring (or segment of a vortex ring). To estimate the net
effect, one can therefore add up
in a random walk fashion deflections from patches of size $\lambda_{\rm c}$
along the line of sight through a galaxy, similar to what we did for the magnification itself. 
Using $\sqrt{R_{\rm halo}/\lambda_{\rm c}} \sim 20$, we find
$\dot\theta_I \sim 8 \times 10^{-10} {\,\rm arcsec}/{\,\rm yr}$.\footnote{This estimate of the apparent image velocity
does not crucially rely on details of the vortices; all it assumes is
the existence of order unity density fluctuations on scale of
$\lambda_{\rm c}$.
It is also worth noting that the net image displacement after the passage
of a single $\lambda_{\rm c}$ size blob or vortex 
is roughly $8\pi G\bar\rho \lambda_{\rm c}^2 \sim 10^{-4} {\,\rm
  arcsec}$ assuming the same parameters as in
eq.~\eqref{smallthetadot}. The smallness of $\dot\theta_I$ is due to the fact that the passage
takes such a long time.
}

Alternatively, an optimistic scenario goes like this: we have a vortex
ring of size order $\lambda_{\rm c}$, but imagine there is a segment of it that has high
curvature and is therefore moving at high speed (and crossing the line of
sight to a distant source). The resulting estimate for the time
variation of the lensing deflection is:
\be
\dot\theta_I \sim 5 \times 10^{-10} {\,\rm arcsec} / {\rm yr.} 
\left({\bar\rho \over 0.3 {\,\rm GeV\,cm^{-3}}}\right) 
\left({r_0 \over 0.24 {\,\rm kpc}}\right) 
\left({v_{\rm vortex} \over 10^3 {\,\rm km/s}}\right) \, .
\label{eq:optimisticdeflectiontime}
\ee
The fact that this is roughly at the same level as the random walk
estimate means both---deflection by a random sum of $\lambda_{\rm c}$
patches and deflection by an especially high velocity vortex---
should be taken into account.
The estimate in eq. (\ref{eq:optimisticdeflectiontime})
might be overly optimistic: 
even though
the vortex ring is of size $\lambda_{\rm c}$, the high curvature around
the segment of interest (so it has a high velocity) likely means the
generic $r_\perp^2$ density profile extends to
a smaller distance than $\lambda_{\rm c}$.\footnote{This is why we refrain from choosing an even higher $v_{\rm
    vortex}$ in eq. (\ref{eq:optimisticdeflectiontime}).
Note that a large vortex velocity (or for that matter large fluid
velocity) is made possible by the nonlinear combination of derivatives 
of $\Psi$ that go into it
(eq. (\ref{velocitymatrix})). Interesting observables typically involve 
multiplying this velocity by other quantities (such as density or
gradient of density), which tends to counteract the effects of a large velocity.
}
On the other hand,
the density $\bar\rho$ could be quite a bit higher than the solar
neighborhood value assumed here.

Even the optimistic $\dot\theta_I$~\eqref{eq:optimisticdeflectiontime} is challenging to observe, for two
reasons. One is technological: the {\it Gaia}
experiment has a precision of about $10^{-6}$ arcsec at best, though
this will improve with future experiments.
The other challenge is proper motion: sources move so that 
$\dot\theta_S \ne 0$. Even a source at cosmological distance has
$\dot\theta_S \sim 2 \times 10^{-8} {\,\rm arcsec/yr.}$, assuming
source velocity of $200$ km/s and distance of $2$ Gpc.
One would therefore need a significantly higher density $\bar\rho$ for
$\dot\theta_I$ to rise safely above the proper motion limit. 
It was suggested by~\cite{VanTilburg:2018ykj,Mondino:2020rkn,Mishra-Sharma:2020ynk} that 
if many sources were present, their correlated apparent motions due to
lensing by an extended object (such as a vortex) can be used to
look for very small effects.
The expected $r_\perp^2$ density profile of a vortex, and its
effective gravitational repulsion instead of attraction, might
be helpful signposts in this endeavor.

\subsection{Experimental implications}
\label{sec:experiment}

The small-scale features of light dark matter that we have been discussing also 
have implications for direct detection experiments.
Different axion (or axion-like-particle ALP) detection experiments are sensitive to
a range of masses: all the way from $10^{-15} - 10^{-3}~{\rm eV}$~\cite{Graham:2015ouw,Irastorza:2018dyq}.
Interference substructures, including vortices, should be just
as ubiquitous as they are in the ultra-light  ({\it i.e.}, fuzzy) regime. The only difference is
the overall scale (see eq.~\eqref{eq:characteristicscales}). Recall that the axion field $\phi$ is
related to the wavefunction $\Psi$ by
\be
\begin{aligned}
\label{Psitophi}
 \phi &= {1\over \sqrt{2m}} \left(\Psi e^{-i m t} + \Psi^* e^{i m t}
  \right) \\
	&= {\sqrt{2\over m}} \left( \Psi_1 
  {\,\rm cos\,}(mt) + \Psi_2
  {\,\rm sin\,}(mt) \right)  \\
	&= {\sqrt{2\over m}} \lvert\Psi\rvert
  {\,\rm cos\,}(mt-\theta)  \, ,
\end{aligned}
\ee
where we exhibit different ways to express $\Psi = \Psi_1 + i
\Psi_2 = |\Psi| e^{i\theta}$ that will be helpful in our
discussion.

The behavior of $\phi$ is governed by two different time scales. One
is the fast oscillation time scale associated with $m$ (Compton): $t_{\rm osc.} \sim 2
\times 10^{-9}~{\rm s}$ for $m \sim 10^{-6}~{\rm eV}$.
The other is a slower coherence time scale associated with $\Psi$ (de Broglie):
$t_{\rm c} \sim 3 \times 10^{-3}~{\rm s}$ for $m \sim 10^{-6}~{\rm eV}$. Moreover,
$\Psi$ varies in space on the scale
$\lambda_{\rm c} \sim 7.4 \times 10^4~{\rm cm}$. In other words, at a given
location (say of the detector), the rapid $t_{\rm osc.}$ oscillations of $\phi$
are modulated slowly on scale $t_{\rm c}$. The slow modulation is due to the
interference substructures of interest. A recent paper that emphasizes
this aspect of the signal is~\cite{Centers:2019dyn}.

Why are interference substructures interesting for axion detection
experiments? Let us give a brief preview:
1. At a minimum, they
are relevant for properly interpreting experimental results: limits on
the local axion density must account for the probability distribution
of density \cite{Centers:2019dyn}. 2. We will see that correlation
functions of $\phi$ are expected to have definite shapes; the
associated templates can be fruitfully applied to noisy data to
enhance detection significance. 3. The probability distribution for
$\phi$ contains information about the halo velocity dispersion
which can be cross-checked against astronomical
measurements. 4. Axion detection experiments largely focus on the
amplitude $|\Psi|$ of axion oscillations. It is worth exploring
the phase $\theta$ as another potential observable. 

Two types of interactions are often used by
axion detection experiments, which couple the axion either to photons or
to fermions.\footnote{There is also a possible coupling to gluons. See
  {\it e.g.}, \cite{Graham:2013gfa}.}
The corresponding couplings are given by
\be
{\cal L}_{\rm int.}^{\gamma} \sim g_{\gamma} \, \phi \, \epsilon_{\mu\nu\alpha\beta} \,
  F^{\mu\nu} F^{\alpha\beta} ~, \qquad\quad
{\cal L}_{\rm int.}^{\psi} \sim g_\psi \, \partial_\mu \phi \, \bar\psi
  \gamma^\mu \gamma_5 \psi \, ,
\ee
where $\phi$ is the axion, $F^{\mu\nu}$ is the photon field strength,
$\psi$ is a fermion (for instance it can be the electron or a nucleon), $g_\gamma$ and $g_\psi$
are dimension-ful couplings, generically of the order of $1/f$ where $f$ is the
axion decay constant. Some general features of the physics of these couplings can be found in Appendices~\ref{sec:axionphoton} and~\ref{sec:axionfermion}.
There are many axion detection experiments and it is beyond our scope
to discuss them systematically (see \cite{Graham:2015ouw,Sikivie:2020zpn} for
reviews). Here, we pick a few examples to illustrate the main points.

A cavity experiment such as ADMX~\cite{Du:2018uak} seeks 
photons produced in the presence of the axion and an external
magnetic field, via the ${\cal L}_{\rm int.}^\gamma$ interaction.
The signal is often phrased in terms of the power output in a
microwave cavity which is proportional to $\phi^2$, or $\phi^2$
averaged over the rapid oscillations~\cite{Sikivie:1983ip}.
Another type of experiment such as ABRACADABRA~\cite{Kahn:2016aff,Ouellet:2018beu} seeks a time varying magnetic field produced by an
oscillating axion in the presence an external magnetic field.
Its signal is proportional to $\dot\phi$.
The typical size of these experiments is small compared to $\lambda_{\rm c}$,
thus one can think of the data stream as a time series of $\phi^2 (t)$
or $\dot\phi(t)$ at a single location.
The key point is that:
{\it $\phi$ is a stochastic variable because it consists of a
superposition of waves.} The expectation value $\langle \phi^2 \rangle$
tells us the average dark matter density 
$\bar\rho = m^2 \langle \phi^2 \rangle$. 
In the random phase model, the
probability distribution of $\rho$
is exponential
$\propto e^{-\rho/\bar\rho}$ (see eq.~\eqref{expP}).
Equivalently,
the probability distribution for the amplitude of
rapid oscillations of $\phi$ (essentially $\lvert\Psi\rvert$) obeys the Rayleigh
distribution (see eq.~\eqref{RayleighP}), as emphasized
by~\cite{Centers:2019dyn}.
In what follows, we will use the random phase model to
explore several correlation functions involving $\phi$. 
But it is worth keeping in mind that while the random phase model captures the probability at low $\lvert\Psi\rvert$ well (destructive
interference), it does not do a good job at high $\lvert\Psi\rvert$ in a halo
formed from gravitational collapse, as
evidenced by Figure \ref{psipdf}. 
It would be useful to revisit the computation of~\cite{Centers:2019dyn}, who pointed
out
that the constraints on the average local dark matter density
are weakened in light of the Rayleigh distribution, largely because of
the non-negligible probability of destructive interference. 
The high $\lvert\Psi\rvert$ tail above the Rayleigh distribution (as seen in Figure
\ref{psipdf}) acts in the opposite direction. 

Recognizing that $\phi$ is a stochastic
variable suggests new observables.
One option is to measure
the two point correlation function of $\phi$:\footnote{Here, we focus
  on correlators at a single spatial location but at two different
  times. More broadly, one could be interested in general space-time dependent correlators of the $\phi$ variables. We provide explicit formulas for these in Appendix~\ref{eq:axioncorrelators}.}
\be
\langle \phi (t) \phi (t') \rangle = {1\over 2m} 
\left( \langle \Psi (t) \Psi^* (t') \rangle e^{-im(t-t')} + {\,\rm
  c.c.} \right) \, ,
\ee
where ${\rm c.c.}$ represents complex conjugate.\footnote{The 2-point correlation function of $\dot \phi$ can be simply
  obtained by differentiation and making the non-relativistic
  approximation {\it i.e.} ignoring $\dot\Psi$ in comparison to
  $m\Psi$, $\langle \dot \phi (t) \dot \phi (t') \rangle = m^2 \langle
  \phi (t) \phi (t') \rangle$.}
We assume $\langle \Psi (t) \Psi (t') \rangle = \langle \Psi^* (t)
\Psi^* (t') \rangle = 0$, consistent with (but more general than) the
random phase model.\footnote{The assumption of $\langle \Psi (t) \Psi (t') \rangle = 
\langle \Psi^* (t) \Psi^* (t') \rangle = 0$ is equivalent to
$\langle \Psi_1 (t) \Psi_1 (t') \rangle = \langle \Psi_2 (t) \Psi_2
(t') \rangle$ and $\langle \Psi_1 (t) \Psi_2 (t')\rangle = - \langle
\Psi_2 (t) \Psi_1 (t') \rangle$. See eq.~\eqref{2pointPsi}. 
}  
Using the notation of Section
\ref{sec:randomphase} the $\phi$ two-point function can be rewritten in terms of real and imaginary parts $\Psi_1$ and
$\Psi_2$ as
\be
\langle \phi (t) \phi (t') \rangle 
= {2\over m} \Big[ \langle \Psi_1 (t) \Psi_1 (t') \rangle {\,\rm
  cos\,}(m(t-t')) + \langle \Psi_2 (t) \Psi_1 (t') \rangle {\,\rm
  sin\,}(m(t-t')) \Big] \, .
\ee
In the random phase model discussed in Section \ref{rpm}, the $\Psi_i$ correlation functions are given by
\begin{align}
\langle \Psi_1 (t) \Psi_1 (t') \rangle &= {1\over 2} \sum_{\vec k}
  A_{\vec k}^2 \, {\,\rm cos}[\omega_k (t-t')] \\
\langle \Psi_2 (t) \Psi_1 (t') \rangle &= -{1\over 2} \sum_{\vec k}
  A_{\vec k}^2 \,{\,\rm sin}[\omega_k (t-t')]  \, , \\
 \langle \Psi (t)
\Psi^* (t') \rangle &= 2 \langle \Psi_1 (t) \Psi_1 (t') \rangle + 2 i \langle
\Psi_2 (t) \Psi_1 (t') \rangle
\, .
\end{align}

The $\Psi$ correlation functions are ``slow'' in the sense
that they deviate from a constant when $t-t'$ approaches
the coherence time $t_{\rm c}$ or larger. 
The standard approach of considering output power ($\propto \phi^2$) is
essentially the same as focusing on equal times, $t=t'$. 
For $t \ne t'$, the oscillations on the timescale $1/m$ provide a way to
measure the axion mass. For large time separations $\lvert t-t'\rvert$, the rapid
oscillations would make it difficult to measure
$\langle \phi(t) \phi(t') \rangle$.
An alternative approach is to measure correlation functions of power
itself. 
This correlation function also has pieces that oscillate with
frequency $m$, but it has a piece that survives averaging
over the rapid oscillations:
\be
\langle \phi(t)^2 \phi(t')^2 \rangle - \langle \phi^2 \rangle^2 \simeq 
{1\over m^2} \lvert\langle \Psi (t) \Psi^* (t') \rangle\rvert^2  \, ,
\label{powercorr}
\ee
where we assume $\Psi$ and $\Psi^*$ are correlated Gaussian random
fields.\footnote{Typically, the power is already an averaged quantity over
the rapid oscillations {\it i.e.} $\phi^2(t)$ is implicitly averaged,
in which case eq. (\ref{powercorr}) is exact. Note that the averaged
output power is determined by $|\Psi|^2$ as well as the size of the
external magnetic field and volume of the cavity.
}
This is a nice non-oscillatory quantity to consider, and
the characteristic time separation on which it varies is the coherence time $t_{\rm c}$. This correlator can be computed exactly in the random
phase model, assuming $A_{\vec k} \propto e^{-k^2/k_0^2}$, which leads to
\be
\xi (t-t') \equiv {\langle \phi(t)^2 \phi(t')^2 \rangle - \langle \phi^2 \rangle^2
} \simeq \langle \phi^2 \rangle^2 \left(1 + {k_0^4 (t-t')^2 \over
  16m^2}
\right)^{-3/2} = {\bar\rho^2 \over m^4} \left(1 + {\pi^2 (t-t')^2 \over
  9 t_{\rm c}^2}
\right)^{-3/2} \, .
\label{eq:xi2pt}
\ee
The power correlation function, $\xi$, is also essentially the density correlation
function. 
It is interesting that $\xi$ has a power law decay at large $\lvert t-t'\rvert$ as
opposed to an exponential one. 
Keep in mind that this is the unequal time correlation at a single
location. The spatial correlation at equal time looks very
different:
\be
{\langle \phi(\vec x)^2 \phi(\vec x \, {}')^2 \rangle - \langle \phi^2 \rangle^2
} \simeq {4 \over m^2} {\langle \Psi_1 (\vec x) \Psi_1 (\vec
  x \, {}') \rangle^2} = \langle \phi^2 \rangle^2 {\,\rm
  exp}\,
\left(- {1\over 4} k_0^2 \lvert\vec x - \vec x \, {}'\rvert^2 \right) \, ,
\label{eq:phi2xphi2y}
\ee
in particular it falls off exponentially with distance.
The difference is due to the fact $\omega_k$, the frequency for a
Fourier mode, goes as $k^2$ rather than $k$. In Appendix~\ref{eq:axioncorrelators} for completeness 
we present the $\phi^2$ two-point function as a general space-time function, which reduces to~\eqref{eq:xi2pt} and~\eqref{eq:phi2xphi2y} in the appropriate limits.

Axion detection experiments often have durations longer than $t_{\rm
  c}$, especially for the higher masses. 
It is thus possible to measure $\xi(t-t')$ for time separations
exceeding $t_{\rm c}$. It is worth investigating
to what extent the characteristic $\lvert t-t'\rvert^{-3}$ decay can be exploited
to help pull signal out of noise.

The axion-photon coupling is exploited in a different way in
birefringence experiments, for instance
ADBC~\cite{Liu:2018icu}. There is no external
magnetic field, and thus no photon creation.
It seeks to detect, by interferometry, the phase difference
between photons of opposite helicities propagating in an axion
background. It can be shown the phase difference is proportional to
$\phi(t, \vec x) - \phi(t', \vec x \, {}')$, where $t, \vec x$ and
$t', \vec x \, {}'$ denote the beginning and end of the photon
trajectory. This sets up the possibility, if suitable
separations are chosen, to measure correlation functions in the full
space-time sense (Appendix~\ref{eq:axioncorrelators}), rather like interferometry in radio astronomy. Similar proposals were considered in~\cite{Derevianko:2016vpm,Savalle:2019jsb,Martynov:2019azm}.

So far, our discussion has been about general interference
substructures and what they imply for axion detection.
What about vortices in particular? Given that the typical size of
experiments are much smaller than the de Broglie wavelength, the
likelihood of encountering a vortex ring is small. Nonetheless, this
doesn't preclude one passing close by, leading to a drop in the signal
without going all the way to zero. Indeed, the fact that the one-point
probability distribution of the density is exponential 
$\propto e^{-\rho/\bar\rho}$~\eqref{expP} tells us destructive
interference is not all that uncommon. It is also worth noting that if
a vortex is actually encountered, it manifests as a phase jump 
(a wall) since the data stream in time is effectively one
dimensional. To truly identify a vortex would require detectors at
multiple locations to measure the telltale winding of the phase
$\theta$ in the oscillation of the axion~\eqref{Psitophi}. 

Another type of experiments
make use of the coupling of the axion to fermions (such as nucleons)
${\cal L}^\psi_{\rm int.}$. An example is CASPEr~\cite{Budker:2013hfa,JacksonKimball:2017elr}. The coupling means
the Hamiltonian for the nucleon has a term:
\be
H \sim g_\psi \vec \nabla \phi \cdot \vec \sigma \, ,
\ee
where $\sigma$ is the nucleon spin operator. 
The concept is similar to nuclear magnetic resonance. 
Set up a magnetic field so the nucleon spin is aligned with it.
The vector $\vec \nabla\phi$ (which oscillates with frequency $m$) 
is in general not aligned with the magnetic field, and causes the spin
to precess. The effect is most pronounced when the strength of the
magnetic field is such that the precession frequency matches $m$. 
Another type of experiment that utilizes the same coupling is the torsion pendulum experiment~\cite{Terrano:2019clh}. 

Our discussion of the correlation functions for $\phi$ and
$\phi^2$ can be translated in a straightforward way to correlation functions for $\vec \nabla
\phi$. One interesting feature of this technique is that encountering
a vortex does not mean the signal goes to zero. At a vortex, 
$\phi$ vanishes, but $\vec \nabla \phi$ does not. This can be checked
by studying the simple solution $\Psi \propto x + i y$ of a vortex
extending in the $z$ direction (see Section \ref{sec:general} 
on why this is a good local approximation to a realistic vortex). This
gives $\nabla_x \phi \propto {\,\rm cos\,} (mt)$ and $\nabla_y \phi
\propto {\,\rm sin\,} (mt)$. This example also vividly shows that
$\vec\nabla \phi$ is not necessarily described by $\sqrt{\bar\rho}$ times
the velocity of the Earth through the dark matter halo. Wave interference
inevitably causes space and time fluctuations in $\vec
\nabla \phi$. The random phase model provides a useful starting point
in characterizing them.

The existence of vortices tells us there is interesting spatial
structure, {\it i.e.} winding, in the phase of $\Psi$. 
The phase manifests itself as the phase $\theta$ of axion oscillations: 
$\phi \propto |\Psi| {\,\rm cos\,}(mt - \theta)$. 
What observable can one use to access $\theta$? 
A possibility is:
\begin{eqnarray}
\phi \vec \nabla \dot \phi - \dot\phi \vec \nabla \phi
= 2 |\Psi|^2 \vec \nabla \theta \, ,
\end{eqnarray}
where the equality holds in the non-relativistic regime ($|\dot\Psi|
\ll m |\Psi|$). This is essentially the fluid momentum density
$\propto \rho \vec v$. It is not clear what the best way is to measure
this quantity, or its correlation function. It could be a combination
of experiments that use the axion-photon coupling and the
axion-fermion coupling. It could also be a series of detectors in
different locations. It would be interesting to explore whether
this quantity can be fruitfully combined with other quantities
to isolate $\vec\nabla \theta$, and under what conditions the phase
winding is measurable. Perhaps there might be a way to connect
this with some kind of Berry phase. We leave this for future work.

\section{Conclusions}
\label{sec:conclude}

Dark matter consisting of particles lighter than about $\sim 30~{\rm eV}$ behaves essentially like a classical wave. Under the influence of gravity, haloes form with a central cored density profile surrounded by an an outer halo with a time-averaged density profile of the NFW form. Although the average density profile is similar to that of massive cold dark matter particles, the outer halo is a dynamical place: it can be thought of as a bound configuration of superposed waves interfering in a highly complex manner.

The classical superposition of wave-like dark matter in the outer halo region leads to a wide variety of interesting phenomena on small scales. In fact many of the interesting small-scale features that have been identified in models of ultra-light dark matter can be understood in this context. In this paper we have investigated these small-scale interference phenomena, focusing in particular on the formation of defects like vortex strings. The presence and dynamics of these objects are 
sufficiently generic that they can be characterized in a universal way.

We have investigated the formation and characteristics of vortices using a combination of analytical arguments and numerical simulations and have identified a number of robust features that wave-like dark matter vortices have: (1) The mere existence of vortices is generic in three spatial dimensions, where vorticity of the dark matter fluid is localized on closed loops where the dark matter density vanishes. (2) The number density of vortex rings is approximately one per de Broglie volume. (3) The winding (or velocity circulation) around such vortex lines is generically $\pm1$ in units of $2\pi/m$, where $m$ is the dark matter mass.
(4) The dark matter density profile in the vicinity of the vortices
scales like $\sim r_\perp^2$ (once circularly averaged) where
$r_\perp$ is the distance from the vortex, while the (circularly
averaged) velocity profile scales as $\sim 1/r_\perp$. We further have argued for the following dynamical features of vortices: that rings should move with characteristic velocity inversely proportional to their size $v\sim (m\ell)^{-1}$ and that vortices should reconnect rather than frustrate. We have not verified these latter expectations in detail with numerics, and it would be very interesting to do so.

A natural question is how the existence of vortices might relate to angular
momentum. The angular momentum of a halo is given by:
\begin{eqnarray}
\vec J = \int \rd^3 x \, \rho \, \vec x \, \times \vec v
= -i \int \rd^3 x \, \vec x \,\times \Psi^* \vec \nabla \Psi
\end{eqnarray}
assuming the origin is at the halo center of mass.
The numerical examples in 
Section~\ref{sec:numerics} show that even in a halo with zero net
angular momentum, vortex rings appear anyway, as a result of chance
interference. This suggests there is not a tight relation
between halo angular momentum and the existence of vortices.
Moreover, it is not difficult to construct solutions 
to the Schr\"odinger equation that carry angular momentum, but do not
have vortices. Consider a superposition of the solutions discussed
in~\eqref{eq:hydrogenatomlikewfs}. For example, adding together an
$s$-wave and a
solution of the form 
$\Psi_{l,l}$, we can arrange for a solution that
globally carries $l$ units of angular momentum, but the
wavefunction nowhere vanishes
(by making the $s$-wave component large enough). 
This is not to say one cannot find solutions where angular momentum
and vortices co-exist: for instance, a pure $\Psi_{l,l}$ solution on its own
{\it i.e.}, essentially an angular momentum eigenstate.\footnote{By this we mean a field configuration 
$\Psi$ such that $\partial_\varphi \Psi \propto \Psi$, where $\varphi$
is the azimuthal angle, for angular momentum in the $z$ direction.}
Solutions of this kind that include gravity---rotating
solitons---were investigated by~\cite{Hertzberg:2018lmt}. 
Realistic haloes are mainly supported by velocity dispersion rather
than rotation {\it i.e.} they are not described by an angular momentum 
eigenstate.

Throughout this paper we have focused on the non-relativistic regime,
where the dynamics is well-captured by the Schr\"odinger--Poisson
equations. In this context, the winding around a vortex is a winding
of the U$(1)$ phase of the non-relativistic wavefunction. 
This corresponds to a winding in the phase space of the real scalar $\phi$.
In the infrared, the total number of vortices is fixed once one imposes boundary conditions of fixed winding at infinity---vortices can only be produced in vortex-anti-vortex pairs, or as rings. It would be interesting to elucidate the properties of the symmetry that protects this number conservation and to investigate its fate in the ultraviolet.

The presence of small-scale features in general (and vortices in
particular) in a dark matter halo can have a number of consequences
for observations and experiments. 
We discussed both astrophysical signals for an ultra-light scalar
(fuzzy dark matter) and experimental signals for a light, but not necessarily
ultra-light, axion or axion-like particle.
On the astrophysical side, a natural possibility is to detect 
the interference-induced ${\cal O}(1)$ density fluctuations via gravitational
lensing. We estimated that interference substructures could induce
$5 - 10 \%$ anomalous flux ratios in strongly lensed systems, in
agreement with the recent study~\cite{2020arXiv200210473C}.
It would be useful to investigate in detail whether the $\sim 10\%$ flux anomaly
detected by \cite{Dai:2020rio} can be explained in this way, and how
this explanation might be differentiated from one invoking small
sub-haloes---such as by measuring the flux anomaly as a function of
image separation. We took a first step in examining how the rapid
motion of some vortices might impact on gravitational lensing. 
It would be useful to investigate further the motion of
interference substructures in numerical simulations.
This might have relevance beyond gravitational lensing.
Stars in the galaxy can get heated by gravitational encounters
with interference substructures
\cite{Hui:2016ltb,Amorisco:2018dcn,Church:2018sro,Bar-Or:2018pxz,Marsh:2018zyw}. 
If some fraction of such encounters were at a high relative speed, as
suggested by the example of vortices, is the heating efficiency
thereby suppressed? 

The wave-like nature of dark matter also has important consequences
for local direct searches for axions or axion-like particles. The stochastic
variability of the axion field suggests that density
correlations either in time or space will be useful probes of models
of this type. We advocate in particular the measurement of $\langle
\phi^2 (t) \phi^2 (t') \rangle$ which has a characteristic
$1/|t-t'|^3$ decay~\eqref{eq:xi2pt}. It would be interesting to see if this feature
can be exploited to extract a signal out of noisy data.
We investigated how vortices impact experiments
that exploit the axion-photon coupling versus those that use the
axion-fermion coupling.
It would be useful to improve upon the random phase model, to arrive
at a more accurate description of the one-point probability distribution of
the amplitude $|\Psi|$ of axion oscillations (Figure \ref{psipdf}).
More work needs to be done to investigate how to
probe the phase of axion oscillations ($\theta$ in~\eqref{Psitophi}).
Axion detection experiments largely focus on measuring the amplitude
$\lvert\Psi\rvert$, for good reasons. The example of vortices illustrates there is interesting
structure in the phase that can potentially be measured too.
We hope to address these issues in the near future.

\newpage
\noindent
{\bf Acknowledgements:} We would like to thank Mustafa Amin, Matteo
Biagetti, Greg Bryan, Andy Cohen, Liang Dai, Neal Dalal, Sergei Dubovsky, Benjamin Elder, Angelo Esposito,
Matt Kleban, Andrey Kravtsov, Denis Martynov, Jerry Ostriker,
Enrico Pajer, Alessandro Podo, Luca Santoni, Sergey Sibiryakov, Pierre
Sikivie, Dam Son,
John Stout, and Cora Uhlemann for helpful
discussions. We wish to thank especially Alberto Nicolis for
collaboration at an early stage, and Riccardo Rattazzi and Tanmay
Vachaspati for raising questions that motivated this project.
LH thanks Andy Cohen, Henry Tye, and the HKUST IAS for hospitality,
and the Munich Institute for Astro- and Particle Physics (MIAPP) for
a workshop where part
of the research was done.  Our work was supported in part by the NASA
NXX16AB27G and the DOE DE-SC011941.  AJ is supported in part by the Delta ITP consortium, a program of the Netherlands Organisation for Scientific Research (NWO) that is funded by the Dutch Ministry of Education, Culture and Science (OCW).
XL thanks the Center for Computational Astrophysics at Flatiron Institute for computational resources.
Research at Perimeter Institute is supported in part by the Government of Canada through the Department of Innovation, Science and Economic Development Canada and by the Province of Ontario through the Ministry of Colleges and Universities.

\newpage

\appendix

\section{Solution-generating technique for the Schr\"odinger equation}
\label{sec:solngenerating}
In~\cite{2000PhRvA..61c2110B} a useful procedure is given to produce solutions to the free Schr\"odinger equation with configurations of vortex lines. 
The basic idea is to take a solution to the time-dependent Schr\"odinger equation, $\Psi(\vec x, t)$ that at some initial time has the form of a plane wave times some spatial profile:
\be
\Psi(\vec x, 0) = e^{i\vec k\cdot\vec x}\phi(\vec x),
\label{eq:t0planewave}
\ee
and from this construct a solution to the full Schr\"odinger equation.
The key trick is to notice that $\Psi(\vec x, t)$ solves the Schr\"odinger equation for arbitrary $\vec k$, so we can differentiate $\Psi(\vec x, t)$ with respect to $\vec k$ arbitrarily to generate new solutions. At $t=0$ these solutions look like polynomials in $\vec x$ times $e^{i\vec k\cdot\vec x}\phi(\vec x)$. Recall that vortices lie at the intersection between the places where the real and imaginary parts of the wavefunction vanish. So, by suitably choosing the differential operators, we can arrange for any configuration of vortex lines that can be described by the intersection of surfaces defined by polynomials. The time dependence of these configurations are then fixed by the solution $\Psi(\vec x, t)$.

All of the examples that we will need can be generated from the plane-wave solution to the Schr\"odinger equation:
\be
\Psi_k(\vec x, t) = e^{i \vec k\cdot\vec x - \frac{i\vec k^2}{2 m} t}.
\label{eq:planewavesol}
\ee
At $t=0$, this takes the form of a plane wave $\Psi(\vec x,0) = e^{-\vec k\cdot\vec x}$. We can generate a configuration of vortex lines at $t=0$ by differentiating with respect to $\vec k$. If we then take the same differential operator and act on $\Psi(\vec x, t)$, this will generate a solution to the time-dependent Schr\"odinger equation which has the desired vortex configuration at $t=0$.

We may also be interested in situations with more than one vortex line. With gravity turned off, we are solving the free Schr\"odinger equation---which is linear---so superpositions will still solve the equation. But, they will typically not vanish anywhere, so blindly adding vortex solutions will generically remove the vortices. In order to get solutions with multiple vortices we have to be a little more careful. However, given two vortex configurations at $t=0$, their product will vanish at the locations of both of the vortices individually, and so will describe a configuration with two vortices. 
If we then take the differential operator that generates the $t=0$ configuration from~\eqref{eq:t0planewave} and act on~\eqref{eq:planewavesol} we can generate a time-dependent solution for the dynamics of this configuration.

\newpage

\section{Gravitational lensing by a vortex}
\label{sec:lensvortex}

In this appendix, we compute the deflection induced by a vortex with the density
fluctuation profile considered in Section~\ref{sec:deflect}
\be
\delta\rho = \begin{cases}
  ~\bar\rho \left( {r_\perp^2 \over r_0^2} - 1 \right)
  {\quad \rm for \quad} r_\perp < r_0 \\[5pt]
~0 ~\qquad\qquad\quad\,{\rm otherwise}\
\end{cases}
,
\ee
with $\nabla^2 \Phi_{\rm pert.} = 4\pi G \delta\rho$.
The lensing equation is
\be
\label{lenseqtApp}
\vec \theta_S = \vec \theta_I - 2 \int_0^{r_S}\rd r_L {r_S - r_L \over r_S} \,
  \vec\nabla_\perp (\Phi_{\rm smooth} + \Phi_{\rm pert.}) \,  \, ,
\ee
where $\vec \theta_S$ and $\vec \theta_I$  are are the source and
image positions. Assuming a thin lens, we write this as
\begin{eqnarray}
\vec \theta_S = \vec \theta_I + {r_S - r_L \over r_S} \vec \alpha_{\rm
  smooth} + {r_S - r_L \over r_S} \vec \alpha_{\rm pert.} \, .
\end{eqnarray}
Suppose there is some source position that maps to some image position
with $\vec \alpha_{\rm pert.}$ switched off. The same source position
then maps to a slightly different image position with $\vec
\alpha_{\rm pert.}$ turned on. Taking the difference between the corresponding
lens equations, and relabeling $\vec \theta_I$ as the shift in image
position, we have
\begin{eqnarray}
0 = \vec \theta_I +  {r_S - r_L \over r_S} \vec \alpha_{\rm pert.} +
  O(\theta_I \, \partial_{\theta_I} \, \alpha_{\rm smooth}) \, ,
\end{eqnarray}
where the term that involves derivative of $\alpha_{\rm smooth}$ can be
dropped assuming it is much less than unity. 

The deflection $x$ component of $\vec \alpha_{\rm pert.}$  is
\begin{eqnarray}
\alpha_{\rm pert.} = {\rm sgn\,}(\theta_I - \theta_V)\, {4\pi G\bar\rho
  r_0^2} \,\begin{cases}
{\pi \over 2} \qquad\qquad \qquad \qquad \qquad {\quad \rm if \quad} b > r_0
\\[5pt]
{\pi\over 2} -\bar\theta + {b^2 \over r_0^2} (2\bar\theta - {\,\rm
  sin}\bar\theta \, {}^2) \,{\quad \, \, \rm otherwise \quad}
\end{cases} \, ,
\end{eqnarray}
where $b \equiv r_L |\theta_I - \theta_V|$ and ${\,\rm
  cos\,}\bar\theta \equiv b/r_0$. We have assumed $z$ is the line of
sight direction, the vortex extends
in the $y$ direction and $\theta_V$ represents its $x$ coordinate,
$\vec\theta_I$ has vanishing $y$ component and $\theta_I$ exhibited
above is its $x$ component. Substituting $\theta_V = v_{\rm vortex}
t/r_L$ allows one to deduce how $\theta_I$ changes with time $t$,
for a vortex line that is moving in the $x$ direction (assuming
$v_{\rm vortex} > 0$). 
For $t \rightarrow -\infty$, $\alpha_{\rm pert.} \rightarrow 2\pi^2 G
\bar\rho r_0^2$, which tells us $\theta_I \rightarrow -(r_S-r_L) 2\pi^2 G
\bar\rho r_0^2 /r_S$. The negative sign is correct, reflecting the effective
repulsion of photons by the vortex (recall the source is at the
origin). Conversely, for $t \rightarrow \infty$,
$\alpha_{\rm pert.} \rightarrow - 2\pi^2 G
\bar\rho r_0^2$, which tells us $\theta_I \rightarrow (r_S-r_L) 2\pi^2 G
\bar\rho r_0^2 /r_S$. 

The vortex model used above could be made more realistic by allowing
for a non-circularly symmetric density profile, and an arbitrary
orientation of the vortex with respect to the line of sight and the
direction of motion. Further, the vortex is generally part of a ring rather than an
infinitely long string. 

\newpage
\section{Axion-photon interaction}
\label{sec:axionphoton}

Here, we
summarize a few results relevant to our discussion in Section \ref{sec:experiment}. 
A comprehensive review can be found in \cite{Sikivie:2020zpn}.
The Lagrangian for the axion and the photon (ignoring axion
self-interaction) is
\begin{eqnarray}
{\cal L} = -{1\over 4} F_{\mu\nu} F^{\mu\nu} + A_\mu J^\mu - {1\over
  2} \partial_\mu \phi \, \partial^\mu \phi - {1\over 2} m^2 \phi^2 +
  {1\over 4} g_\gamma \phi F_{\mu\nu} \tilde F^{\mu\nu} \, ,
\end{eqnarray}
where $\tilde F^{\mu\nu} \equiv \epsilon^{\mu\nu\alpha\beta}
F_{\alpha\beta} / 2$. The photon equation of motion is
\begin{eqnarray}
\partial_\mu F^{\mu\nu} = -J^\nu + g_\gamma \partial_\mu ( \phi\tilde
  F^{\mu\nu}) \, ,
\end{eqnarray}
which is equivalent to
\begin{eqnarray}
\label{dEdivE}
-\partial_t \vec E + \vec \nabla \times \vec B = \vec J - g_\gamma \partial_t
  (\phi \vec B) - g_\gamma \vec \nabla \times (\phi \vec E) \quad , \quad 
\vec \nabla \cdot \vec E = J^0 + g_\gamma \vec \nabla \cdot (\phi \vec B) \, .
\end{eqnarray}
The setup for ADMX or ABRACADABRA, for instance, is to first generate some constant external
magnetic field using $\vec J$, let's call it $\vec B_0$. With $J^0 =
0$, and assuming $\dot\phi \gg \nabla \phi$, we have
\begin{eqnarray}
-\partial_t \vec E + \vec \nabla \times \vec B = - g_\gamma \dot\phi \vec B_0
  \quad , \quad 
\vec \nabla \cdot \vec E = 0 \, .
\end{eqnarray}
The production of photons, or the generation of a magnetic field, can
be deduced from these equations. 

For photon propagation in an axion background (without an external
magnetic field), eq.~\eqref{dEdivE} can be used with $J^\mu$ set to
zero. The standard relation between $\vec E$ and $\vec B$,
following from $\partial_t \vec B + \vec \nabla \times \vec E = 0$,
continues to
hold:
\begin{eqnarray}
\vec \beta = {\vec k \over \omega} \times \vec \epsilon \, ,
\end{eqnarray}
where 
$\vec B = \vec \beta e^{-i\omega t + i
  \vec k \cdot \vec x}$ and $\vec E = \vec \epsilon e^{-i\omega t + i
  \vec k \cdot \vec x}$. 
The wave equation is modified by the presence of the axion:
\begin{eqnarray}
\label{wavephi}
(\partial_t^2 - \nabla^2) \vec E = g_\gamma \, (\partial_t \phi \, \partial_t
  \, \vec B
  - \partial_i \phi \, \partial_i \vec B ) \, ,
\end{eqnarray}
and a similar equation holds with $\vec E \rightarrow \vec B$
and $\vec B \rightarrow -\vec E$. With the source turned off, 
$\vec \epsilon = \pm i \vec \beta$ gives the usual circular
polarizations. Substituting this into eq. (\ref{wavephi}), one can
deduce the modified dispersion relation to first order in $g_\gamma$:
\begin{eqnarray}
\omega = |\vec k| \pm {1\over 2} g_\gamma (\dot\phi + \hat k \cdot \vec
  \nabla \phi) \, .
\end{eqnarray}
Alternatively, a WKB analysis
$\vec E = \vec \epsilon e^{i S}$
can be done to show the phase $S$ to first order in $g_\gamma$ is given by
\begin{eqnarray}
S = -|\vec k| t + \vec k \cdot \vec x \, \mp \, {g_\gamma\over 2} \int
  \rd t \,
  {D\phi \over Dt} \, ,
\end{eqnarray}
where $D/Dt$ is a total time derivative: $\partial_t + \hat k \cdot
\vec \nabla$. Thus the phase is sensitive only to $\phi$ at the
beginning and at the end. Having a long path length does not
confer an advantage except to make $\phi$ differ between start and finish.
It is worth noting the various contributions to this difference:
the dominant contribution is the Compton scale $m^{-1}$ time
variation; the next contribution comes through the {\it spatial}
variation of $\Psi$, giving the time scale $(mv)^{-1}$ (this is the de
Broglie scale divided by the speed of light, relevant because we are
interested in the propagation of photons); the smallest
contribution comes through the time dependence of $\Psi$ with
a time scale of $t_{\rm c} \sim (mv^2 )^{-1}$.

\newpage
\section{Axion-fermion interaction}
\label{sec:axionfermion}

We summarize a few results relevant for the discussion in Section~\ref{sec:experiment}. Comprehensive reviews can be found in \cite{Graham:2015ouw,Sikivie:2020zpn}.  Let us start with the interaction
Hamiltonian associated with the coupling of an axion to the spin of a fermion such as some nucleus:
\begin{eqnarray}
H = -\mu B \sigma_3 + g_\psi \vec \nabla \phi \cdot \vec \sigma \, ,
\end{eqnarray}
where $B$ is an external magnetic field assumed to point in the $z$
direction, $\mu$ is the magnetic moment and $\vec \sigma$ represents the Pauli
matrices. Assuming $B$ is large, we ignore the component of $\vec\nabla
\phi$ in the $z$ direction, and keep only $\nabla_x \phi$ and
$\nabla_y \phi$. Thus, the Hamiltonian takes the form
\begin{eqnarray}
H =
\left( 
\begin{array}{cc}
-\mu B & g_\psi (\partial_x \phi - i \partial_y \phi) \\
g_\psi (\partial_x \phi + i \partial_y \phi) & \mu B
\end{array}
\right) \, .
\end{eqnarray}
This can be rewritten as
\begin{equation}
\small
H = \left( 
\begin{array}{cc}
-\mu B & \eta e^{imt} + \gamma e^{-imt} \\
\eta^* e^{-imt} + \gamma^* e^{imt} & \mu B
\end{array}
\right) ,
\end{equation}
where $\gamma \equiv g_\psi (\partial_x \Psi - i \partial_y \Psi)/\sqrt{2m}$ and
$\eta \equiv g_\psi (\partial_x \Psi^* - i \partial_y \Psi^*)/\sqrt{2m}$.
Assuming the state of the system is 
\begin{equation}
\left(
\begin{array}{c}
e^{imt/2} g_1(t) \\ e^{-imt/2} g_2(t)
\end{array}
\right) \, ,
\end{equation}
the Schr\"odinger equation takes the form
\begin{eqnarray}
i \left(
\begin{array}{c}
\dot g_1 \\ \dot g_2
\end{array}
\right) =
\left(
\begin{array}{cc}
-\mu B + {m \over 2} & \eta \\
\eta^* & \mu B - {m\over 2}
\end{array}
\right)
\left(
\begin{array}{c}
g_1 \\ g_2
\end{array}
\right)
+ \left(
\begin{array}{cc}
0 & \gamma \, e^{-2i mt} \\
\gamma^* e^{2imt} & 0
\end{array}
\right)
\left(
\begin{array}{c}
g_1 \\ g_2
\end{array}
\right) \, .
\end{eqnarray}
At or close to resonance, $\mu B = m/2$, we can ignore the oscillatory 
second term on the right, and a general state evolves as
\begin{eqnarray}
  {1\over \sqrt{2}} 
\left(
\begin{array}{c}
(c_+ e^{-i|\eta|t} + c_- e^{i|\eta|t}) e^{+imt/2} \\
(c_+ e^{-i|\eta|t} - c_- e^{i|\eta|t}) e^{-imt/2} \eta^*/|\eta|
\end{array}
\right) \, .
\end{eqnarray}
where $c_+$ and $c_-$ are arbitrary constants. 
For illustration, choosing $c_+ = c_- = 1/\sqrt{2}$ (corresponding to
spin up at $t=0$), we find
\begin{eqnarray}
\langle \sigma_3 \rangle = {\,\rm cos\,}(2 |\eta| t) \quad , \quad
\langle \sigma_1 \rangle = - {\,\rm sin\,} (2 |\eta| t) \, {\,\rm
  sin\,}(mt + \theta_\eta) \quad , \quad \langle \sigma_2 \rangle = - {\,\rm sin\,} (2 |\eta| t) \, {\,\rm
  cos\,}(mt + \theta_\eta) \, ,
\end{eqnarray}
where $\eta \equiv |\eta| e^{i\theta_\eta}$. 
Thus, measuring the ($x$, $y$ and $z$) magnetization from a set of spins provides
information on $\eta$, related to a particular combination of
derivatives on $\Psi$.

\newpage
\section{Real scalar correlations from the random phase model}
\label{eq:axioncorrelators}
Here we collect some formulas that are useful in the analysis presented in Section~\ref{sec:experiment}. The observable quantities of interest are local correlation functions for the real scalar field, $\phi$, which is related to the non-relativistic wavefunction through the formula
\be
\phi(\vec x,t) = {1\over \sqrt{2m}} \left(\Psi(\vec x,t) e^{-i m t} + \Psi^*(\vec x,t) e^{i m t}\right).
\ee
Axion direct detection experiments are sensitive to correlation functions of operators built out of $\phi$. In order to characterize the statistics of these fluctuations, we employ the random phase model introduced in Section~\ref{sec:randomphase}. The key equation we require from the random phase model is the two-point correlation function
\be
\langle \Psi^*(\vec x, t)\Psi(\vec y, t')\rangle = \sum A_{k}^2\,\exp\left(i\vec k\cdot (\vec x-\vec y) - i\omega_k (t-t')\right).
\label{eq:randomphasepsipsiS}
\ee
Assuming a dispersion relation $\omega_k = k^2/2m$ and assuming a statistical distribution for $A_k$:
\be
A_k \propto e^{-k^2/k_0^2} ,
\ee
we can approximate the sum in~\eqref{eq:randomphasepsipsiS} by an integral to obtain
\be
\langle \Psi^*(\vec x, t)\Psi(\vec y, t')\rangle \propto k_0^3 \left(1+\frac{ik_0^2(t-t')}{4m}\right)^{-\frac{3}{2}} \exp\left(-\frac{k_0^2\lvert \vec x-\vec y\rvert^2}{8+\frac{2ik_0^2(t-t')}{m}}\right).
\label{eq:phiSphi2pt}
\ee
Here we have not kept track of the overall normalization. 
A final fact we require (which is true in the random phase model) is that the following correlators vanish
\be
\langle \Psi(\vec x, t)\Psi(\vec y, t')\rangle = \langle \Psi^*(\vec x, t)\Psi^*(\vec y, t')\rangle = 0.
\ee
By combining these formulas, we can now compute the statistics of operators built out of the field $\phi$. For example, we have
\be
\langle \phi(\vec x, t)\phi(\vec y, t')\rangle = \frac{1}{2m}\langle \Psi^*(\vec x, t)\Psi(\vec y, t')\rangle e^{im(t-t')}+ \frac{1}{2m}\langle \Psi(\vec x, t)\Psi^*(\vec y, t')\rangle e^{-im(t-t')}.
\label{eq:phixtphiytp}
\ee
One can plug~\eqref{eq:phiSphi2pt} into this formula to get an explicit expression. Note that when we set $\vec x = \vec y$ and $t=t'$, this becomes the density one-point function:
\be
m^2\langle \phi^2\rangle = \bar\rho \propto m\,k_0^3.
\ee
The correlator~\eqref{eq:phixtphiytp} is challenging to measure because of the overall oscillatory factors, which vary on a timescale of order $m^{-1}$. In order to overcome this difficulty, we can consider the two-point function of the composite operator $\phi^2$, subtracting off the average of $\phi^2$ as
\be
\langle \phi(\vec x,t)^2\phi(\vec y,t')^2 \rangle -\langle\phi^2\rangle^2= \frac{1}{m^2}\left\lvert\langle \Psi^*(\vec x, t)\Psi(\vec y, t')\rangle\right\rvert^2+\frac{1}{2m^2}\left(\langle \Psi^*(\vec x, t)\Psi(\vec y, t')\rangle e^{2im(t-t')}+{\rm c.c.}\right),
\ee
where ${\rm c.c.}$ means the complex conjugate of the last term.
Plugging in~\eqref{eq:phiSphi2pt} we get
\be
\begin{aligned}
\frac{\langle \phi(\vec x,t)^2\phi(\vec y,t')^2 \rangle}{\langle\phi^2\rangle^2} -1  =~&\left(1+\frac{k_0^4 (t-t')^2}{16m^2}\right)^{-3/2} \exp\left(-\frac{4k_0^2m^2\lvert \vec x-\vec y\rvert^2}{16m^2+k_0^4(t-t')^2}\right)\\
~&+\left[\frac{32m^3 e^{-2im(t-t')}}{[4m-ik_0(t-t')]^3}\exp\left(-\frac{imk_0^2\lvert\vec x-\vec y\rvert^2}{4im+k_0^2(t-t')}\right)+{\rm c.c.}\right].
\end{aligned}
\ee
The second line of this correlator oscillates with frequency $m^{-1}$, but if we average over times longer than this (but still small compared to the de Broglie time) we can isolate the first line, which is responsible for the behaviors discussed in Section~\ref{sec:experiment}.

\renewcommand{\em}{}
\bibliographystyle{utphys}
\addcontentsline{toc}{section}{References}
\bibliography{references}

\end{document}